\documentclass[citeautoscript,floatfix,aps,prx,twocolumn,superscriptaddress,longbibliography]{revtex4-1}  %
\usepackage{graphicx}  %
\usepackage{dcolumn}   %
\usepackage{bm}        %
\usepackage{amssymb}   %
\usepackage{amsmath}
\usepackage{amsthm}
\newtheorem{theorem}{Theorem}

\newtheorem{remark}{Remark}
\newtheorem{definition}{Definition}

\usepackage{braket}
\usepackage{amsfonts}
\usepackage{multirow}
\usepackage{dsfont}
\usepackage{hyperref}
\usepackage{xcolor}
\usepackage{subcaption}
\usepackage{latexsym}

\usepackage{pgf}
\newcommand*{\Parallelogramm}[1][]{%
	\pgfpicture\pgfsetroundjoin
	\pgftransformxslant{.6}%
	\pgfpathrectangle{\pgfpointorigin}{\pgfpoint{.50em}{.50em}}%
	\pgfusepath{stroke,#1}%
	\endpgfpicture}

\usepackage{stmaryrd}
\usepackage{tensor}
\usepackage{tikz}
\usepackage{tikz-network}
\usepackage{fusioncategory}

\usepackage{fp}
\usepackage{QCAdrawing}

\newcommand{\id}{\mathrm{id}}

\newcommand{\C}{\mathbb{C}}

\newcommand{\A}{\mathcal{A}}

\newcommand{\Tr}{\mathrm{Tr}}
\newcommand{\Hil}{\mathcal{H}}
\newcommand{\blhd}{\blacktriangleleft}
\newcommand{\brhd}{\blacktriangleright}

\newcommand{\bowtiewb}{\mathrel{\rhd\mkern-10mu\blhd}}

\newcommand{\bltimes}{\mathrel{\brhd\mkern-5mu<}}

\newcommand{\mytriangledown}{\mathbin{\text{\rotatebox[origin=c]{180}{$\triangle$}}}}
\renewcommand{\Diamond}{\mathbin{\text{\rotatebox[origin=c]{45}{$\square$}}}}

\newcommand{\deltad}{\tilde{\delta}} 
\newcommand{\deltak}{\delta}
\newcommand{\N}{L}
\newcommand{\NN}{N}
\begin{document}

\title{Hopf algebras and solvable unitary circuits}
\author{Zhiyuan Wang}
\affiliation{Max-Planck-Institut f{\"{u}}r Quantenoptik, Hans-Kopfermann-Str. 1, 85748 Garching, Germany}
\date{\today}

\begin{abstract}
Exactly solvable models in quantum many body dynamics provide
valuable insights into many interesting physical phenomena, and serve as
platforms to rigorously investigate fundamental theoretical questions.
Nevertheless, they are extremely rare and existing solvable models and
solution techniques have serious limitations. In this paper we introduce a new
family of exactly solvable unitary circuits which model quantum many body
dynamics in discrete space and time. Unlike many previous solvable models, one can
exactly compute the full quantum dynamics initialized from any matrix product
state in this new family of models. The time evolution of local observables
and correlations, the linear growth of Renyi entanglement entropy,
spatiotemporal correlations, and out-of-time-order correlations are all exactly
computable. 
A key property of these models enabling the exact solution is that any time evolved local operator is an exact matrix product operator with finite bond dimension, even at arbitrarily long time,  
which we prove using the underlying $\C^*$-(weak) Hopf algebra
structure along with tensor network techniques. %
We lay down the general framework for the construction and solution of this family of models, and give several explicit examples.
In particular, we study in detail  a model constructed out of a $\C^*$-weak Hopf algebra that is very close to a floquet version of the PXP model, and the exact results we obtain may shed light on  the phenomenon of quantum many body scars, and more generally, floquet quantum dynamics in constrained systems.  
\end{abstract}

\maketitle

\section{Introduction}\label{sec:intro}
Quantum many body dynamics is an exciting area of research~\cite{eisert2015quantum} that hosts a plethora of novel physical phenomena, involving both fundamental theoretical interests and practical applications.
A fundamental theme %
in this area concerns quantum thermalization~\cite{nandkishore:many_2015,DAlessio2016ChaosETH,deutsch2018eigenstate,RMP2019MBL}, with the goal of understanding how generic isolated quantum systems thermalize at long time, and why some special classes of quantum systems fail to reach thermal equilibrium, such as integrable systems~\cite{Rigol2007,Prosen2011OpenXXZ,Pozsgay2014Corr,Ilievski2015Complete,Langen2015GGEExp,Vidmar2016,Alvaredo2016Emergent,Vasseur2016Nonequilibrium}, many-body localized phases~\cite{Prosen2008Many, nandkishore:many_2015,RMP2019MBL}, quantum many body scars~\cite{Turner2018QMBS,choi2019emergent,Khemani2019Signature,Serbyn2021QMBS,Moudgalya2022QMBS,chandran2023QMBS}, and systems with Hilbert space fragmentation~\cite{Sala2020HSF,Khemani2020Localization,Moudgalya2022QMBS}. Equally important is to understand universal behaviors in the process of quantum thermalization, with interesting questions concerning the growth of correlation and entanglement~\cite{bravyi2006,Nahum2017EntanglementGrowth,Keyserlingk2018OperatorHydrodynamics,Bertini2019Entanglement,Foligno2023Growth}, quantum information spreading~\cite{bravyi2006,Foligno2024QIS}, and quantum chaos~\cite{Bertini2018SFF,Kos2018MBQC,Bertini2020OE,Prosen2020Chao,dowling2023scrambling}.  

A particularly interesting type of quantum many body dynamics is floquet quantum dynamics~\cite{Moessner2017,Eckardt2017RMP}, i.e. quantum systems subject to a periodic drive.  
Floquet quantum dynamics are natural to realize in experiment~\cite{Zhang2017,Choi2017ObservationTC,bordia2017periodically}, and host a wealth of novel physical phenomenon. 
Intriguingly, such quantum systems can host  novel dynamical phases that are absent in equilibrium, such as  floquet topological insulator~\cite{Lindner2011Nphys,Gu2011Floquet,Rudner2013Anomolous,nathan2015topological},
floquet time crystals~\cite{Else2016FTC,Else2017Prethermal,Ho2017Critical,Yao2017Discrete,Choi2017ObservationTC,Zhang2017},
and many-body localized phases that are exclusive to non-equilibrium~\cite{Khemani2016Phase,Keyserlingk2016FloquetMBL,bordia2017periodically}. %

Unfortunately, quantum many body dynamics is notoriously hard. A straightforward numerical approach by brute force exact diagonalization is hindered by the exponential growth of Hilbert space dimension, and is therefore limited to finite systems with small sizes, whose dynamics can qualitatively deviate from the thermodynamic limit at long time due to finite size errors~\cite{Wang2021Bounding}. 
Some tensor network algorithms~\cite{white2004real,schollwoeck:density_2011} %
can simulate much larger system sizes or even directly simulate the thermodynamic limit, however, in general, it is still hard to simulate long time dynamics with tensor network algorithms due to the linear growth of entanglement entropy, and the time complexity of these algorithms typically grow exponentially in evolution time.  %
There are also various approximate analytic methods  for specific types of systems, but they come with no guarantee and it is generally hard to rigorously study the long time dynamics of strongly interacting quantum many body systems. 

Exactly solvable models are therefore extremely valuable in that they provide viable platforms to rigorously investigate fundamental theoretical questions as well as benchmarking numerical algorithms and approximation techniques. A particularly prominent family of solvable quantum many body dynamics is 
solvable unitary circuits generated by local quantum gates, %
which can be viewed as models of quantum many body dynamics in discretized space and time.  
To date, there are quite a few families of solvable unitary circuits, such as Clifford circuits, free-fermion solvable circuits~\cite{Valiant2002Matchgate,Terhal2002Matchgate,Jozsa2008Matchgate,vona2024exact}, Yang-Baxter integrable circuits~\cite{Vanicat2018Integrable,Gombor2021Integrable,hubner2024generalized}, the quantum rule 54~\cite{Friedman2019Rule54,Klobas2019TDMPA,Klobas2021Rule54,Buca2021Rule54,Alba2019Rule54LOMPO}, 
dual unitary circuits~\cite{Prosen2019DU,Piroli2020DUinitialMPS,Bertini2020OE,Claeys2021Ergodic,suzuki2022computational,borsi2022construction,Foligno2023Growth} and their generalizations~\cite{jonay2021triunitary,yu2024hierarchical,Foligno2024QIS}, and random unitary circuits~\cite{brandao2016local,Nahum2017EntanglementGrowth,Nahum2018RUC,khemani2018operator}. 
However, some of these families of models have rather limited solvability, along  with other kinds of limitations. For example,
physical quantities in quench dynamics generated by Yang-Baxter integrable circuits %
are often hard to compute except in the limit of large time and system size. Solving quench dynamics in dual unitary circuits requires special initial states~\cite{Piroli2020DUinitialMPS} which generally do not include the experimentally  most relevant class of product states. 
Clifford circuits and free fermion solvable models do not display the universal behavior of generic interacting systems, and it is often hard to exactly compute the growth of entanglement entropy in these models. 
Random unitary circuits display generic chaotic behaviors, but the solution requires averaging over disorder, and therefore cannot be used to accurately predict the behavior of specific systems without disorder. 

In this paper we introduce a family of solvable unitary circuits with examples that lie beyond all previously known families of solvable models. %
The techniques we use to construct and solve this family of models %
are very different from existing techniques in this area, and our techniques are motivated by the deep connection between Hopf algebras and  tensor networks that was recently introduced to study topologically-ordered tensor network states~\cite{molnar2022matrix,Ruiz-de-Alarcon2024MPOAHopfII}. %
Starting from an arbitrary  finite dimensional $\C^*$-(weak) Hopf algebra, we define a two qudit unitary gate $\hat{U}$ using algebraic data, and show that the full evolution dynamics of the brickwork unitary circuit generated by $\hat{U}$ can be exactly solved for any initial matrix product state~(MPS). The exact solution is enabled by the underlying Hopf algebra structure which reduces a two dimensional tensor network problem into a one dimensional problem.  Such a dimensional reduction is similar in spirit to the exact computation of spatiotemporal correlations in dual unitary circuits, but enabled by a very different technique. In our models, time evolved local observables in the Heisenberg picture are exact matrix product operators~(MPOs) with finite bond dimension~(even at arbitrary long times), and consequently their expectation values and correlation functions can all be computed exactly. Renyi entanglement entropy can also be exactly computed for any subsystem size and any time, and exhibits a typical linear growth as in generic quantum systems. Remarkably, our models are solvable even in finite systems with periodic boundary condition~(PBC), a feature that is absent in most known integrable circuits and dual unitary circuits. 

In our construction, solvable gates constructed from  $\C^*$-weak Hopf algebras generally describe floquet quantum dynamics in locally constrained systems. Quantum dynamics in constrained systems have recently gained tremendous attention since they display rich collective behaviors and are natural to realize in experiments, for example in Rydberg atom arrays~\cite{bernien2017probing,Bluvstein2021Controlling}. Prominent examples are the floquet versions~\cite{wilkinson2020exact,giudici2023unraveling,Vijay2020Nonergodic,mizuta2020exact,Hudomal2022QMBSPXP,Sellapillay2022EntanglementDynamicsQCA} of the PXP model~\cite{Turner2018QMBS,Serbyn2021QMBS}, which models the floquet quantum dynamics in systems with Rydberg-blockade-type constraint, and has been used as a platform to investigate quantum many body scars. This model includes a special  point which is integrable and deterministic called the RCA201~\cite{wilkinson2020exact}, and its solution has shed light on the physical origin of many body scar states in the PXP model~\cite{giudici2023unraveling}. In this paper we construct and study a constrained solvable floquet dynamics using our construction, which can be mapped to a model that is very close to the integrable floquet PXP model, yet our model is non-deterministic~(purely quantum) and still allows a full exact solution.  

Our paper is organized as follows. In Sec.~\ref{sec:general_formalism} we present the fundamental building block of our construction and show that every finite dimensional bialgebra defines a family of solvable 2D tensor network. In Sec.~\ref{sec:solvableQCA} we specialize to the case of finite dimensional $\C^*$-(weak) Hopf algebras and show that our construction leads to solvable unitary circuits, and give the exact 1D tensor network representation for physical quantities.  
We then present several explicit examples of solvable unitary circuits constructed out of finite dimensional $\C^*$-Hopf algebras~(Sec.~\ref{sec:HAexamples}) and  $\C^*$-weak Hopf algebras~(Sec.~\ref{sec:GoldenModel}), and present the exact results for physical quantities. 
In Sec.~\ref{sec:concl} we conclude our work. In the appendices we provide the mathematical foundation of our construction and fill in some technical details of the main text. In Apps.~\ref{app:BAandTN}-\ref{app:WHA} we review the basic concepts of bialgebras, Hopf algebras, and weak Hopf algebras, and prove the main theorems in Secs.~\ref{sec:general_formalism} and \ref{sec:solvableQCA}, and also present the algebraic structures underlying the explicit examples in Secs.~\ref{sec:HAexamples} and \ref{sec:GoldenModel}. %
In App.~\ref{app:TMRenyi} we give the transfer matrices  for the computation of the  Renyi entanglement entropy  in Sec.~\ref{sec:RenyiEE}. In App.~\ref{app:PBCevol} we derive the PBC evolution operator which we use in the PBC solution in Sec.~\ref{sec:PBCsolution}. 

\section{Bialgebras and solvable tensor networks}\label{sec:general_formalism}
In this section we show that every finite dimensional bialgebra defines a family of solvable 2D tensor network, which is the foundation for the solvable unitary circuits we introduce in Sec.~\ref{sec:solvableQCA}.  %
We construct the solvable 2D tensor network  using data obtained from the algebraic structure and derive an efficient MPO representation for it. 

The fundamental building block of our construction are three four-index tensors 
$
\begin{tikzpicture}[baseline={([yshift=-0.5ex]current bounding box.center)}, scale=0.63]
	\Wgatered{0}{0}
\end{tikzpicture}$,
$
\begin{tikzpicture}[baseline={([yshift=-0.5ex]current bounding box.center)}, scale=0.63]
	\rhotensor{0}{0}%
\end{tikzpicture}$, 
$
\begin{tikzpicture}[baseline={([yshift=-0.5ex]current bounding box.center)}, scale=0.63]
	\vtensor{0}{0}%
\end{tikzpicture}$, and	vectors $\begin{tikzpicture}[baseline={([yshift=-0.5ex]current bounding box.center)}, scale=0.63]
\unittensor{0}{0}
\end{tikzpicture}$, $\begin{tikzpicture}[baseline={([yshift=-0.5ex]current bounding box.center)}, scale=0.63]
\counittensor{0}{0}
\end{tikzpicture}$
satisfying the following diagrammatic equations
\begin{eqnarray}\label{eq:MUpentagon}
	&&\begin{tikzpicture}[baseline={([yshift=-0.5ex]current bounding box.center)}, scale=0.63]
		\rhotensor{-0.5}{1}%
		\vtensor{0.5}{1}%
		\hindices{{a,i}}{-0.5}{1.5}{above}
		\hindices{{b,j}}{-0.5}{0.5}{below}
		\paraindices{-0.5}{1}{y}{}
		\paraindices{0.5}{1}{}{x}
	\end{tikzpicture}	
	=
	\begin{tikzpicture}[baseline={([yshift=-0.5ex]current bounding box.center)}, scale=0.63]
		\rhotensor{0.5}{1}%
		\vtensor{-0.5}{1}%
		\paraindices{-0.5}{1}{y}{}
		\paraindices{0.5}{1}{}{x}
		\Wgatered{0}{0}
		\hindices{{i,a}}{-0.5}{1.5}{above}
		\hindices{{b,j}}{-0.5}{-0.5}{below}
	\end{tikzpicture},\\
	&&\begin{tikzpicture}[baseline={([yshift=-0.5ex]current bounding box.center)}, scale=0.63]
		\MYsquareB{-0.5}{0}
		\vtensor{0}{0}%
	\end{tikzpicture}=
	\begin{tikzpicture}[baseline={([yshift=-0.5ex]current bounding box.center)}, scale=0.63]
		\deltatensor{0}{0}{\colorR}{}{}
		\MYsquareB{0.3}{0}
		\draw[ultra thick] (0.35, 0 ) -- (0.7,0);
	\end{tikzpicture}, \quad
	\begin{tikzpicture}[baseline={([yshift=-0.5ex]current bounding box.center)}, scale=0.63]
		\rhotensor{0}{0}%
		\MYsquare{0.5}{0}
	\end{tikzpicture}=
	\begin{tikzpicture}[baseline={([yshift=-0.5ex]current bounding box.center)}, scale=0.63]
		\deltatensor{0}{0}{\colorL}{}{}
		\draw[ultra thick] (-0.35, 0 ) -- (-0.7,0);
				\MYsquare{-0.3}{0}
	\end{tikzpicture},\quad 
	\begin{tikzpicture}[baseline={([yshift=-0.5ex]current bounding box.center)}, scale=0.63]
	\draw[ultra thick] (-0.25, 0 ) -- (0.25,0);
	\MYsquareB{-0.25}{0}
	\MYsquare{0.25}{0}
	\end{tikzpicture}=1.\nonumber %
\end{eqnarray}
Here we use standard tensor network notations where  a contracted leg indicates a summation over the corresponding index. The indices $a,b$ take values in $\{1,2\ldots d_\rho\}$, the indices $i,j$ take values in $\{1,2\ldots d_v\}$, and the indices $x,y$ take values in $\{1,2\ldots d_\A\}$,
where $d_\rho,d_v$ and $d_\A$ are integers to be defined soon. 

Eq.~\eqref{eq:MUpentagon} is the starting point for all the rest of the derivations, including  the construction and solution of the 2D tensor network and calculation of physical observables. %
The key and the only role played by the algebraic structures in our construction is that they provide a systematic way of finding solutions to Eq.~\eqref{eq:MUpentagon}, 
as stated in the following theorem:
\begin{theorem}\label{thm:bialgebraTN}
Let $\A$ be a finite dimensional bialgebra, and let $\rho$ be a $d_\rho$-dimensional representation of $\A$ and $\{v_{ij}\in \A\}_{1\leq i,j\leq d_v}$ be a $d_v$-dimensional corepresentation~($d_\rho$ and $d_v$ are positive integers).
We use $\rho_{ab}(x)$ to denote the matrix elements of $\rho(x)$, where $1\leq a,b\leq d_\rho$ and $x\in\A$.
Such a triple $\{\A,\rho,v\}$ of algebraic data automatically generates a solution to Eq.~\eqref{eq:MUpentagon} via
\begin{eqnarray}\label{def:Urhovbialg}
	\begin{tikzpicture}[baseline={([yshift=-0.5ex]current bounding box.center)}, scale=0.63]
		\WgateredInd{0}{0}{$i$}{$a$}{$b$}{$j$}
	\end{tikzpicture}&=&\rho_{ab}(v_{ij}),\nonumber\\
\begin{tikzpicture}[baseline={([yshift=-0.5ex]current bounding box.center)}, scale=0.63]
	\rhotensor{0}{0}%
	\paraindices{0}{0}{y}{x}
	\quantumindices{0}{0}{a}{b}
\end{tikzpicture}&=&(\deltad_y\otimes\rho_{ab} )\Delta(x), \nonumber\\ %
\begin{tikzpicture}[baseline={([yshift=-0.5ex]current bounding box.center)}, scale=0.63]
	\vtensor{0}{0}%
	\paraindices{0}{0}{y}{x}
	\quantumindices{0}{0}{i}{j}
\end{tikzpicture}&=&\deltad_y(x\cdot v_{ij}),\nonumber\\ %
	 \begin{tikzpicture}[baseline={([yshift=-0.5ex]current bounding box.center)}, scale=0.63]
		\unittensor{0}{0}
		\node at (-0.3,0) [left] {\footnotesize $x$};
	\end{tikzpicture}&=&\deltad_x(1_\A), \quad
	\begin{tikzpicture}[baseline={([yshift=-0.5ex]current bounding box.center)}, scale=0.63]
		\counittensor{0}{0}
		\node at (0.3,0) [right] {\footnotesize $x$};
	\end{tikzpicture}=\epsilon(x).
\end{eqnarray}
where $\cdot$ denotes the multiplication of $\A$, $\Delta$ the comultiplication of $\A$, $1_\A,\epsilon$ are the unit and counit of $\A$, respectively.  $x,y\in \A$ denote basis elements of $\A$~(so the dimension of the indices $x,y$ are equal to the dimension $d_\A$ of the algebra $\A$), and $\deltad_x,\deltad_y$ denotes the corresponding dual basis element of the dual vector space $\A^*$ satisfying $\deltad_x(y)=\deltak_{x,y}$.
\end{theorem}
We prove this theorem in App.~\ref{app:BAandTN}, where we also review the necessary basic concepts of bialgebras and their (co)representations. 

Once we have a solution to  Eq.~\eqref{eq:MUpentagon}, we no longer need the bialgebra structure, since all the following derivations only make use of Eq.~\eqref{eq:MUpentagon} along with elementary tensor network manipulations. 
Of course, one can use alternative methods to find solutions to Eq.~\eqref{eq:MUpentagon}, such as parametrized ansatz or numerical techniques, and equally well obtain solvable tensor networks~\footnote{However, we conjecture that the converse of Thm.~\ref{thm:bialgebraTN} is also true:  any solution to Eq.~(\ref{eq:MUpentagon}) must be constructable from a bialgebra via Eq.~(\ref{def:Urhovbialg}). 
In other words, finite dimensional bialgebras generate all possible solutions to  Eq.~(\ref{eq:MUpentagon}).}. 
In the following we assume that a solution to Eq.~\eqref{eq:MUpentagon} is already given and use it to construct solvable tensor networks.  
Specifically, we show that the 2D tensor network generated by the four index tensor 
\begin{tikzpicture}[baseline={([yshift=-0.5ex]current bounding box.center)}, scale=0.63]
\Wgatered{0}{0}
\end{tikzpicture}
has an efficient MPO representation, 
as expressed by the following equation
\begin{eqnarray}\label{eq:triangularTNMPO}
[\hat{U}^{(n)}_{\triangle}]^{\mathbf{a},\mathbf{i}}_{\mathbf{b},\mathbf{j}}
&:=&
\begin{tikzpicture}[baseline={([yshift=-0.5ex]current bounding box.center)}, scale=0.63]
		\lowertriangularTN{-1}{1}{\Wgatered}{4}
		\legindices{{i_1,i_2,\ldots,i_n}}{-3.5}{-2.5}{1}{1}{above left=-0.1}
		\legindices{{a_1,a_2,\ldots,a_n}}{0.5}{0.5}{1}{-1}{above right=-0.1}
		\hindices{{b_1,j_1,b_2,j_2,\ldots,\ldots,b_n,j_n}}{-3.5}{-3.5}{below}
	\end{tikzpicture}\nonumber\\
	&=&
	\begin{tikzpicture}[baseline={([yshift=-0.5ex]current bounding box.center)}, scale=0.63]
\manyrhov{-3.5}{-3}{4}
		\MYsquare{4}{-3}
		\MYsquareB{-4}{-3}
		\hindices{{a_1,i_1,a_2,i_2,\ldots,\ldots,a_n,i_n}}{-3.5}{-2.5}{above}%
		\hindices{{b_1,j_1,b_2,j_2,\ldots,\ldots,b_n,j_n}}{-3.5}{-3.5}{below}%
	\end{tikzpicture},
\end{eqnarray}
where we use bold $\mathbf{a}$ to collectively denote the indices $\{a_1,a_2,\ldots,a_n\}$, and similarly for $\mathbf{b},\mathbf{i},\mathbf{j}$. 
Eq.~\eqref{eq:triangularTNMPO} can be proved using elementary tensor network manipulations by repeatedly applying Eq.~\eqref{eq:MUpentagon}:
\begin{eqnarray}\label{eq:triangularTNMPOderive}
	&&\begin{tikzpicture}[baseline={([yshift=-0.5ex]current bounding box.center)}, scale=0.63]
		\rhotriangB{-3.5}{-3}
		\vtriangB{3.5}{-3}
		\manyvrho{-2.5}{-3}{3}
		\hindices{{a_1,i_1,a_2,i_2,\ldots,\ldots,a_n,i_n}}{-3.5}{-2.5}{above}%
		\hindices{{b_1,j_1,b_2,j_2,\ldots,\ldots,b_n,j_n}}{-3.5}{-3.5}{below}%
	\end{tikzpicture}\nonumber\\
	&=&
	\begin{tikzpicture}[baseline={([yshift=-0.5ex]current bounding box.center)}, scale=0.63]
		\rhotriangB{-2.5}{-4}
		\manyvrho{-1.5}{-4}{2}
		\vtriangB{2.5}{-4}
		\foreach \i in {0,...,3}{\Wgatered{2*\i-3}{-4-1}}
		\hindices{{a_1,i_2,a_2,\ldots,\ldots,i_n}}{-2.5}{-3.5}{above}%
		\hindices{{i_1, , , , , , ,a_n}}{-3.5}{-4.5}{above}
		\hindices{{b_1,j_1,b_2,j_2,\ldots,\ldots,b_n,j_n}}{-3.5}{-5.5}{below}
	\end{tikzpicture}\nonumber\\
	&=&
	\begin{tikzpicture}[baseline={([yshift=-0.5ex]current bounding box.center)}, scale=0.63]
		\rhotriangB{-1.5}{-4}
		\manyvrho{-0.5}{-4}{1}
		\vtriangB{1.5}{-4}
		\foreach \i in {0,...,2}{\Wgatered{2*\i-2}{-5}}
		\foreach \i in {0,...,3}{\Wgatered{2*\i-3}{-6}}
		\hindices{{a_1,\ldots,\ldots,i_n}}{-1.5}{-3.5}{above}
		\hindices{{i_1~, , , , , , ,~~~a_n}}{-3.5}{-5.5}{above}
		\hindices{{i_{2}~, , , , ,~~~a_{n-1}}}{-2.5}{-4.5}{above}
		\hindices{{b_1,j_1,b_2,j_2,\ldots,\ldots,b_n,j_n}}{-3.5}{-6.5}{below}
	\end{tikzpicture}\nonumber\\
	&=&\ldots\nonumber\\
	&=&
	\begin{tikzpicture}[baseline={([yshift=-0.5ex]current bounding box.center)}, scale=0.63]
		\lowertriangularTN{-1}{1}{\Wgatered}{4}
		\legindices{{i_1,i_2,\ldots,i_n}}{-3.5}{-2.5}{1}{1}{above left=-0.1}
		\legindices{{a_1,a_2,\ldots,a_n}}{0.5}{0.5}{1}{-1}{above right=-0.1}
		\hindices{{b_1,j_1,b_2,j_2,\ldots,\ldots,b_n,j_n}}{-3.5}{-3.5}{below}
	\end{tikzpicture}
\end{eqnarray}
where we use the first relation in Eq.~\eqref{eq:MUpentagon} whenever we swap a $\rho$ tensor and a $v$ tensor, and in each step we apply the first two relations in the second line of Eq.~\eqref{eq:MUpentagon} at the left and right boundaries, and in the final step we use $%
\begin{tikzpicture}[baseline={([yshift=-0.5ex]current bounding box.center)}, scale=0.63]
	\draw[ultra thick] (-0.25, 0 ) -- (0.25,0);
	\MYsquareB{-0.25}{0}
	\MYsquare{0.25}{0}
\end{tikzpicture}=1$. This completes the  proof of Eq.~\eqref{eq:triangularTNMPO}. %

Eq.~\eqref{eq:triangularTNMPO} is the foundation for the exact solution of the unitary circuit we construct in the next section, since it reduces a hard 2D tensor network problem into a 1D MPO problem with finite bond dimension $d_\A$, which can be computed efficiently~\cite{schollwoeck:density_2011,Cirac2021Matrix}. Such a dimensional reduction through tensor network manipulations is similar in spirit to dual unitary circuits~\cite{Prosen2019DU,Piroli2020DUinitialMPS} and their generalizations~\cite{jonay2021triunitary,Piroli2022Temporal,yu2024hierarchical,Foligno2024QIS,wang2024exact}; however, in our case, this reduction is enabled by the underlying  bialgebra structure of the elementary tensors, which allows much more solvability, as we will see in the next section. We also mention here that the MPO representation in Eq.~\eqref{eq:triangularTNMPO} can be generalized to 2D tensor networks of several different shapes, including a diamond shape~(App.~\ref{app:squareshape}) and an inverted triangle shape~(App.~\ref{app:invertedtriangleshape}), both of which are useful for the exact solution of our model in PBC, which we present in Sec.~\ref{sec:PBCsolution}. 
Finally, we remark that Thm.~\ref{thm:bialgebraTN} can be generalized to prebialgebras~(which are generalizations of bialgebras), which is important for the construction of solvable unitary circuits based on weak Hopf algebras in Sec.~\ref{sec:solvableQCA}. The tensors constructed from a prebialgebra via Eq.~\eqref{def:Urhovbialg} only satisfy a weaker version of Eq.~\eqref{eq:MUpentagon}~[see Eq.~\eqref{eq:MUpentagon-PreBA} in App.~\ref{app:SolvableTNfrompreBA}], however, the MPO representation of the 2D tensor network in Eq.~\eqref{eq:triangularTNMPO} and its derivation in Eq.~\eqref{eq:triangularTNMPOderive} are still valid. We present the details of this generalization in App.~\ref{app:BAandTN}.

\section{$\C^*$-(weak) Hopf algebras and solvable unitary circuits}\label{sec:solvableQCA}
In the previous section we showed how every finite dimensional (pre)bialgebra defines a solvable 2D tensor network. In this section we specialize to the case of $\C^*$-(weak) Hopf algebras, and we will see in the following that in this case our construction gives rise to %
a 1+1D solvable unitary circuit. Specifically, we first define the brickwork unitary circuit in Sec.~\ref{sec:circuitdef},  then  we derive an exact MPO representation for time evolved local observables in Sec.~\ref{sec:LOMPO} and use this MPO representation to compute their expectation values in quench dynamics in Sec.~\ref{sec:quench_obs}. In Sec.~\ref{sec:RenyiEE} we show how to exactly compute the Renyi entanglement entropy  using an exact 1D tensor network representation of the reduced density matrix. In Sec.~\ref{sec:st_corr_func} mention some other solvable physical quantities. Finally, in Sec.~\ref{sec:PBCsolution} we show how to exactly solve the dynamics in PBC. 

\subsection{Circuit construction}\label{sec:circuitdef}
We begin with the following theorem
\begin{theorem}\label{thm:HopfUgate}
	Let $\A$ be a finite dimensional $\C^*$-Hopf algebra, $\rho$ a finite dimensional $*$-representation of $\A$ and $v$ is a finite-dimensional unitary corepresentation. 
	Then the tensor~[Eq.~\eqref{def:Urhovbialg}] $\rho_{ab}(v_{ij})=
	\begin{tikzpicture}[baseline={([yshift=-0.5ex]current bounding box.center)}, scale=0.63]
		\WgateredInd{0}{0}{$i$}{$a$}{$b$}{$j$}
	\end{tikzpicture}$ constructed from the triple $\{\A,\rho,v\}$ is unitary if we group the indices $(b,j)$ as input indices and $(a,i)$ as output indices.
\end{theorem}

We denote the representation space as $\Hil_\rho$ and the corepresentation space as $\Hil_v$, and define a unitary map $\hat{U}: \Hil_\rho\otimes \Hil_v\to \Hil_v\otimes \Hil_\rho$ through its matrix elements
\begin{equation}\label{eq:gate_def}
	\langle i,a |\hat{U}|b,j\rangle\equiv
	\begin{tikzpicture}[baseline={([yshift=-0.5ex]current bounding box.center)}, scale=0.63]
		\WgateredInd{0}{0}{$i$}{$a$}{$b$}{$j$}
	\end{tikzpicture}. %
\end{equation}
In the rest of this paper, we call a blue leg in the northwest-southeast direction a $v$-leg, labeled by indices $i,j$ with dimension $d_v$, and we call a red leg in the northeast-southwest direction a $\rho$-leg, labeled by indices $a,b$ with dimension $d_\rho$. 
Since our main motivation of this paper is to construct solvable unitary circuits as models of quantum many body dynamics in discrete space and time, we will mainly focus on the case $d_\rho=d_v=d$, and we identify $\Hil_\rho\cong\Hil_v=\Hil$. In this case $\hat{U}$ can be viewed as a local quantum gate acting on two neighboring qudits. 
However, in principle, Thm.~\ref{thm:HopfUgate} and most of the main results in this paper apply to arbitrary $d_\rho$ and $d_v$. 

Using tensor network notation, we can express unitarity of $\hat{U}$ graphically as
\begin{equation}\label{eq:Unitarygraphical}
	\begin{tikzpicture}[baseline={([yshift=-0.5ex]current bounding box.center)}, scale=0.63]
		\WgateblueInv{0}{1}
		\Wgatered{0}{0}
	\end{tikzpicture}=	
	\begin{tikzpicture}[baseline={([yshift=-0.5ex]current bounding box.center)}, scale=0.63]
		\draw[very thick, draw=red] (0,-1) -- (0,+1);
		\draw[very thick, draw=blue] (1,-1) -- (1,+1);
	\end{tikzpicture}.
\end{equation}
This unitary gate generates a 1+1D brickwork unitary circuit whose unitary evolution operator is $\hat{\mathbf{U}}(t)=\hat{\mathbf{U}}^t$, where  
\begin{eqnarray}\label{eq:Uevolutionoperator}
	\hat{\mathbf{U}}&=&\hat{\mathbf{U}}_e \hat{\mathbf{U}}_o\equiv\prod^\N_{j=1} \hat{U}_{j,j+1/2}\prod^\N_{j=1} \hat{U}_{j-1/2,j}\\
	&=&
	\begin{tikzpicture}[baseline={([yshift=-0.5ex]current bounding box.center)}, scale=0.63]
		\foreach \i in {0,...,4}
		{\foreach \j in {0}
			{\Wgatered{2*\i}{2*\j}}}
		\foreach \i in {0,...,4}
		{\foreach \j in {0}
			{\Wgatered{2*\i+1}{2*\j+1}}}
	\end{tikzpicture}\nonumber
\end{eqnarray}
is the evolution operator for one period, defined on a %
$d^{2\N}$-dimensional Hilbert space $ \Hil^{\otimes 2\N}$ of a 1D chain of $2\N$ qudits, and we assume PBC.  %
In the following we will use the formalism presented in the previous section to solve this unitary circuit exactly, and calculate various physical quantities including local observables and their correlation functions, and Renyi entanglement entropy. In most part of this paper, we consider the thermodynamic $\N\to \infty$, the only exception is in Sec.~\ref{sec:PBCsolution}, where we study the dynamics of a finite chain with PBC. 

Actually, an important theorem in Hopf algebra theory~\cite{LarsonRadford1988} implies that the gate constructed from finite-dimensional $\C^*$-Hopf algebras, as defined in  Eq.~\eqref{eq:gate_def}, is always a dual-unitary gate. Although certain properties of dual unitary dynamics are already known to be exactly computable, such as spatiotemporal correlation functions~\cite{Prosen2019DU} and spectral form factor~\cite{Bertini2018SFF}, important physical quantities in quench dynamics~(such as local observables, correlation functions, and  Renyi entanglement entropy) are only known to be computable for very special initial MPS~\cite{Piroli2020DUinitialMPS}. %
In particular, generic dual unitary dynamics is not known to be solvable for the most experimentally relevant case of initial product states.
By contrast, for dual unitary gates constructed from finite dimensional $\C^*$-Hopf algebras, all the aforementioned quantities are exactly computable for any initial MPS. Therefore finite dimensional $\C^*$-Hopf algebras define a special class of dual unitary circuits that are much more solvable than a generic dual unitary circuit. 

Exactly solvable unitary circuits that are not dual unitary can be constructed by generalizing the above construction
to finite-dimensional $\C^*$-weak Hopf algebras, with only one additional technical subtlety.
If $\A$ in Thm.~\ref{thm:HopfUgate} is only a $\C^*$-weak Hopf algebra, then the tensor 
\begin{tikzpicture}[baseline={([yshift=-0.5ex]current bounding box.center)}, scale=0.63]
	\Wgatered{0}{0}
\end{tikzpicture} constructed from Eq.~\eqref{def:Urhovbialg} is generally not unitary, but is always an isometry of the form
\begin{equation}\label{eq:WHAUP}
	\begin{tikzpicture}[baseline={([yshift=-0.5ex]current bounding box.center)}, scale=0.63]
		\WgateredInd{0}{0}{$i$}{$a$}{$b$}{$j$}
	\end{tikzpicture}=\langle i,a |\hat{P}\hat{U}|b,j\rangle=\langle i,a |\hat{U}\hat{Q}|b,j\rangle,
\end{equation}
where $\hat{U}$ is unitary and $\hat{P},\hat{Q}$ are projection operators. In  App.~\ref{app:WHAprojector} we define  $\hat{P},\hat{Q}$ using the algebraic structure of $\A$. 
Importantly, these projection operators mutually commute %
\begin{equation}\label{eq:UPcommute}
	[\hat{P}_{i,i+1/2},\hat{Q}_{i-1/2,i}]=0=[\hat{Q}_{i,i+1/2},\hat{P}_{i-1/2,i}],~\forall i\in\frac{1}{2}\mathbb{Z}. %
\end{equation}
We prove Eqs.~\eqref{eq:WHAUP} and \eqref{eq:UPcommute} for an arbitrary $\C^*$-weak Hopf algebra in App.~\ref{app:WHAprojector}. 

As in the case of $\C^*$-Hopf algebras, we define the unitary evolution operator $\hat{\mathbf{U}}$ using the first line in Eq.~\eqref{eq:Uevolutionoperator}. Due to Eq.~\eqref{eq:WHAUP}, the first and second line of Eq.~\eqref{eq:Uevolutionoperator} are not equal for $\C^*$-weak Hopf algebras in general. Fortunately, they are still equal when restricted to a suitable subspace that we define below. We first define two global projection operators
\begin{eqnarray}\label{eq:total_projector}
	\hat{\mathbf{P}}&=&\hat{\mathbf{P}}_e\hat{\mathbf{Q}}_o=\prod_{j=1}^\N \hat{P}_{j,j+1/2}\prod_{j=1}^\N \hat{Q}_{j-1/2,j},\nonumber\\
	\hat{\mathbf{P}}_{\frac{1}{2}}&=&\hat{\mathbf{Q}}_e\hat{\mathbf{P}}_o=\prod_{j=1}^\N \hat{Q}_{j,j+1/2}\prod_{j=1}^\N \hat{P}_{j-1/2,j}.
\end{eqnarray}
Note that $\hat{\mathbf{P}}$ and $\hat{\mathbf{P}}_{\frac{1}{2}}$ are transformed into each other by a translation by half a unit cell. %
In App.~\ref{app:WHAprojector} we prove the following important identities
\begin{eqnarray}\label{eq:UPCommuteInv}
\hat{\mathbf{P}}_{\frac{1}{2}}\hat{\mathbf{U}}_o\hat{\mathbf{P}}&=&\hat{\mathbf{U}}_o\hat{\mathbf{P}},\nonumber\\
\hat{\mathbf{P}}\hat{\mathbf{U}}_e\hat{\mathbf{P}}_{\frac{1}{2}}&=&\hat{\mathbf{U}}_e\hat{\mathbf{P}}_{\frac{1}{2}}.
\end{eqnarray}
Let $\mathrm{P}$ denote the space of states $\ket{\psi}$ satisfying  $\hat{\mathbf{P}}\ket{\psi}=\ket{\psi}$, or equivalently, the common eigenspace of all $\hat{P}_{j,j+1/2} $ and $ \hat{Q}_{j-1/2,j}$ with eigenvalue $+1$, and similarly let $\mathrm{P}_{\frac{1}{2}}$ denote the space of states $\ket{\psi}$ satisfying  $\hat{\mathbf{P}}_{\frac{1}{2}}\ket{\psi}=\ket{\psi}$. Eq.~\eqref{eq:UPCommuteInv} implies that
\begin{eqnarray}\label{eq:UPCommuteInvSpace}
\hat{\mathbf{U}}_o\mathrm{P}&=&\mathrm{P}_{\frac{1}{2}},\nonumber\\
\hat{\mathbf{U}}_e\mathrm{P}_{\frac{1}{2}}&=&\mathrm{P},
\end{eqnarray}
therefore
\begin{eqnarray}\label{eq:UPCommuteInvSpaceCombine}
	\hat{\mathbf{U}}\mathrm{P}=\mathrm{P},
\end{eqnarray}
i.e. the floquet evolution operator $\hat{\mathbf{U}}$ leaves the subspace $\mathrm{P}$ invariant. 
Therefore, if the initial state $\ket{\psi(0)}$ lies in $\mathrm{P}$, then $\ket{\psi(t)}$ will remain in $\mathrm{P}$ at all time $t\in\mathbb{Z}$, and $\ket{\psi(t)}\in \mathrm{P}_{\frac{1}{2}}$ for $t\in\mathbb{Z}+1/2$.  
In the rest of this paper we call the space $\mathrm{P}$ the solvable subspace, and $\mathrm{P}_{\frac{1}{2}}$ the solvable subspace at half-integer time. 
For the explicit example we construct later in Sec.~\ref{sec:GoldenModel}, we happen to have $\hat{P}=\hat{Q}$ , implying that $\hat{\mathbf{P}}_{\frac{1}{2}}=\hat{\mathbf{P}}$, and therefore $\ket{\psi(t)}\in \mathrm{P}$ at all integer and half-integer time.
The dimension of the solvable subspace $D_\N=\mathrm{dim}[\mathrm{P}]=\Tr[\hat{\mathbf{P}}]$ for a length-$\N$ chain grows exponentially as $D_\N\sim (D_\rho D_v)^\N$, where $D_\rho$ and $ D_v$ are the quantum dimensions of the representations $\rho$ and $v$ which we define in App.~\ref{app:dimension_solvable_space}.  
Within the  subspace $\mathrm{P}$, 
the first and second lines of Eq.~\eqref{eq:Uevolutionoperator} are equal, and we show in the following sections that the dynamics generated by $\hat{\mathbf{U}}$ is solvable in the subspace $\mathrm{P}$.  %
Importantly, this construction contains non-dual unitary examples, as we demonstrate explicitly in Sec.~\ref{sec:GoldenModel}.

\subsection{MPO representation for time-evolved local observables}\label{sec:LOMPO}
We start with the time evolution of local observables in the Heisenberg picture. 
Using Eq.~\eqref{eq:Uevolutionoperator}, we obtain the following tensor network representation of %
$\hat{O}(t)=\hat{\mathbf{U}}(t)^\dagger \hat{O}\hat{\mathbf{U}}(t)$: %
\begin{equation}\label{eq:OTHeisenbergTN}
	\hat{O}(t)~~=~~
	\begin{tikzpicture}[baseline={([yshift=-0.5ex]current bounding box.center)}, scale=0.63]
	\foreach \i in {0,...,4}
	{\foreach \j in {0,...,1}
		{\WgateblueInv{2*\i}{2*\j+1}}}
	\foreach \i in {0,...,3}
	{\foreach \j in {0,...,1}
		{\WgateblueInv{2*\i+1}{2*\j+2}}}
	
	\foreach \i in {0,...,4}
	{\foreach \j in {0,...,1}
		{\Wgatered{2*\i}{-2*\j-1}}}
	\foreach \i in {0,...,3}
	{\foreach \j in {0,...,1}
		{\Wgatered{2*\i+1}{-2*\j-2}}}
	\foreach \i in {0,...,4}{
		\draw[very thick, draw=\colorR] (8-0.5-2*\i, -0.51) -- (8-0.5-2*\i,+0.51);
		\draw[very thick, draw=\colorL] (+.5+2*\i, -0.51) -- (+.5+2*\i,+0.51);
	}
	\MYcircleB{3.5}{0}
	\node at (3,0){\small $\hat{O}$ };
\end{tikzpicture}.
\end{equation}
Here we have assumed that the operator $\hat{O}$ sits on a $v$-leg, the other case can be treated in an identical way. 
A small caveat here is that if one computes $\braket{\psi(t)|\hat{O}|\psi(t)}=\braket{\psi|\hat{O}(t)|\psi}$ for   an observable $\hat{O}$ at a fixed position, then $\hat{O}$ will alternate between $\rho$-legs and $v$-legs at integer and half integer times. We will take care of this issue later in Sec.~\ref{sec:quench_obs}.

The MPO representation~\eqref{eq:triangularTNMPO} of triangular tensor networks immediately leads to an MPO representation of %
$\hat{O}(t)$:
\begin{eqnarray}\label{eq:local_observableMPO}
\hat{O}(t)	&=&\quad
	\begin{tikzpicture}[baseline={([yshift=-0.5ex]current bounding box.center)}, scale=0.63]
\lowertriangularTN{-1}{0}{\Wgatered}{4}
\uppertriangularTN{-1}{0}{\WgateblueInv}{4}		
		\foreach \i in {0,...,3}{
			\draw[very thick, draw=\colorR] (-0.5-\i, -0.51-\i) -- (-0.5-\i,+0.51+\i);
			\draw[very thick, draw=\colorL] (+0.5+\i, -0.51-\i) -- (+0.5+\i,+0.51+\i);
		}
		\MYcircleB{-0.5}{0}
		\node at (-1,0){\small $\hat{O}$};
	\end{tikzpicture}\nonumber\\
	&=&
	\begin{tikzpicture}[baseline={([yshift=-0.5ex]current bounding box.center)}, scale=0.63]
		\manyrhov{-3.5}{-4}{4}
		\MYsquare{4}{-4}
		\MYsquareB{-4}{-4}
		\manyrhovb{-3.5}{-3}{4}
		\MYsquare{4}{-3}
		\MYsquareB{-4}{-3}
		\MYcircleB{3.5}{-3.5}
		\node at (4,-3.5){\small $\hat{O}$};
	\end{tikzpicture},
\end{eqnarray}
where in the first line we use the unitarity condition~\eqref{eq:Unitarygraphical} to simplify the 2D tensor network in Eq.~\eqref{eq:OTHeisenbergTN}, and we define
\begin{equation}\label{eq:rhovconjugate}
	\begin{tikzpicture}[baseline={([yshift=-0.5ex]current bounding box.center)}, scale=0.63]
		\rhotensorb{0}{0 }%
		\paraindices{0}{0}{\beta}{\alpha}
		\quantumindices{0}{0}{a}{b}
	\end{tikzpicture}=\left[
	\begin{tikzpicture}[baseline={([yshift=-0.5ex]current bounding box.center)}, scale=0.63]
	\rhotensor{0}{0}%
	\paraindices{0}{0}{\beta}{\alpha}
	\quantumindices{0}{0}{b}{a}
	\end{tikzpicture}\right]^*,~
	\begin{tikzpicture}[baseline={([yshift=-0.5ex]current bounding box.center)}, scale=0.63]
		\vtensorb{0}{0 }%
		\paraindices{0}{0}{\beta}{\alpha}
		\quantumindices{0}{0}{i}{j}
	\end{tikzpicture}=
	\left[%
	\begin{tikzpicture}[baseline={([yshift=-0.5ex]current bounding box.center)}, scale=0.63]
	\vtensor{0}{0}%
	\paraindices{0}{0}{\beta}{\alpha}
	\quantumindices{0}{0}{j}{i}
	\end{tikzpicture}\right]^*.
\end{equation}
For single site operators sitting on $\rho$-legs, we have a similar expression
\begin{eqnarray}\label{eq:local_observableMPOhi}
	\hat{O}(t)	=
	\begin{tikzpicture}[baseline={([yshift=-0.5ex]current bounding box.center)}, scale=0.63]
		\manyrhov{-3.5}{-4}{4}
		\MYsquare{4}{-4}
		\MYsquareB{-4}{-4}
		\manyrhovb{-3.5}{-3}{4}
		\MYsquare{4}{-3}
		\MYsquareB{-4}{-3}
		\MYcircleB{-3.5}{-3.5}
		\node at (-4,-3.5){\small $\hat{O}$ };
	\end{tikzpicture}.
\end{eqnarray}
\subsection{Local observables and correlations in quench dynamics from an initial MPS}\label{sec:quench_obs}
In this section we compute local observables and their correlation functions in quench dynamics generated by the unitary circuit in Eq.~\eqref{eq:Uevolutionoperator}, with initial state $|\psi(0)\rangle$ being  an arbitrary MPS  with finite bond dimension, represented in tensor graphical notation as
\begin{eqnarray}
	|\psi(0)\rangle~~&=&~~
	\begin{tikzpicture}[baseline={([yshift=-0.5ex]current bounding box.center)}, scale=0.63]
		\MPS{0}{0}{4 }
	\end{tikzpicture},%
\end{eqnarray}
where each triangle denotes a 3-index tensor, with one physical index~(vertical) and two virtual indices~(horizontal). %
Note that we do not require  $\ket{\psi(0)}$ to be translationally invariant, i.e., the triangle tensors  are allowed to be different along the chain. Furthermore, for solvable unitary circuits constructed from $\C^*$ weak Hopf algebras, we assume that $\ket{\psi(0)}$ belongs to the solvable subspace, i.e.
\begin{equation}
\hat{\mathbf{P}}\ket{\psi(0)}=\ket{\psi(0)}. 
\end{equation}

The MPO representation of local operators in Eq.~\eqref{eq:local_observableMPO} allows us to exactly compute the expectation value of local observables and correlation functions in the time evolved state $|\psi(t)\rangle=\hat{\mathbf{U}}(t)|\psi(0)\rangle$. A caveat here is that even though $\hat{O}(t)$ is always an MPO with finite bond dimension, $\ket{\psi(t)}$ is generally not an MPS with finite bond dimension at large $t$:
	\begin{eqnarray}\label{eq:psitTN}
	|\psi(t)\rangle&=&	
	\begin{tikzpicture}[baseline=(current bounding box.center), scale=0.55]
		\foreach \i in {0,...,4}
		{\foreach \j in {0,...,1}
			{\Wgatered{2*\i}{2*\j}}}
		\foreach \i in {0,...,3}
		{\foreach \j in {0,...,1}
			{\Wgatered{2*\i+1}{2*\j+1}}}
		\MPS{-0.5}{-1}{5 }
	\end{tikzpicture}\nonumber\\
	&\neq&\text{MPS with finite BD, as }t\to\infty.%
\end{eqnarray}
Indeed, we will see later in this paper that the bipartite Renyi entanglement entropy of $\ket{\psi(t)}$ in our models generally grows linearly in time. Nevertheless, going to the Heisenberg picture and utilizing Eq.~\eqref{eq:local_observableMPO} still allows us to efficiently and  exactly compute expectation values of local observables in $\ket{\psi(t)}$. For example, let $x$ be the position of the local observable $\hat{O}$. If $x-t\in\mathbb{Z}$, then we have
\begin{eqnarray}\label{eq:OtMPS}
\langle\hat{O}\rangle_{\psi(t)}&=&\braket{\psi(t)|\hat{O}|\psi(t)}\\
	&=&\braket{\psi|\hat{O}(t)|\psi}\nonumber\\
	&=&%
	\begin{tikzpicture}[baseline={([yshift=-0.5ex]current bounding box.center)}, scale=0.63]
\manyrhov{-3.5}{-4}{4}
\MYsquare{4}{-4}
\MYsquareB{-4}{-4}
\manyrhovb{-3.5}{-3}{4}
\MYsquare{4}{-3}
\MYsquareB{-4}{-3}
\MYcircleB{3.5}{-3.5}
\node at (3,-3.5){\small $\hat{O}$};
\MPS{-3.5}{-5}{4 }
\MPSb{-3.5}{-2}{4}
\LeftBoundary{-4.5}{-3.5}{1.5}
\RightBoundary{4.5}{-3.5}{1.5}
	\end{tikzpicture},\nonumber
\end{eqnarray}
where $\Lambda_L$ and $\Lambda_R$ are vectors that one obtains after contracting semi-infinite subchains of $\ket{\psi(0)}$, see, e.g. Ref.~\cite{Perez2007matrix}. 
The case  $x-t\in\mathbb{Z}+1/2$ can be treated in a similar way, where $\hat{O}(t)$ in Eq.~\eqref{eq:local_observableMPOhi} should be used instead. 
The last line in Eq.~\eqref{eq:OtMPS} can be efficiently computed using the transfer matrix formalism for 1D MPS~\cite{Cirac2021Matrix}, the overall bond dimension~(i.e. the dimension of the transfer matrix) which measures the computational complexity is $d_\A^2d_\psi^2$, where $d_\A$ is the dimension of the underlying Hopf algebra and $d_\psi$ is the bond dimension of the initial MPS $\ket{\psi(0)}$. For example, if $\ket{\psi(0)}$ is translationally invariant, then
\begin{equation}\label{eq:OtTM}
\langle\hat{O}(t)\rangle_\psi=\llparenthesis K_L| (T_\rho T_v)^{2t-1}T_\rho T^O_v |K_R\rrparenthesis,
\end{equation}
where the transfer matrices are defined as
\begin{equation}\label{eq:OtTMTTKK}
	T^O_{\rho}=
	\begin{tikzpicture}[baseline={([yshift=-0.5ex]current bounding box.center)}, scale=0.7]
		\rhotensor{0}{0}
		\rhotensorb{0}{1}
\rhoMPS{0}{-1}{}
\rhoMPSb{0}{2}{}
	\MYcircleB{0}{0.5}
\node at (-0.5,0.5){\small $\hat{O}$};
	\end{tikzpicture},~
		T^O_{v}=
	\begin{tikzpicture}[baseline={([yshift=-0.5ex]current bounding box.center)}, scale=0.7]
		\vtensor{0}{0}
		\vtensorb{0}{1}
		\vMPS{0}{-1}{}
		\vMPSb{0}{2}{}
		\MYcircleB{0}{0.5}
		\node at (-0.5,0.5){\small $\hat{O}$};
	\end{tikzpicture},~%
	\llparenthesis K_L|=	
	\begin{tikzpicture}[baseline={([yshift=-0.5ex]current bounding box.center)}, scale=0.7]
		\LeftBoundary{0}{0}{1.5}
		\counittensorVL{0.3}{0.5}{0.3}
		\counittensorVL{0.3}{-0.5}{0.3}
	\end{tikzpicture},~
	|K_R\rrparenthesis=	
	\begin{tikzpicture}[baseline={([yshift=-0.5ex]current bounding box.center)}, scale=0.7]
	\RightBoundary{0}{0}{1.5}
	\unittensorVL{-0.3}{0.5}{0.3}
	\unittensorVL{-0.3}{-0.5}{0.3}
	\end{tikzpicture},
\end{equation}
and $T_{\rho}=T^1_{\rho},T_{v}=T^1_{v}$. 
In this case, $\Lambda_L$ and $\Lambda_R$ are the left and right principal eigenvectors of the transfer matrix of $\ket{\psi(0)}$, respectively. 

The above method can be straightforwardly generalized to compute any $n$-point equal time correlation function of local observables in $\ket{\psi(t)}$. %
In this paper we focus on a simple special case of 2-point correlations: let $\hat{O}$ be a single site operator lying on an integer site, say the site $0$, and $\hat{O}'$ a single site operator on a half-integer site $x+1/2$ with $x\geq 0, x\in\mathbb{Z}$. Then at an integer time $t$, we have
\begin{equation}
	\hat{O}(t)\hat{O}'(t)=
	\begin{tikzpicture}[baseline={([yshift=-0.5ex]current bounding box.center)}, scale=0.63]
		\foreach \i in {0,...,4}
		{\foreach \j in {0,...,1}
			{\WgateblueInv{2*\i}{2*\j+1}}}
		\foreach \i in {0,...,3}
		{\foreach \j in {0,...,1}
			{\WgateblueInv{2*\i+1}{2*\j+2}}}
		
		\foreach \i in {0,...,4}
		{\foreach \j in {0,...,1}
			{\Wgatered{2*\i}{-2*\j-1}}}
		\foreach \i in {0,...,3}
		{\foreach \j in {0,...,1}
			{\Wgatered{2*\i+1}{-2*\j-2}}}
		\foreach \i in {0,...,4}{
			\draw[very thick, draw=\colorR] (8-0.5-2*\i, -0.51) -- (8-0.5-2*\i,+0.51);
			\draw[very thick, draw=\colorL] (+.5+2*\i, -0.51) -- (+.5+2*\i,+0.51);
		}
		\MYcircleB{1.5}{0}
		\node at (1,0){\small $\hat{O}$};
		\MYcircleB{6.5}{0}
		\node at (7,0){\small $\hat{O}'$};
	\end{tikzpicture}.
\end{equation}
As before, we first use the unitarity condition~\eqref{eq:Unitarygraphical} to simplify the 2D tensor network, and then apply Eq.~\eqref{eq:triangularTNMPO} to obtain 
\begin{eqnarray}\label{eq:MPOOrOv}
	&&\hat{O}(t)\hat{O}'(t)\nonumber\\
	&=&\quad
	\begin{tikzpicture}[baseline={([yshift=-0.5ex]current bounding box.center)}, scale=0.63]
		\lowertriangularTN{-1}{0}{\Wgatered}{6}
		\uppertriangularTN{-1}{0}{\WgateblueInv}{6}		
		\foreach \i in {0,...,5}{
			\draw[very thick, draw=\colorR] (-0.5-\i, -0.51-\i) -- (-0.5-\i,+0.51+\i);
			\draw[very thick, draw=\colorL] (+0.5+\i, -0.51-\i) -- (+0.5+\i,+0.51+\i);
		}
			\MYcircleB{-2.5}{0}
		\node at (-3,0){\small $\hat{O}$};
		\MYcircleB{2.5}{0}
		\node at (3,0){\small $\hat{O}'$};
	\end{tikzpicture}\nonumber\\
	&=&
	\begin{tikzpicture}[baseline={([yshift=-0.5ex]current bounding box.center)}, scale=0.63]
		\manyrhov{-5.5}{-4}{6}
		\MYsquare{6}{-4}
		\MYsquareB{-6}{-4}
		\manyrhovb{-5.5}{-3}{6}
		\MYsquare{6}{-3}
		\MYsquareB{-6}{-3}
		
		\MYcircleB{-1.5}{-3.5}
	\node at (-2,-3.5){\small $\hat{O}'$};
	\MYcircleB{1.5}{-3.5}
	\node at (2,-3.5){\small $\hat{O}$};
	\end{tikzpicture}.\nonumber\\
\end{eqnarray}
Note that in the second line of Eq.~\eqref{eq:MPOOrOv} we added some gates between $\hat{O}$ and $\hat{O}'$ using Eq.~\eqref{eq:Unitarygraphical} to obtain a triangular shape such that Eq.~\eqref{eq:triangularTNMPO} can be applied. 
Using similar derivations as in Eqs.~(\ref{eq:OtMPS},\ref{eq:OtTM}), we obtain 
\begin{eqnarray}\label{eq:OOtTMexpresionlatetime}
		\langle\hat{O}(t)\hat{O}'(t)\rangle_\psi&=&\llparenthesis K_L|(T_\rho T_v)^{x}  T^{O'}_\rho (T_v T_\rho)^{2t-x-1}\nonumber\\
		&&{}\cdot T^O_v(T_\rho T_v)^{x}|K_R\rrparenthesis,
\end{eqnarray}
if $x\leq 2t-1$, and 
\begin{eqnarray}\label{eq:OOtTMexpresionearlytime}
	\langle\hat{O}(t)\hat{O}'(t)\rangle_\psi&=&\llparenthesis K_L|(T_\rho T_v)^{2t-1} T_\rho T_v^O (T_\rho T_v)^{x-2t}\nonumber\\
	&&{}\cdot T_\rho^{O'} T_v(T_\rho T_v)^{2t-1}|K_R\rrparenthesis,
\end{eqnarray}
if $x\geq 2t$. 

Eqs.~(\ref{eq:OtTM},\ref{eq:OOtTMexpresionlatetime},\ref{eq:OOtTMexpresionearlytime}) are used in the accompanying Mathematica code~\cite{HAQCACode} for the exact numerical computation of these quantities.  
\subsection{Renyi entanglement entropy}\label{sec:RenyiEE}
In this section we show how to exactly compute the Renyi entanglement entropy for the time evolved state $\ket{\psi(t)}$ at arbitrary time $t$. This computation also gives us an exact expression for the equilibration time $t^*$. %

We first recall the definition of the Renyi entanglement entropy. Let $\hat{\rho}$ be a density matrix representing a mixed quantum state. The Renyi entropy for $\hat{\rho}$  at index $\alpha$ is defined as
\begin{equation}\label{def:RenyiEE}
	H_\alpha[\hat{\rho}]=\frac{1}{1-\alpha}\ln \Tr[\hat{\rho}^\alpha].
\end{equation}
In this paper we only consider the case $\alpha\geq 2$ being a positive integer. For any pure quantum state $\ket{\psi}$ of the 1D chain, let $A$ be a subsystem~(a subset of qudits of the 1D chain) and let $\bar{A}$ denote the complement of $A$. The Renyi entanglement entropy of the subsystem $A$ in the state $\ket{\psi}$ is defined as $H_\alpha[\hat{\rho}_A]$, where $\hat{\rho}_A$ is the reduced density matrix for the subsystem $A$ defined as
\begin{equation}\label{def:rhoAt}
	\hat{\rho}_A=\mathrm{Tr}_{\bar{A}}\left[\ket{\psi}\bra{\psi}\right].
\end{equation}
Our goal in the following is to compute the Renyi entanglement entropy of the subsystem $A$ in the time evolved state $\ket{\psi(t)}$, which we denote as
\begin{equation}
	H_{\alpha}(A,t):=H_\alpha[\hat{\rho}_A(t)].
\end{equation} 
In Sec.~\ref{sec:RenyiFinite} we consider a finite subsystem $A$ and in Sec.~\ref{sec:RenyiSemiinfinite} we take $A$ to be a semi-infinite half chain. 
\subsubsection{Renyi entanglement entropy for finite subsystem}\label{sec:RenyiFinite}
In this section, we compute the Renyi entanglement entropy of a finite subsystem $A$ of length $L_A$ consisting of a contiguous block of $2L_A$ qudits. 
Throughout this section, when drawing tensor network diagrams, we assume the initial state $\ket{\psi(0)}$ to be a product state for simplicity, and the generalization to MPS is straightforward. We represent $\ket{\psi(0)}$ as
\begin{eqnarray}
|\psi(0)\rangle=
\begin{tikzpicture}[baseline={([yshift=-0.5ex]current bounding box.center)}, scale=0.55]
	\foreach \i in {0,...,4}{
		\draw[very thick, draw=\colorL] (4-0.5-2*\i, -4.5) -- (4-0.5-2*\i,-4.2);
		\draw[very thick, draw=\colorR] (5-0.5-2*\i, -4.5) -- (5-0.5-2*\i,-4.2);
		\MYtriangle{-0.5-\i}{-4.6}
		\MYtriangle{+0.5+\i}{-4.6}
	}
\end{tikzpicture}~,
\end{eqnarray}
where each black triangle represents a vector in the corresponding local Hilbert space. Inserting the tensor network representation of $\ket{\psi(t)}$ in  Eq.~\eqref{eq:psitTN} into Eq.~\eqref{def:rhoAt}, and using the  unitarity condition~\eqref{eq:Unitarygraphical} to simplify the 2D tensor network as before, we obtain
\FPeval{\numm}{clip(3)}
\FPeval{\num}{clip(\numm+2)}
\FPeval{\numt}{clip(\numm+3)}
\FPeval{\numx}{clip(2*\num+1)}
\FPeval{\realnum}{clip(\numm+1.6)}
\begin{equation}\label{eq:rhoAtfinTN}
\rho_A(t)=
\begin{tikzpicture}[baseline={([yshift=-0.5ex]current bounding box.center)}, scale=0.6]
	\foreach \j in {2,...,\num}
{\foreach \i in {0,...,\j}
		{\Wgatered{2*\i-\j}{-\j+1}}}
\foreach \j in {2,...,\num}
{\foreach \i in {0,...,\j}
		{\WgateblueInv{2*\i-\j}{\j-1}}}
\foreach \i in {0,...,\numm}{
		\draw[very thick, draw=\colorR] (-2.5-\i, -0.51-\i) -- (-2.5-\i,+0.51+\i);
		\draw[very thick, draw=\colorL] (2.5+\i, -0.51-\i) -- (2.5+\i,+0.51+\i);
	}
\foreach \i in {0,...,\num}{
		\MYtriangle{-0.5-\i}{-\numm-1.6}
		\MYtriangle{+0.5+\i}{-\numm-1.6}
		\MYtriangleI{-0.5-\i}{\numm+1.6}
		\MYtriangleI{+0.5+\i}{\numm+1.6}
	}
\end{tikzpicture}.
\end{equation}
At this step we cannot directly apply Eq.~\eqref{eq:triangularTNMPO} since the 2D circuit tensor network in Eq.~\eqref{eq:rhoAtfinTN} does not have a triangular shape as in Eq.~\eqref{eq:triangularTNMPO}. However, we can complete the triangle by doing a suitable unitary transformation on the finite subsystem $A$. 
Notice that the Renyi entropy of the transformed density matrix $\hat{\rho}^\prime_A(t)=\hat{U}_A\hat{\rho}_A(t)\hat{U}^\dagger_A$ is the same as $\hat{\rho}_A(t)$, for any unitary transformation $\hat{U}_A$ supported on  $A$. We choose  $\hat{U}_A$ to transform the trapezoids in Eq.~\eqref{eq:rhoAtfinTN}  into triangles and then apply Eq.~\eqref{eq:triangularTNMPO} to obtain an MPO representation for $\hat{\rho}^\prime_A(t)$: %
\begin{eqnarray}\label{eq:DMMPOFinSubs}
	\hat{\rho}^\prime_A(t)&=&\hat{U}_A\hat{\rho}_A(t)\hat{U}^\dagger_A\\
	&=&~~
	\begin{tikzpicture}[baseline={([yshift=-0.5ex]current bounding box.center)}, scale=0.6]
\lowertriangularTN{-1}{0}{\Wgatered}{\numt}
\uppertriangularTN{-1}{0}{\WgateblueInv}{\numt}	
\legindices{{b_1,b_2}}{0.5}{0.5}{1 }{1}{below right=-0.1}
\legindices{{a_1,a_2}}{0.5}{-0.5}{1 }{-1}{above right=-0.1}
\legindices{{i_2,i_1}}{-0.5}{-0.5}{-1 }{-1}{above left=-0.15}
\legindices{{j_2,j_1}}{-0.5}{0.5}{-1 }{1}{below left=-0.15}
		\foreach \i in {0,...,\numm}{
			\draw[very thick, draw=\colorR] (-2.5-\i, -2.51-\i) -- (-2.5-\i,+2.51+\i);
			\draw[very thick, draw=\colorL] (2.5+\i, -2.51-\i) -- (2.5+\i,+2.51+\i);
		}
		\foreach \i in {0,...,\num}{
			\MYtriangle{-0.5-\i}{-\num-1.6}
			\MYtriangle{+0.5+\i}{-\num-1.6}
			\MYtriangleI{-0.5-\i}{\num+1.6}
			\MYtriangleI{+0.5+\i}{\num+1.6}
		}
	\end{tikzpicture}\nonumber\\
	&=&\!
	\begin{tikzpicture}[baseline={([yshift=-0.5ex]current bounding box.center)}, scale=0.6]
		\manyrhov{-0.5}{-4}{\numt}
		\manyrhovb{-0.5}{-2}{\numt}
		\legindices{{b_1,b_2}}{-0.5}{-2.5}{2 }{0}{below =-0.1}
		\legindices{{a_1,a_2}}{-0.5}{-3.5}{2 }{0}{above =-0.1}
		\legindices{{i_1,i_2}}{\numx-2.5}{-3.5}{2 }{0}{above =-0.1}
		\legindices{{j_1,j_2}}{\numx-2.5}{-2.5}{2 }{0}{below =-0.1}
			\MYsquare{\numx}{-2}
			\MYsquareB{-1}{-2}
		\MYsquare{\numx}{-4}
		\MYsquareB{-1}{-4}
		
		\foreach \i in {0,...,\numx}{
			\MYtriangle{-0.5+\i}{-4.6}
			\MYtriangleI{-0.5+\i}{-1.4}
		}
		\foreach \i in {0,...,\numm}{
			\draw[very thick, draw=\colorR] (2*\numm+1-0.5-2*\i, -3.51) -- (2*\numm+1-0.5-2*\i,+0.51-3);
			\draw[very thick, draw=\colorL] (3.5+2*\i, -3.51) -- (3.5+2*\i,+0.51-3);
		}
	\end{tikzpicture}.\nonumber
\end{eqnarray}
To compute $H_{\alpha}(A,t)$, we need to compute  $\Tr[\hat{\rho}_A(t)^\alpha]=\Tr[\hat{\rho}^\prime_A(t)^\alpha]$, i.e., we need to evaluate the matrix power of the MPO in the last line of Eq.~\eqref{eq:DMMPOFinSubs}. %
In this paper we focus on the following two cases:\\
1. If the subsystem $A$ is small, then we can directly evaluate the last line of Eq.~\eqref{eq:DMMPOFinSubs} using the transfer matrix method as before, and obtain the exact matrix representation for $\hat{\rho}^\prime_A(t)$ which we use to compute  $H_{\alpha}(A,t)$. The computational cost of this method is independent of  $\alpha$ and $t$, but grows exponentially in the size of $A$.\\
2. If the Renyi index $\alpha$ is small, then  the last line of Eq.~\eqref{eq:DMMPOFinSubs} leads to a 1D tensor network representation for $\Tr[\hat{\rho}^\prime_A(t)^\alpha]$ with bond dimension $d_\A^{2\alpha}$, which can again be evaluated efficiently using the transfer matrix method. The computational cost of this method is independent of $t$ and the size of $A$, but grows exponentially with the Renyi index $\alpha$.\\

In App.~\ref{app:TMRenyi} we provide explicit expressions for $\Tr[\hat{\rho}_A(t)^\alpha]$ using the transfer matrix formalism, which are implemented in the accompanying Mathematica code~\cite{HAQCACode}, and in Secs.~\ref{sec:HAexamples} and \ref{sec:GoldenModel} we use these expressions to obtain
exact results for both cases  when studying explicit examples of solvable unitary circuits constructed using our framework.
\subsubsection{Equilibration time $t^*$} %
The exact expression~\eqref{eq:DMMPOFinSubs} for the reduced density matrix allows us to exactly compute the system equilibration time $t^*$. Specifically, the transfer matrix $T=T_\rho T_v$ has spectral radius equal to $1$ and has at least an eigenvalue equal to 1, due to the normalization condition
\begin{equation}
1=\mathrm{Tr}[\hat{\rho}_A(t)]=\llparenthesis K_L|(T_\rho T_v)^{2t}|K_R\rrparenthesis,
\end{equation}
for all $t\geq 0$. 
Let $\lambda_1$ be the largest~(in absolute value) eigenvalue of $T$ whose magnitude is strictly smaller than 1.  Then at large time $t\geq L_A/2$ we have
\begin{eqnarray}\label{eq:DMMPOFinSubs-longtime}
	[\hat{\rho}'_A(t)]^{\mathbf{a},\mathbf{i}}_{\mathbf{b},\mathbf{j}}&=&\llparenthesis K_L^{\mathbf{a},\mathbf{b}}|(T_\rho T_v)^{2t-L_A}|K_R^{\mathbf{i},\mathbf{j}}\rrparenthesis\nonumber\\
	&=&\llparenthesis K_L^{\mathbf{a},\mathbf{b}}|[P_0+O(\lambda_1^{2t-L_A})]|K_R^{\mathbf{i},\mathbf{j}}\rrparenthesis %
\end{eqnarray}
where $\llparenthesis K_L^{\mathbf{a},\mathbf{b}}|$ and $|K_R^{\mathbf{i},\mathbf{j}}\rrparenthesis$ are some unimportant vectors independent of $t$, and $P_0$ is the projector to the principal eigenvector of $T$. 
Therefore $\hat{\rho}'_A(t)$ and the entanglement entropy of the system reaches the equilibrium value at time
\begin{equation}\label{eq:eqltime}
t^*=\frac{1}{2}\N_A+O[\log \lambda_1^{-1}].
\end{equation}
This result is verified in the exact results of the explicit examples in Sec.~\ref{sec:HAexamples} and Sec.~\ref{sec:GoldenModel}.

\subsubsection{Renyi entanglement entropy for semi-infinite subsystem}\label{sec:RenyiSemiinfinite}
Now consider the case when the subsystem $A$ is the half chain extending infinitely to the right~(the whole system is still the entire 1D chain as before). We have the following tensor network representation for $\hat{\rho}_A(t)$ analogous to Eq.~\eqref{eq:rhoAtfinTN}:
\begin{equation}\label{eq:rhosemiinf_TN}
\hat{\rho}_A(t)=
\begin{tikzpicture}[baseline={([yshift=-0.5ex]current bounding box.center)}, scale=0.63]
	\Wgatered{2}{-1}\Wgatered{3}{-2}
	\WgateblueInv{2}{1}\WgateblueInv{3}{2}
	\lowertriangularTN{-1}{0}{\Wgatered}{4}
	\uppertriangularTN{-1}{0}{\WgateblueInv}{4}	
	\foreach \i in {0,...,3}{
		\draw[very thick, draw=\colorR] (-0.5-\i, -0.51-\i) -- (-0.5-\i,+0.51+\i);
		\MYtriangle{-0.5-\i}{-4.6}
		\MYtriangle{+0.5+\i}{-4.6}
		\MYtriangleI{-0.5-\i}{4.6}
		\MYtriangleI{+0.5+\i}{4.6}
	}
\end{tikzpicture}~.
\end{equation}
As before, we apply a unitary transformation $\hat{U}_A$ on the subsystem $A$ to transform the circuit tensor network in Eq.~\eqref{eq:rhosemiinf_TN} into a triangular shape, and then apply Eq.~\eqref{eq:triangularTNMPO} to obtain an MPO representation for the transformed density matrix %
\begin{eqnarray}\label{eq:reducedDMMPO}
	\hat{\rho}^\prime_A(t)&=&\hat{U}_A\hat{\rho}_A(t)\hat{U}^\dagger_A\\
	&=&\begin{tikzpicture}[baseline={([yshift=-0.5ex]current bounding box.center)}, scale=0.63]
\lowertriangularTN{-1}{0}{\Wgatered}{4}
\uppertriangularTN{-1}{0}{\WgateblueInv}{4}	
\legindices{{b_1,b_2,\ldots,b_{2t}}}{0.5}{0.5}{1 }{1}{below right=-0.1}
\legindices{{a_1,a_2,\ldots,a_{2t}}}{0.5}{-0.5}{1 }{-1}{above right=-0.1}
		\foreach \i in {0,...,3}{
			\draw[very thick, draw=\colorR] (-0.5-\i, -0.51-\i) -- (-0.5-\i,+0.51+\i);
			\MYtriangle{-0.5-\i}{-4.6}
			\MYtriangle{+0.5+\i}{-4.6}
			\MYtriangleI{-0.5-\i}{4.6}
			\MYtriangleI{+0.5+\i}{4.6}
		}
	\end{tikzpicture}\nonumber\\
	&=&
	\begin{tikzpicture}[baseline={([yshift=-0.5ex]current bounding box.center)}, scale=0.63]
			\manyrhov{-3.5}{-4}{4}
		\manyrhovb{-3.5}{-2}{4}
\legindices{{b_1,b_2,\ldots,b_{2t}}}{-3.5}{-2.5}{2 }{0}{below=-0.1}
\legindices{{a_1,a_2,\ldots,a_{2t}}}{-3.5}{-3.5}{2 }{0}{above=-0.1}
			\MYsquare{4}{-2}
			\MYsquareB{-4}{-2}

		\MYsquare{4}{-4}
		\MYsquareB{-4}{-4}
		
		\foreach \i in {0,...,3}{
			\draw[very thick, draw=\colorR] (4-0.5-2*\i, -3.51) -- (4-0.5-2*\i,+0.51-3);
			\MYtriangle{-0.5-\i}{-4.6}
			\MYtriangle{+0.5+\i}{-4.6}
			\MYtriangleI{-0.5-\i}{-1.4}
			\MYtriangleI{+0.5+\i}{-1.4}
		}
	\end{tikzpicture}~~.\nonumber
\end{eqnarray}
This allows exact computation of $H_{\alpha}(A,t)$ at any values of $t$ for small $\alpha$, using the 1D tensor network representation for $\Tr[\hat{\rho}^\prime_A(t)^\alpha]$, similar to the second case in the previous section, and the explicit expression is given at the end of App.~\ref{app:TMRenyi}. 

\subsection{Spatiotemporal correlation functions and out-of-time-order correlations}\label{sec:st_corr_func}
The MPO representation of time-evolved local observables in Eq.~\eqref{eq:local_observableMPO} allows us to calculate many other physical quantities involving local observables and their correlations, for example, the
spatiotemporal correlation functions defined as
\begin{equation}\label{def:STcorr}
C(\hat{A},\hat{B},x,t)=\mathrm{Tr}[\hat{\mathbf{P}}\hat{A}_0(t)\hat{B}_x(0)],
\end{equation}
and the out-of-time-order correlation~(OTOC) function defined as
\begin{equation}\label{def:OTOC}
F(\hat{V},\hat{W},x,t)=\frac{1}{d} \operatorname{tr}\left[\hat{\mathbf{P}} \hat{W}^{\dagger}_0 \hat{V}_x^{\dagger}(t) \hat{W}_0 \hat{V}_x(t)\right], 	
\end{equation}
where $\hat{A}_0, \hat{B}_x$ are local observables at positions $0,x$, respectively, and $\hat{W}_0, \hat{V}_x$ are local unitary operators acting at positions $0,x$, respectively. $\hat{\mathbf{P}}$ is the projection operator to the solvable subspace~($\hat{\mathbf{P}}=\mathds{1}$ for models constructed from $\C^*$-Hopf algebras) defined in Eq.~\eqref{eq:total_projector}, which is by construction a 
 matrix product density operator. Both quantities can be exactly computed using Eq.~\eqref{eq:local_observableMPO} and the transfer matrix method for 1D tensor networks, similar to the computation of $\langle\hat{O}(t)\rangle_\psi$ as explained in Sec.~\ref{sec:quench_obs}. We present the exact results for a specific model in Secs.~\ref{sec:GoldenModel:st_corr_func} and \ref{sec:FibOTOC}.
 
\subsection{Solution in finite systems with PBC}\label{sec:PBCsolution}
One important feature of our models is that the quench dynamics is still exactly solvable even in finite systems with PBC, a feature that many existing family of solvable models~(e.g. dual unitary circuits and integrable circuits) do not have.

The key to solve the dynamics in PBC is the following MPO representation of the PBC evolution operator at $t=t_k=(k \N+1)/2,k\in \mathbb{Z}_{\geq 1}$:
\begin{equation}\label{eq:MPOPBCUt0}
\hat{\mathbf{U}}(t_k)\!=\!\hat{\mathcal{T}}^k_{L/2}\!\sum_{x,y}%
c^{k-1}_{xy} 
	\!\!\begin{tikzpicture}[baseline={([yshift=-0.5ex]current bounding box.center)}, scale=0.61]
		\foreach \i in {0,...,3}
		{
			\rhorhootimesvv{2*\i-3}{0}
		}
		\MYsquare{4}{-0.2}
		\MYsquareB{-4}{0.2}

		\MYsquare{4}{-0.6}
		\MYsquareB{-4}{-0.6}
		
		\node at (4,0.2) [right=-0.1] {\footnotesize $x$};
		\node at (-4,-0.2) [left=-0.1] {\footnotesize $y$};
		
		\MYsquareB{4}{0.6}
		\MYsquare{-4}{0.6}
	\end{tikzpicture},
\end{equation}
where $\hat{\mathcal{T}}_{L/2}$ is the  operator that translates the 1D chain by $L/2$ unit cells~(i.e. $L$ sites), satisfying $\hat{\mathcal{T}}_{L/2}^2=1$. 
In App.~\ref{app:MPOPBCUt} we give the derivation for Eq.~\eqref{eq:MPOPBCUt0},
where we also define the coefficient $c^{k-1}_{xy}$ and the MPO tensor in Eq.~\eqref{eq:MPOPBCUt0}. 

From Eq.~\eqref{eq:MPOPBCUt0} we see that $\hat{\mathbf{U}}(t_k)$ is an exact MPO with finite bond dimension~(up to a translation $\hat{\mathcal{T}}_{L/2}$), which implies that the revival time for the bipartite entanglement entropy~(for any initial MPS) is $\N/2$:
\begin{equation}
t_{\mathrm{rev,EE}}=\N/2.
\end{equation}
The finite system revival time $t_{\mathrm{rev}}$  is defined as the smallest $t$ such that 
\begin{equation}
	\hat{\mathbf{U}}(t)=\mathds{1},
\end{equation}
which is the revival time for all physical quantities, such as local observables and correlations functions, and is often larger than the entanglement entropy revival time. In our models, $t_{\mathrm{rev}}$ can be exactly computed using  Eq.~\eqref{eq:MPOPBCUt0}~(see App.~\ref{app:recurrencetime}), the result is
\begin{equation}\label{eq:trevHA-0}
	t_{\mathrm{rev}}=\eta \N,
\end{equation}
where $\eta\in \mathbb{Z}$ is the exponent of the underlying $\C^*$-(weak) Hopf algebra~\cite{etingof1999exponent}. 

To exactly compute local observables and correlation functions at arbitrary $t$, let $t=t_k+t_0$, where $k\in \mathbb{Z}$ and $0\leq t_0<\N/2$. Then we have $\hat{\mathbf{U}}(t)=\hat{\mathbf{U}}(t_0)\hat{\mathbf{U}}(t_k)$, and therefore
\begin{eqnarray}\label{eq:PBCOt}
\hat{O}(t)&=&\hat{\mathbf{U}}(t)^\dagger \hat{O}\hat{\mathbf{U}}(t)\nonumber\\
&=&\hat{\mathbf{U}}(t_k)^\dagger \hat{O}(t_0)\hat{\mathbf{U}}(t_k).
\end{eqnarray}
Since $t_0<\N/2$, the MPO representation in Eq.~\eqref{eq:local_observableMPO} applies to $\hat{O}(t_0)$. Using Eq.~\eqref{eq:MPOPBCUt0}, we obtain an exact MPO representation for $\hat{O}(t)$ at all values of $t$, allowing  exact computation of $\braket{\hat{O}(t)}_\psi$ for any initial MPS $\ket{\psi}$, and similarly for other physical quantities such as $\braket{\hat{O}(t)\hat{O}'(t)}_\psi$. Unfortunately, the bond dimension of the MPO representation of $\hat{O}(t)$ given in Eq.~\eqref{eq:PBCOt} is much larger than that in the thermodynamic limit in Eq.~\eqref{eq:OtMPS}, and appears to be beyond our current computational power, so we will not explicitly compute any of these physical quantities in PBC in this paper. 

\section{Examples: solvable gates from $\C^*$-Hopf algebras}\label{sec:HAexamples}
In this section we present several families of  solvable unitary circuits constructed from finite dimensional $\C^*$-Hopf algebras. We start with finite group algebras~(Sec.~\ref{sec:GroupHA}) which provide the simplest examples of $\C^*$-Hopf algebras, and solvable gates constructed this way already display non-trivial behaviors despite their simplicity, as demonstrated by the exact results we present in Sec.~\ref{sec:result-CD3} for the specific example of the dihedral group $D_3$. 
In particular, we will see that %
the quench dynamics of this model for an initial product state display significantly different behavior compared to that of the solvable initial states of general dual unitary circuits~\cite{Piroli2020DUinitialMPS}. %
Then in Sec.~\ref{sec:GroupExt}, we present more sophisticated solvable gates constructed from non-trivial $\C^*$- Hopf algebras, which are not group algebras, but the solvable gates constructed here can still be described and understood within finite group theory. 
\subsection{Solvable gates from finite group algebras}\label{sec:GroupHA}
In the following we show how to construct a family of solvable unitary circuits from an arbitrary finite group, using our general formalism in the previous section. Let $G$ be a finite group, and let $\rho$ be a $d$-dimensional representation of $G$~(not necessarily irreducible). Let $g_1,g_2,\ldots, g_d$ be arbitrary elements of $G$~(not necessarily distinct), and define a unitary gate as follow
\begin{equation}\label{eq:GroupUgate}
\hat{U}=\hat{X}\left[\sum_{i=1}^d \rho(g_i)\otimes \ket{i}\!\bra{i}\right],
\end{equation}
where $\hat{X}$ is the $d$-dimensional swap gate. %
This gate is constructed from Eq.~\eqref{def:Urhovbialg} using the representation  $\rho$ and the corepresentation
\begin{equation}\label{def:Gcorep}
v_{ij}=\deltak_{i,j}g_i,\quad 1\leq i,j\leq d.
\end{equation}
From Eq.~\eqref{def:Urhovbialg} we obtain explicit expressions for the other tensors
\begin{eqnarray}\label{def:rhovgroup}
	\begin{tikzpicture}[baseline={([yshift=-0.5ex]current bounding box.center)}, scale=0.63]
		\rhotensor{0}{0}%
		\paraindices{0}{0}{y}{x}
		\quantumindices{0}{0}{a}{b}
	\end{tikzpicture}&=&(\deltad_y\otimes\rho_{ab} )\Delta(x)%
	=\deltak_{x,y}\rho_{ab}(x),\nonumber\\
	\begin{tikzpicture}[baseline={([yshift=-0.5ex]current bounding box.center)}, scale=0.63]
		\vtensor{0}{0}%
		\paraindices{0}{0}{y}{x}
		\quantumindices{0}{0}{i}{j}
	\end{tikzpicture}&=&\deltad_y(x\cdot v_{ij})=\deltak_{i,j}\deltak_{y,x g_i},\nonumber\\
	\begin{tikzpicture}[baseline={([yshift=-0.5ex]current bounding box.center)}, scale=0.63]
		\unittensor{0}{0}
		\node at (-0.3,0) [left] {\footnotesize $x$};
	\end{tikzpicture}&=&\deltak_{x,1}, \quad
	\begin{tikzpicture}[baseline={([yshift=-0.5ex]current bounding box.center)}, scale=0.63]
		\counittensor{0}{0}
		\node at (0.3,0) [right] {\footnotesize $x$};
	\end{tikzpicture}=1,
\end{eqnarray}
where $x,y\in G$, and we use $1$ to denote the group unit of $G$ when there is no possibility of confusion. With the explicit expressions in Eqs.~(\ref{eq:GroupUgate},\ref{def:rhovgroup}), 
Eq.~\eqref{eq:MUpentagon} can be directly verified without using any knowledge of Hopf algebras.

The gate $\hat{U}$ in Eq.~\eqref{eq:GroupUgate} has a particularly simple expression if we take $\rho$ to be the regular representation of $G$ and take %
$\{g_1,\ldots,g_d\}$ to be the set of all group elements of $G$ with $d=|G|$. In this case it is convenient to label the qudit basis states using elements of $G$. Then $\hat{U}$ has the following action on a neighboring pair of qudits:  %
\begin{equation}\label{eq:finitegroupgate}
	\hat{U}\ket{g,h}=\ket{h,hg},\quad \forall g,h\in G.
\end{equation}

We now provide a more explicit description for solvable gates constructed from finite Abelian groups. 
Since any finite dimensional representation of a finite Abelian group $G$ must be a direct sum of one dimensional representations, 
$\hat{U}$ must have the following form
\begin{equation}\label{eq:AbGroupUgate}
	\hat{U}=\hat{X}\hat{\Phi},
\end{equation}
where $\hat{\Phi}$ is a phase gate with all phases being $n$-th roots of unity, where $n$ is the exponent of the group $G$. Conversely, any gate $\hat{U}$ of the form~\eqref{eq:AbGroupUgate} can be constructed from a suitable finite Abelian group, and therefore generates a solvable unitary circuit. If a gate $\hat{U}$ formally has the form in Eq.~\eqref{eq:AbGroupUgate} but with some phases that are not roots of unity, then the unitary circuit generated by $\hat{U}$ is still solvable in polynomial time: for example, it can be proved that Eq.~\eqref{eq:triangularTNMPO} is still formally true but with bond dimension growing as Poly$(t)$. The proof will be given in a future work where we generalize our formalism to certain classes of infinite dimensional Hopf algebras.

Now we consider the simplest family of non-Abelian groups--the dihedral group $D_n$. Let $n\geq 3$ be an integer. The dihedral group $D_n$ is a group of order $2n$ defined by the following presentation
\begin{equation}\label{def:DihedralGrp}
	D_n=\left\langle r, s \mid r^n=s^2=1, \quad s r s^{-1}=r^{-1}\right\rangle.
\end{equation}
To construct a solvable gate, we use the following representation
\begin{equation}\label{eq:dihedralRep}
	\rho(r)=\begin{pmatrix}
		\omega^k & 0 \\
		0 & \omega^{-k} 
	\end{pmatrix},\quad \rho(s)=\begin{pmatrix}
		0 & 1 \\
		1 & 0
	\end{pmatrix},
\end{equation}
for $k=1,2,\ldots, m-1$, where $\omega=e^{2\pi i/n}$ is a root of unity. We obtain a solvable gate by inserting Eq.~\eqref{eq:dihedralRep} into Eq.~\eqref{eq:GroupUgate}, where the elements $\{g_1,g_2\}$ can be chosen arbitrarily. In Sec.~\ref{sec:result-CD3} we present some exact numerical results for this example for the case $n=3$, obtained using our general formalism. 

We finally remark that in Eq.~\eqref{eq:GroupUgate} we allow $\rho$ to be a projective representation of $G$, since using a projective representation of $G$ is equivalent to using an ordinary representation of a central extension $\tilde{G}$ of $G$, and in this case the tensors in Eq.~\eqref{def:rhovgroup} has to be constructed from the corresponding representation of $\tilde{G}$. 
For example, the gate in  Eq.~\eqref{eq:finitegroupgate} can be twisted by a 2-cocycle of $G$:
\begin{equation}\label{eq:finitegroupgatecocycle}
	\hat{U}\ket{g,h}=\omega(g,h)\ket{h,hg},\quad \forall g,h\in G,
\end{equation}
where $\omega\in \mathcal{H}^2(G,U(1))$ satisfies $\omega(1,g)=\omega(g,1)=1$ and
\begin{equation}
	\omega(f,g)\omega(fg,h)=\omega(g,h)\omega(f,gh), ~~\forall f,g,h\in G.
\end{equation}

\subsection{Exact results for the dihedral group}\label{sec:result-CD3}
\begin{figure}
	\includegraphics[width=.8\linewidth]{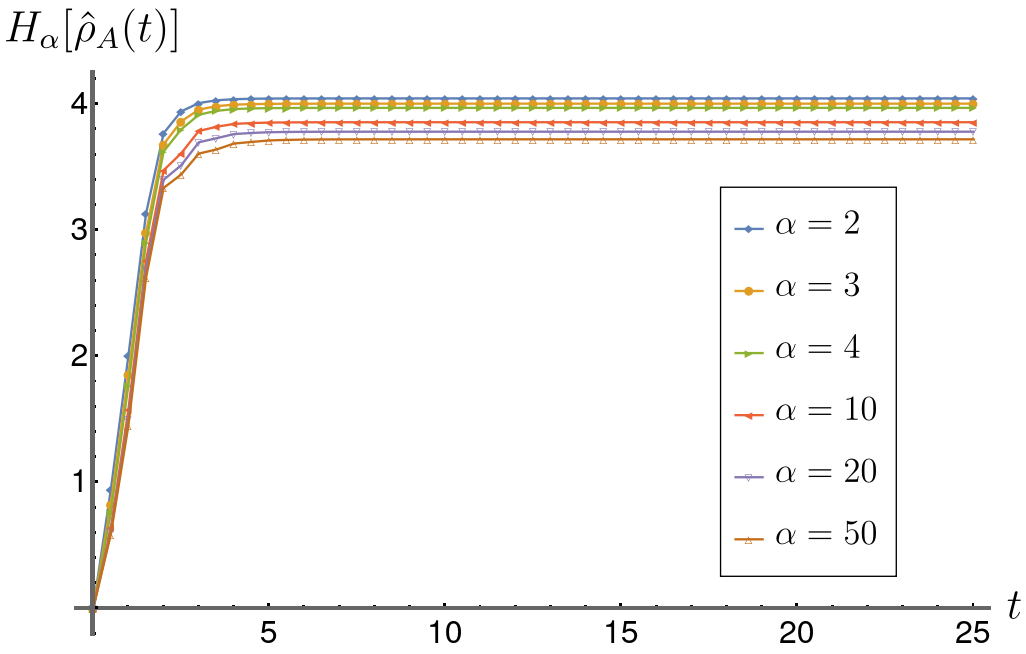}
	\includegraphics[width=.8\linewidth]{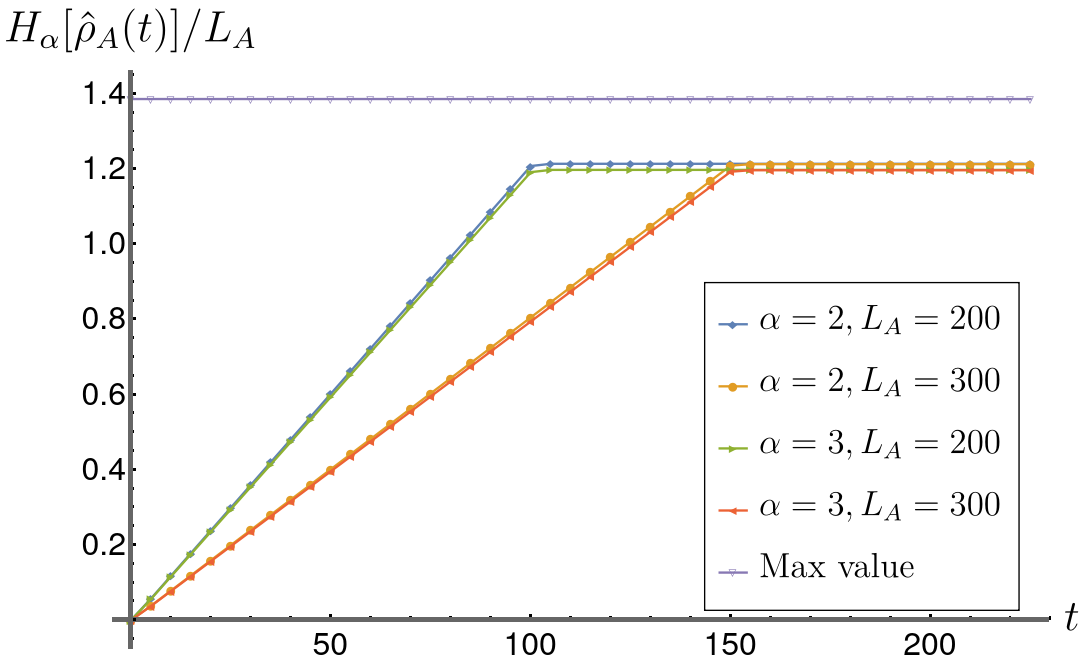}
	\caption{\label{fig:Renyi-CD3} Renyi entanglement entropy $H_\alpha[\hat{\rho}_A(t)]$ for the solvable dynamics based on the dihedral group $D_3$, as defined in Sec.~\ref{sec:result-CD3}:  %
		(top) for a small subsystem of size $\N_A=3$; %
		(bottom) for large subsystems of sizes $\N_A=200$ and $300$. %
		}
\end{figure}

In the following we present some exact results for the gate in Eq.~\eqref{eq:GroupUgate} constructed from the dihedral group $D_3$~(which happens to be isomorphic to the symmetric group $S_3$, the smallest non-Abelian group) using the representation in Eq.~\eqref{eq:dihedralRep}, where $\{g_1,g_2\}$ are chosen as $g_1=s,g_2=r$. We choose the initial state $\ket{\psi_0}$ to be a product state $\ket{\psi_0}=\otimes^{2\N}\ket{+}$, where $\ket{+}=(\ket{1}+\ket{2})/\sqrt{2}$. We exactly compute the Renyi entanglement entropy $H_\alpha[\hat{\rho}_A(t)]$  of the time evolved state $\ket{\psi(t)}$, and the result is shown in Fig.~\ref{fig:Renyi-CD3}. %
For a small subsystem size, we use the first method in Sec.~\ref{sec:RenyiFinite}, while for large subsystem sizes, we use the second method in Sec.~\ref{sec:RenyiFinite}. %
The horizontal line in the second plot of Fig.~\ref{fig:Renyi-CD3} indicates the maximal possible value of entanglement entropy density per unit cell, i.e., the entanglement entropy density of the infinite temperature state,  which is $2\ln(2)$ in this case.
As we can see from these results, the Renyi entanglement entropy of $\ket{\psi(t)}$ clearly depends on the index $\alpha$. This is in contrast to the solvable initial states for a generic dual unitary gate introduced in Ref.~\cite{Piroli2020DUinitialMPS}, where the entanglement spectrum is completely flat and Renyi entropy is independent of $\alpha$. Therefore, for dual unitary gates constructed from a $\C^*$-Hopf algebra, our formalism identifies a strictly larger class of initial states~(i.e. arbitrary MPS) whose dynamics can be exactly solved. 
\subsection{Some extensions to finite group algebras}\label{sec:GroupExt}
In this section we provide more sophisticated examples of solvable gates constructed from a certain class of finite dimensional $\C^*$-Hopf algebras that are not group algebras, but the solvable gates they generate can still be described and understood in purely group-theoretical terms. In particular, we  give explicit expressions for all the tensors in Eq.~\eqref{def:Urhovbialg} with which one can directly verify Eq.~\eqref{eq:MUpentagon} using only knowledge in group theory, without using any knowledge about Hopf algebras.
\subsubsection{Twisting Eq.~\eqref{eq:GroupUgate} by a permutation}\label{sec:twistedperm}
We start with a simple twist of Eq.~\eqref{eq:GroupUgate}. Let $\varphi\in S_d$ be an arbitrary permutation of the set $\{1,2,\ldots,d\}$, and let $g_1,g_2,\ldots, g_d$ be arbitrary elements of $G$ as before. We define a 2-qudit gate as
\begin{equation}\label{eq:GroupUgatePermute}
	\hat{U}=\hat{X}\left[\sum_{i=1}^d  \rho(g_i)\otimes  \ket{\varphi(i)}\bra{i}  \right].
\end{equation}
Notice that taking $\varphi=\id$~(the identity map) gives back Eq.~\eqref{eq:GroupUgate}. %
In the following we consider the cyclic permutation $\varphi(i)=[(i+1){\mod d}]$ as a simple example; arbitrary permutations can be treated in a similar way. The tensors in Eq.~\eqref{def:Urhovbialg} are explicitly constructed as follow.
The horizontal indices $x,y$ have dimension $n|G|^n$, and are labeled by a list $(k,h_1,h_2,\ldots,h_n)$ where $h_1,h_2,\ldots,h_n\in G$ and $k=0,1,\ldots,n-1$ is understood modulo $n$. 
The nonzero elements of the tensors are given by
\begin{eqnarray}\label{def:UrhovTP}	
	\begin{tikzpicture}[baseline={([yshift=-0.5ex]current bounding box.center)}, scale=0.8]
		\rhotensor{0}{0}%
		\paraindices{0}{0}{(k-1,h_2,\ldots,h_n,h_1)}{(k,h_1,\ldots,h_n)}
		\quantumindices{0}{0}{a}{b}
	\end{tikzpicture}&=&\rho_{ab}(h_1)  ,\nonumber\\ 	%
	\begin{tikzpicture}[baseline={([yshift=-0.5ex]current bounding box.center)}, scale=0.8]
		\vtensor{0}{0}%
		\paraindices{0}{0}{(k,h_1g_{j},\ldots,h_ng_{j+n-1})}{(k,h_1,\ldots,h_n)}
		\quantumindices{0}{0}{j+k}{j}
	\end{tikzpicture}&=&1,\nonumber\\
	\begin{tikzpicture}[baseline={([yshift=-0.5ex]current bounding box.center)}, scale=0.63]
		\unittensor{0}{0}
		\node at (-0.3,0) [left] {\footnotesize $(k,1,\ldots,1)$};
	\end{tikzpicture}&=&1, \nonumber\\
	\begin{tikzpicture}[baseline={([yshift=-0.5ex]current bounding box.center)}, scale=0.63]
		\counittensor{0}{0}
		\node at (0.3,0) [right] {\footnotesize $(0,h_1,\ldots,h_n)$};
	\end{tikzpicture}&=&1.
\end{eqnarray}
Any unspecified elements of these tensors are equal to zero.
In App.~\ref{app:HA-TP} we give the Hopf algebra structure behind this construction.  However, 
as before, one can directly verify Eq.~\eqref{eq:MUpentagon} without using any knowledge of Hopf algebras.

\subsubsection{A family of deterministic gates}\label{sec:XYXgate}
Let $G$ be a finite group. %
As in Eq.~\eqref{eq:finitegroupgate} the local qudit has dimension $d=|G|$ and we label its basis states by elements of $G$.  Define a 2-qudit gate as
\begin{equation}\label{eq:finitegroupExtSeth}
	\hat{U}\ket{h,g}=\ket{gh^{-1},gh^{-1}g^{-1}},\quad \forall g,h\in G.
\end{equation}

The tensors in Eq.~\eqref{def:Urhovbialg} are defined as follow. The horizontal indices $x,y$ have dimension $2|G|^2$, and are labeled by a triple $(a,g,s)$ where $a,g\in G$ and $s=0,1$. The vertical indices $a,b,i,j$ are labeled by elements of $G$ as before.  The nonzero elements of the tensors are given by
\begin{eqnarray}\label{def:UrhovXYX}	
	\begin{tikzpicture}[baseline={([yshift=-0.5ex]current bounding box.center)}, scale=0.9]
		\rhotensor{0}{0}%
		\paraindices{0}{0}{(a,gh',0)}{(a,g,0)}
		\quantumindices{0}{0}{ah a'}{h}
	\end{tikzpicture}&=&1,\nonumber\\ 	
	\begin{tikzpicture}[baseline={([yshift=-0.5ex]current bounding box.center)}, scale=0.9]
		\rhotensor{0}{0}%
		\paraindices{0}{0}{(a,gh',1 )}{(a,g,1)}
		\quantumindices{0}{0}{agh'g' a'}{h}
	\end{tikzpicture}&=&1,\nonumber\\
	\begin{tikzpicture}[baseline={([yshift=-0.5ex]current bounding box.center)}, scale=0.9]
		\vtensor{0}{0}%
		\paraindices{0}{0}{(agh ,h'g'h ,\bar{s})}{(a,g,s)}
		\quantumindices{0}{0}{gh}{h}
	\end{tikzpicture}&=&1,\nonumber\\
	\begin{tikzpicture}[baseline={([yshift=-0.5ex]current bounding box.center)}, scale=0.63]
		\unittensor{0}{0}
		\node at (-0.3,0) [left] {\footnotesize $(a,g,s)$};
	\end{tikzpicture}&=&\deltak_{s,0}\deltak_{a,1}, \nonumber\\
	\begin{tikzpicture}[baseline={([yshift=-0.5ex]current bounding box.center)}, scale=0.63]
		\counittensor{0}{0}
		\node at (0.3,0) [right] {\footnotesize $(a,g,s)$};
	\end{tikzpicture}&=&\deltak_{g,1}.
\end{eqnarray}
where $\bar{s}=1-s$ and we use prime to indicate inverse, e.g., $g'=g^{-1},h'=h^{-1}$, and any unspecified elements of the tensors are zero. With these tensors  one can directly verify Eq.~\eqref{eq:MUpentagon} using group theory computations. 
In App.~\ref{app:HA-XYX} we give the Hopf algebra structure behind this construction. %

\section{A non-dual unitary example from a  $\C^*$-weak Hopf algebra}\label{sec:GoldenModel} %
In this section we construct a non-dual-unitary solvable circuit  using a  $\C^*$-weak Hopf algebra with Fibonacci anyon fusion rules~\cite{Bohm1996WHALeeYang}, and then we present exact results for all the solvable physical quantities defined in Sec.~\ref{sec:solvableQCA}.
\subsection{Model definition}\label{sec:modeldef_Fib}

We construct the solvable gate using a $13$-dimensional $\C^*$-weak Hopf algebra $\A_\mathrm{Fib}$~\cite{Bohm1996WHALeeYang} along with a $3$-dimensional representation $\rho$ and a $3$-dimensional corepresentation $v$, and we give the relevant algebraic structures  in App.~\ref{app:FibonacciWHA}. According to the general formalism in Sec.~\ref{sec:circuitdef}, since $\A_\mathrm{Fib}$ is a weak Hopf algebra, 
the solvable tensor constructed from Eq.~\eqref{def:Urhovbialg} is not unitary, but an isometry
\begin{equation}
	\begin{tikzpicture}[baseline={([yshift=-0.5ex]current bounding box.center)}, scale=0.63]
		\WgateredInd{0}{0}{$i$}{$a$}{$b$}{$j$}
	\end{tikzpicture}=\langle i,a |\hat{U}\hat{P}|b,j\rangle=\langle i,a |\hat{P}\hat{U}|b,j\rangle,
\end{equation}
where $\hat{U}$ is a unitary gate defined as
\begin{eqnarray}\label{eq:FibUgatedef}
	\hat{U}|3,3\rangle&=&\zeta |2,1\rangle-\zeta^2|3,3\rangle,\nonumber\\
	\hat{U}|2,1\rangle&=&\zeta^2 |2,1\rangle+\zeta|3,3\rangle,\nonumber\\
	\hat{U}|a,b\rangle&=& |a,b\rangle,~~~(a,b)\notin \{(2,1),(3,3)\},
\end{eqnarray}
and $\hat{P}$ is a rank-5 projection operator onto the subspace 
$\mathrm{span}\{\ket{2,1},\ket{3,3},\ket{1,2},\ket{1,3},\ket{3,2}\}$.
The other tensors in Eq.~\eqref{def:Urhovbialg} can also be directly constructed from algebraic data, and we use them later in Sec.~\ref{sec:Fibquenchdynamics} to obtain exact results for physical quantities. %
Note that in this case the projector $\hat{Q}$ in Eq.~\eqref{eq:WHAUP} happens to be equal to $\hat{P}$, as we mentioned at the end of Sec.~\ref{sec:circuitdef}, and in this example, Eqs.~(\ref{eq:WHAUP}-\ref{eq:UPCommuteInv}) of Sec.~\ref{sec:circuitdef} can be checked directly without using the weak Hopf algebra structure. For example,  
it can be straightforwardly checked that $\hat{U}$ and $\hat{P}$ satisfy the commutativity condition: %
\begin{equation}\label{eq:UPcommuteFib}
[\hat{P}_{i,i+1/2},\hat{P}_{j,j+1/2}]=[\hat{P}_{i,i+1/2},\hat{U}_{j,j+1/2}],~\forall i,j,
\end{equation} 
which immediately implies that $\hat{U}_{j,j+1/2}$ leaves the solvable subspace $\mathrm{P}$ invariant. 
The dimension of the solvable subspace $\mathrm{P}$ for an $\NN$-site chain with open boundary condition is $D_\NN=F_{\NN+3}$, where $F_\NN$ is the $\NN$-th Fibonacci number, defined recursively by $F_1=F_2=1$, and $F_{\NN+2}=F_{\NN+1}+F_\NN$, for $\NN\geq 1$. 

\subsection{Relation to Floquet PXP models}
Before we present the exact results, we first mention an equivalent, but simpler version of this model in the form of an interaction-round-a-face~(IRF) gate, which is  closely related to the famous PXP model~\cite{Turner2018QMBS}. 
In the following we first present the IRF version of this model in Sec.~\ref{sec:FibIRF} and then in Sec.~\ref{sec:FibIRFeqv} we present the exact mapping between the qutrit model~\eqref{eq:FibUgatedef} and the IRF model. 
\subsubsection{The IRF model in Rydberg-constraint space}\label{sec:FibIRF}
In the equivalent IRF version of this model, the local Hilbert space of a qubit is spanned by two states which we denote as $\ket{I}, \ket{\tau}$. The many-body solvable subspace $\mathrm{P}'$ 
is spanned by all states in which neighboring qubits cannot simultaneously occupy the state $\ket{I}$, i.e., state $\ket{I}$ satisfies a Rydberg-type constraint, and this subspace is invariant under time evolution. For an $\NN$-site chain with open boundary, $\mathrm{P}'$ has dimension $\mathrm{dim}[\mathrm{P}'_\NN]=F_{\NN+2}$. 
The time evolution operator is defined similar to Eq.~\eqref{eq:Uevolutionoperator} 
\begin{eqnarray}\label{eq:UevolutionoperatorIRF}
	\hat{\mathbf{U}}&=&\hat{\mathbf{U}}_e \hat{\mathbf{U}}_o\equiv\prod^\N_{j=1} \hat{U}_{j-1/2,j,j+1/2}\prod^\N_{j=1} \hat{U}_{j-1,j-1/2,j}\\
	&=&
	\begin{tikzpicture}[baseline={([yshift=-0.5ex]current bounding box.center)}, scale=0.63]
		\foreach \i in {0,...,4}
		{\foreach \j in {0}
			{\FCtensorMF{2*\i}{2*\j}}}
		\foreach \i in {0,...,4}
		{\foreach \j in {0}
			{\FCtensorMF{2*\i+1}{2*\j+1}}}
	\end{tikzpicture}\nonumber
\end{eqnarray}
where in this case the local gate $\hat{U}_{j-1/2,j,j+1/2}$~[for $j\in\mathbb{Z}\cup (\mathbb{Z}+1/2)$] applies a unitary transformation $\hat{U}_{\alpha\beta}$ of the qubit $j$ controlled by the states $\ket{\alpha}$ and $\ket{\beta}$ of the neighboring qubits $j-1/2$ and $j+1/2$, respectively, with $[\hat{U}_{\alpha\beta}]_{ab}=
\begin{tikzpicture}[baseline=(current bounding box.center), scale=.3]
	\FCtensorMF{0}{0}
	\FCObjLabels{0}{0}{\alpha}{\beta}{a}{b}
\end{tikzpicture}$ and the elements of the IRF tensor is given by
\begin{eqnarray}\label{def:FibIRF}
	\begin{tikzpicture}[baseline=(current bounding box.center), scale=.5]
		\FCtensorMF{0}{0}
		\FCObjLabels{0}{0}{\tau}{\tau}{a}{b}
	\end{tikzpicture}&=&\begin{pmatrix}
		\zeta^2 & \zeta \\
		\zeta & -\zeta^2
	\end{pmatrix}_{ab},\quad a,b\in\{I,\tau\},\\
	\begin{tikzpicture}[baseline=(current bounding box.center), scale=.5]
		\FCtensorMF{0}{0}
		\FCObjLabels{0}{0}{I}{\tau}{a}{b}
	\end{tikzpicture}&=&
	\begin{tikzpicture}[baseline=(current bounding box.center), scale=.5]
		\FCtensorMF{0}{0}
		\FCObjLabels{0}{0}{\tau}{I}{a}{b}
	\end{tikzpicture}=
	\begin{tikzpicture}[baseline=(current bounding box.center), scale=.5]
		\FCtensorMF{0}{0}
		\FCObjLabels{0}{0}{I}{I}{a}{b}
	\end{tikzpicture}=\deltak_{a,\tau}\deltak_{b,\tau},\nonumber
\end{eqnarray}
where the rows and columns of the matrix in the first line is arranged in order $I,\tau$. It can be checked directly that the local gates $\hat{U}_{j-1/2,j,j+1/2}$ leave the solvable subspace $\mathrm{P}'$ invariant. 

The model in Eq.~\eqref{def:FibIRF} is very close to the floquet PXP model~\cite{wilkinson2020exact,giudici2023unraveling}. More precisely, we have the following relation
\begin{eqnarray}\label{eq:IRFmodelExpForm}
	\hat{U}_{j-1/2,j,j+1/2}=i \exp\left[-\frac{\pi i}{2} \hat{H}_j\right],
\end{eqnarray}
where $\hat{H}_j$ is a local Hamiltonian acting on three neighboring sites defined as
\begin{equation}\label{eq:PXPzeta}
\hat{H}_j=\hat{p}^\tau_{j-1/2}(\zeta \hat{\sigma}^x_j+\zeta^2\hat{\sigma}_j^z)\hat{p}^\tau_{j+1/2},
\end{equation}
where $\hat{p}^\tau_{j}=(1-\hat{\sigma}^z_j)/2$.
Notice that $\hat{H}_j$ is a special point of the following parametrized Hamiltonian
\begin{equation}\label{eq:PXPtheta}
	\hat{H}_j(\theta)=\hat{p}^\tau_{j-1/2}[\hat{\sigma}^x_j\cos\theta +\hat{\sigma}_j^z \sin\theta ]\hat{p}^\tau_{j+1/2}
\end{equation}
at  $\theta=\arccos(\zeta)$, while the PXP model~\cite{Turner2018QMBS} corresponds to the special point $\theta=0$. In particular, one obtains the integrable  Floquet-PXP cellular automaton by inserting $\hat{H}_j(0)$  into Eq.~\eqref{eq:IRFmodelExpForm}. 
This model can also be viewed as a solvable floquet dynamics of the golden chain model~\cite{Feiguin2007GoldenChain} describing the dynamics of  interacting anyons.

\subsubsection{Mapping between the qutrit model and the IRF model}\label{sec:FibIRFeqv}
We now give an exact mapping between the qutrit model in Eq.~\eqref{eq:FibUgatedef} with $\NN$ sites  and the IRF model in Eq.~\eqref{def:FibIRF} with $\NN-1$ sites. 
In open boundary condition, the solvable subspace of the qutrit model further splits into a direct sum of smaller invariant subspaces, and in the following we focus on the subspace that can be reached from the product state $\ket{33\ldots3}$ in the dynamics generated by $\hat{\mathbf{U}}$, and we denote this subspace by $\mathrm{P}^0$. %
For example, with $\NN=2$, $\mathrm{P}^0$ is spanned by $\ket{21}, \ket{33}$, and with $\NN=3$, $\mathrm{P}^0$ is spanned by $\ket{333}, \ket{213}, \ket{321}$, and one can prove by induction that 
$\mathrm{dim}[\mathrm{P}^0_{\NN}]=\mathrm{dim}[\mathrm{P}'_{\NN-1}]=F_{\NN+1}$. In the following we show that the dynamics of the qutrit model restricted to $\mathrm{P}^0_{\NN}$ exactly maps to the dynamics of the IRF model restricted to $\mathrm{P}'_{\NN-1}$. 

We first define a mapping between the Hilbert spaces of the two models.  
To map a state in $\mathrm{P}^0_{\NN}$ to a state in $\mathrm{P}'_{\NN-1}$,  apply the following rule: on each link of the 1D lattice of the qutrit model, write $I$ if the two neighboring qutrits are in the state $21$, otherwise write $\tau$~(i.e. for the other four cases 12,13,32,33). Then the state of the links gives the corresponding IRF state.  %
Fig.~\ref{fig:IRFQutrittMapping} shows an example of this mapping, so in this case  
we have
\begin{alignat}{32}\label{eq:IRFQutrittMapping}
		&|2&&~&&1&&~&&2&&~&&1&&~&&3&&~&&3&&~&&2&&~&&1&&~&&2&&~&&1&&~&&3&&~&&2&&~&&1&&~&&3&&~&&3&&\rangle&&\in&& \mathrm{P}^0_{15}\nonumber\\
	 \to&~|&&I&&~&&\tau&&~&&I&&~&&\tau&&~&&\tau&&~&&\tau&&~&&I&&~&&\tau&&~&&I&&~&&\tau&&~&&\tau&&~&&\tau&&~&&\tau&&~&&\tau&&\rangle&&\in&&  \mathrm{P}'_{14}.
\end{alignat}
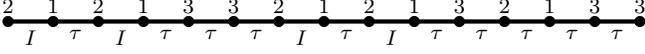
\begin{figure}
\begin{tikzpicture}[baseline={([yshift=-0.5ex]current bounding box.center)}, scale=.6]
	\oneDchain{14}
	\hindices{{2,1,2,1,3,3,2,1,2,1,3,2,1,3,3}}{1}{0}{above}
	\hindices{{I,\tau,I,\tau,\tau,\tau,I,\tau,I,\tau,\tau,\tau,\tau,\tau}}{1.5}{0}{below}
\end{tikzpicture}
\caption{\label{fig:IRFQutrittMapping} Illustration of the mapping in Eq.~\eqref{eq:IRFQutrittMapping} between a  state in $\mathrm{P}^0_{\NN}$ and a state in $\mathrm{P}'_{\NN-1}$. }
\end{figure}

Inverse mapping is also straightforward: for any IRF basis state in $\mathrm{P}'_{N-1}$, %
one first label the local states of $\ket{\psi}$ on the links of a 1D chain, and for any link with $I$, write 2  and 1 to the two adjacent vertices on the left and right of the link, respectively. Note that this can always be done without obstruction, since no two $I$ states are allowed to be neighbors in $\ket{\psi}$. Then one writes $3$ on all the remaining unlabeled sites. It is straightforward to see that this gives the inverse mapping to the above. 

With this mapping between the two Hilbert spaces $\mathrm{P}^0_{\NN}$ and  $\mathrm{P}'_{\NN-1}$, it is straightforward to show that the qutrit gate in Eq.~\eqref{eq:FibUgatedef} is mapped to the IRF gate in Eq.~\eqref{def:FibIRF}.

\subsection{Quench dynamics from an initial product state}\label{sec:Fibquenchdynamics}
\begin{figure}
	\includegraphics[width=.85\linewidth]{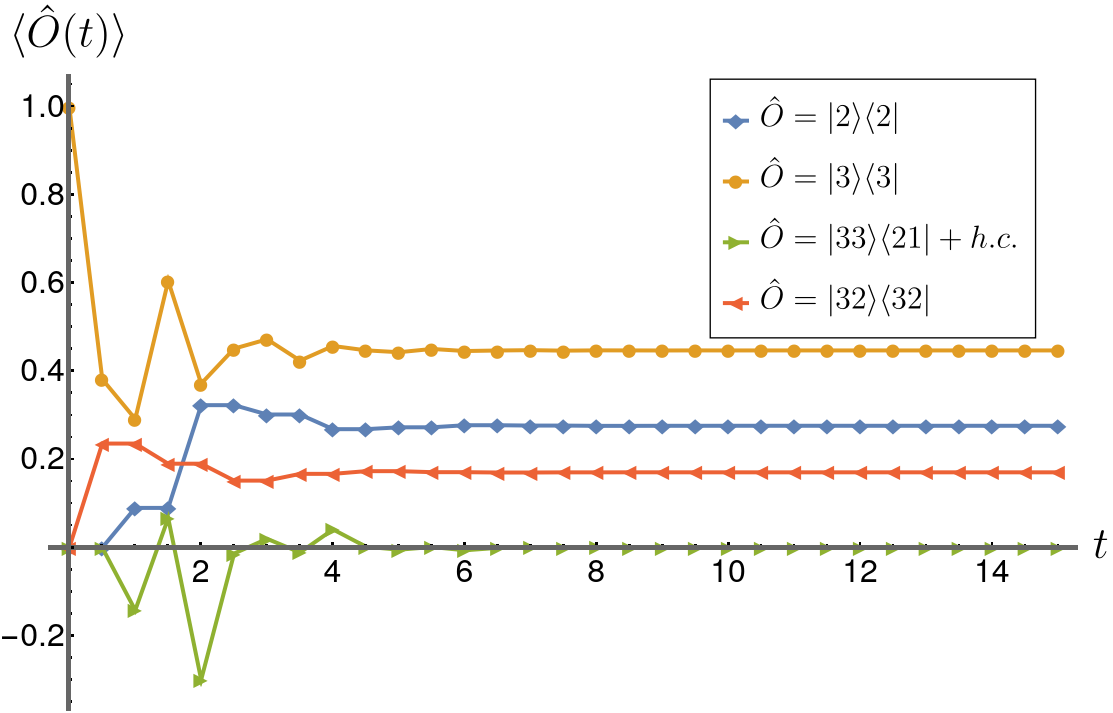}
	\caption{\label{fig:Observables} Time evolution of single site observables, computed in the state $|\psi(t)\rangle$ initialized from the product state $|\psi(0)\rangle=|33\ldots3\rangle$,  for the model defined in Eq.~\eqref{eq:FibUgatedef}. }
\end{figure}

\begin{figure}
	\centering
	\begin{subfigure}[t]{.48\linewidth}
		\centering\includegraphics[width=\linewidth]{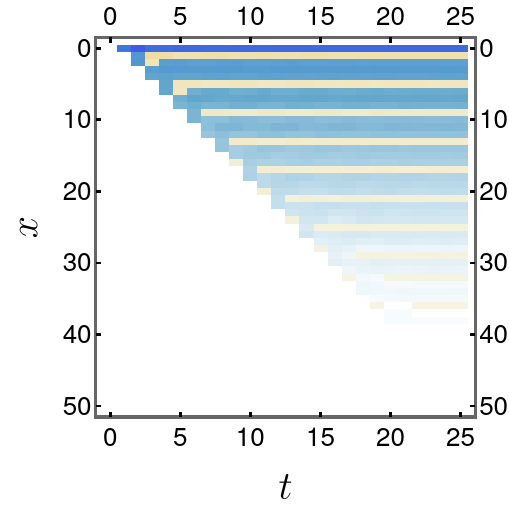}
		\caption{$W^{11}(x,t)$}
	\end{subfigure}
	\begin{subfigure}[t]{.48\linewidth}
		\centering\includegraphics[width=\linewidth]{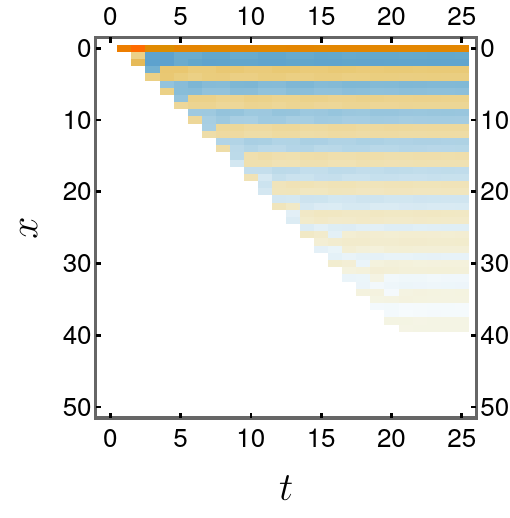}
		\caption{$W^{13}(x,t)$}
	\end{subfigure}
	\begin{subfigure}[t]{.48\linewidth}
		\centering\includegraphics[width=\linewidth]{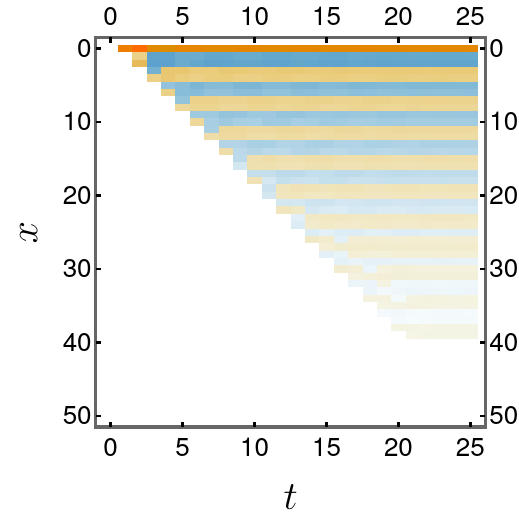}
		\caption{$W^{32}(x,t)$}
	\end{subfigure}
	\begin{subfigure}[t]{.48\linewidth}
		\centering\includegraphics[width=\linewidth]{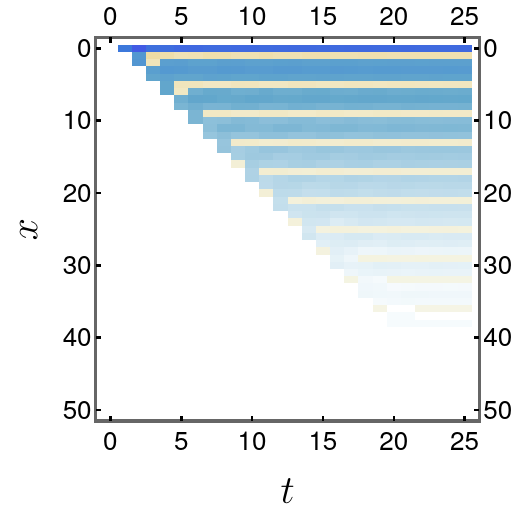}
		\caption{$W^{22}(x,t)$}
	\end{subfigure}

	\caption{\label{fig:Corrfunc} Results for equal time correlation functions $W^{ij}(x,t)$ %
		defined in Eq.~\eqref{eq:def:WijFib},
		for the model defined in Eq.~\eqref{eq:FibUgatedef}, with the initial state $|\psi(0)\rangle=|33\ldots3\rangle$. } %
\end{figure}

\begin{figure}
	\includegraphics[width=.8\linewidth]{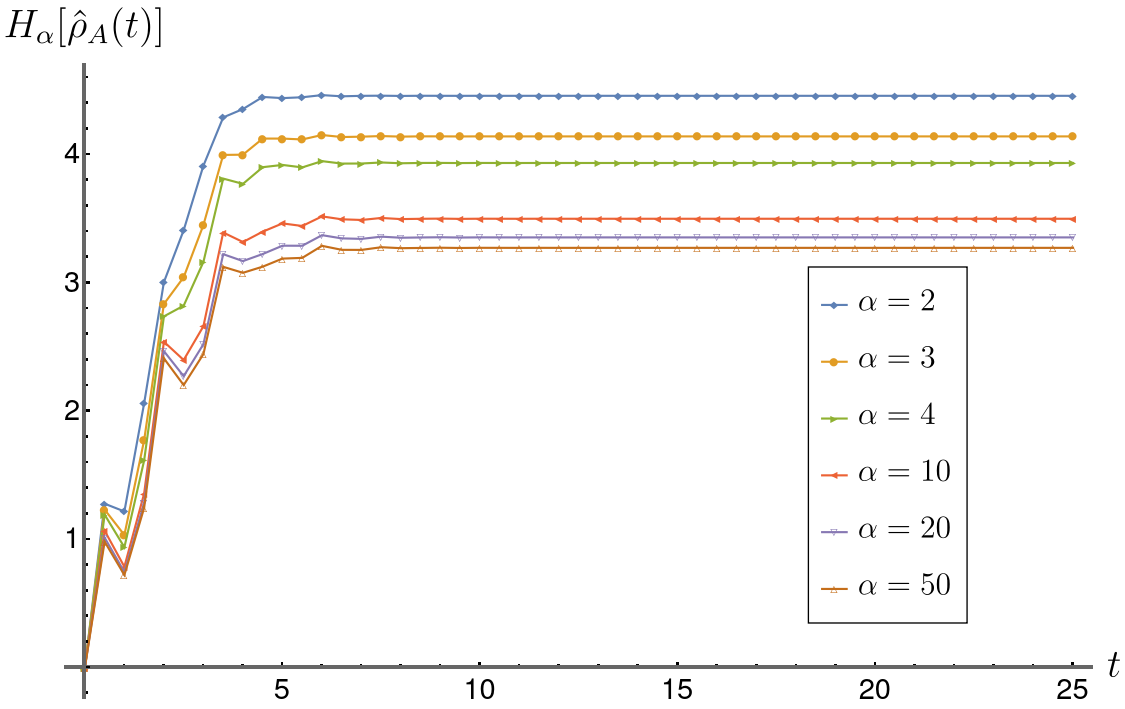}
	\includegraphics[width=.8\linewidth]{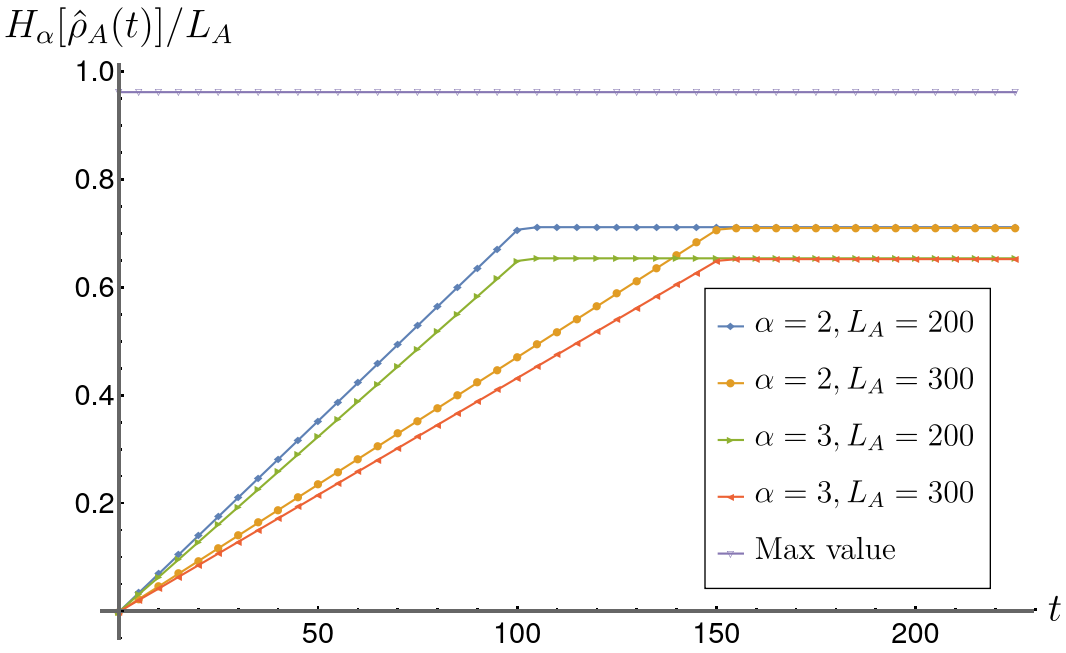}
	\includegraphics[width=.75\linewidth]{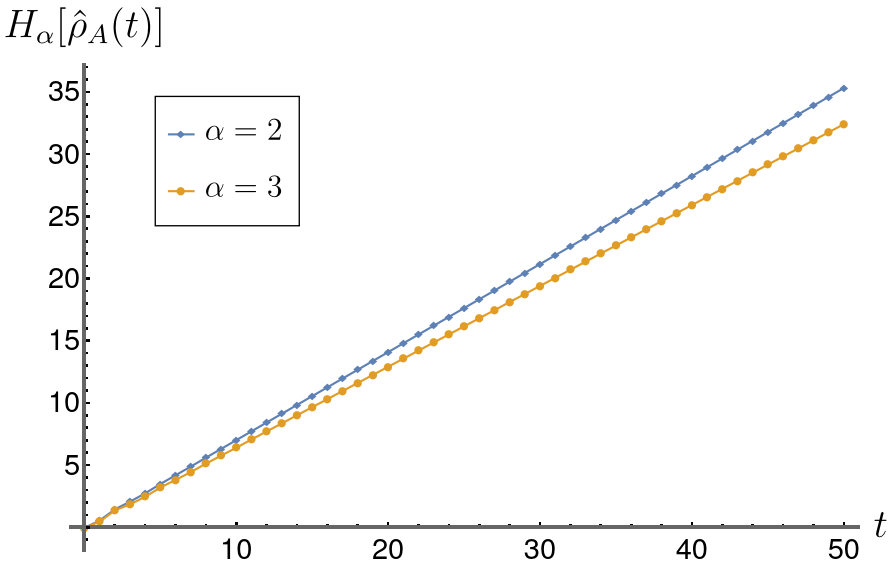}
	\caption{\label{fig:Renyi-Fib} Renyi entanglement entropy $H_\alpha[\hat{\rho}_A(t)]$ for the non-dual unitary solvable gate defined in Eq.~\eqref{eq:FibUgatedef}, with initial state $|\psi(0)\rangle=|33\ldots3\rangle$.  
		(top) for a small subsystem of size $\N_A=5$; %
		(middle) for large subsystems of sizes $\N_A=200$ and $300$; %
		(bottom) for the subsystem  being a semi-infinite half chain. }
\end{figure}

We now compute physical quantities for the time evolved state $|\psi(t)\rangle=\hat{\mathbf{U}}(t)|\psi(0)\rangle$, where the initial state $|\psi(0)\rangle=|33\ldots3\rangle$ is a product state in the solvable subspace $\mathrm{P}_0$. We use 
Eq.~\eqref{eq:OtMPS}
to compute $\braket{\hat{O}(t)}$ for some  local observables $\hat{O}$ and the results are shown in Fig.~\ref{fig:Observables}. 
Note that under the mapping defined in Sec.~\ref{sec:FibIRFeqv}, these local observables are mapped to local observables in the IRF model as 
\begin{eqnarray}
	\hat{e}^{1}_j&\to&  \hat{p}^{I}_{j} \hat{p}^{\tau}_{j+1/2},\nonumber\\
	\hat{e}^2_j&\to& \hat{p}^{\tau}_{j}\hat{p}^{I}_{j+1/2},\nonumber\\ %
	\hat{e}^3_j&\to& \hat{p}^{\tau}_{j} \hat{p}^{\tau}_{j+1/2},\nonumber\\ %
	\hat{e}^{32}_j\hat{e}^{31}_{j+1/2}+\mathrm{h.c.}&\to& \hat{p}^{\tau}_{j}\hat{\sigma}^{x}_{j+1/2} \hat{p}^{\tau}_{j+1},\nonumber\\ 
\end{eqnarray}
where $\hat{e}^{\alpha\beta}=\ket{\alpha}\bra{\beta}$, for $\alpha,\beta=1,2,3$, $\hat{e}^{\alpha}\equiv \hat{e}^{\alpha\alpha}$, and, under the mapping illustrated in Fig.~\ref{fig:IRFQutrittMapping}, we use the convention that the site $j$ of the qutrit model is to the right of the link $j$ of the IRF model. 
We then  use Eq.~\eqref{eq:MPOOrOv} to compute  equal time correlation functions 
\begin{equation}\label{eq:def:WijFib}
W^{ij}(x,t)=\braket{\hat{e}^{i}_0(t)\hat{e}^{j}_x(t)}-\braket{\hat{e}^{i}_0(t)}\braket{\hat{e}^{j}_x(t)},%
\end{equation} 
and the results are shown in Fig.~\ref{fig:Corrfunc}.
We then compute the growth of Renyi entanglement entropy of a subsystem $A$ for  different subsystem sizes $\N_A$ and at different Renyi index $\alpha$, using the tensor network methods  in  Sec.~\ref{sec:RenyiEE}, and the results are shown in Fig.~\ref{fig:Renyi-Fib}. 
The horizontal line in the second plot of Fig.~\ref{fig:Renyi-Fib} indicates the maximal possible value of entanglement entropy density per unit cell, i.e., the entanglement entropy density of the infinite temperature state,  which is $2\ln(\zeta^{-2})$ in this case. All the figures in this section can be reproduced by  the accompanying Mathematica code~\cite{HAQCACode}.

\subsection{Equilibration time and revival time}\label{sec:FibTimescales}
From Fig.~\ref{fig:Renyi-Fib} we can clearly see that the Renyi entanglement entropy of a subsystem of size $\N_A$ equilibrates at time $t^*=\N_A/2+O(1)$, verifying the general result in Eq.~\eqref{eq:eqltime}. The revival time of a periodic system of size $\N$ is given in Eq.~\eqref{eq:trevHA-0}, %
where in this case the underlying weak Hopf algebra $\A_{\mathrm{Fib}}$ has exponent $\eta=5$. %
Eq.~\eqref{eq:trevHA-0} is also verified numerically for this model for small system sizes up to $\N=4$ in the accompanying Mathematica code~\cite{HAQCACode}. 

\subsection{Spatiotemporal correlation functions}\label{sec:GoldenModel:st_corr_func}
Spatiotemporal correlation functions can be evaluated analytically by inserting the MPO representation Eq.~\eqref{eq:local_observableMPO} into the definition in Eq.~\eqref{def:STcorr}, and the result is that for all single site observables that leaves the solvable subspace $\mathrm{P}$ invariant, $C(\hat{A},\hat{B};x,t)$ has the form
\begin{equation}\label{eq:FibSTcorr}
	C(\hat{A},\hat{B},x,t)=c_0(\hat{A},\hat{B})+c_1 (\hat{A},\hat{B}) (-\zeta^4)^{2t+1},
\end{equation}
for   $t\in \frac{1}{2}\mathbb{Z}$, $x-t\in\mathbb{Z}$, and $|x|\leq t$. When $|x|> t$~(outside the lightcone), we have  $C(\hat{A},\hat{B},x,t)=0$ as always. 
The coefficients $c_0(\hat{A},\hat{B})$ and $c_1 (\hat{A},\hat{B})$ potentially depend on whether the operators $\hat{A}$ and $\hat{B}$ lie on $\rho$-legs or $v$-legs, as defined in Sec.~\ref{sec:circuitdef}. %
When $\hat{A}=\hat{e}^i,\hat{B}=\hat{e}^j$ are both single site operators lying on $v$-links~[as in Eq.~\eqref{eq:local_observableMPO}], %
we have  %
\begin{eqnarray}
	c_0(\hat{e}^i,\hat{e}^j)&=& \frac{1}{10}
	\begin{bmatrix}
		3-\sqrt{5} & 3-\sqrt{5} & \sqrt{5}-1 \\
		3-\sqrt{5} & 3-\sqrt{5} & \sqrt{5}-1 \\
		\sqrt{5}-1 & \sqrt{5}-1 & 2 \\
	\end{bmatrix}_{ij},\\
	c_1(\hat{e}^i,\hat{e}^j)&=& \frac{1}{10}
	\begin{bmatrix}
		-3-\sqrt{5} & 2 & 1+\sqrt{5} \\
		7+3 \sqrt{5} & -3-\sqrt{5} & -4-2\sqrt{5} \\
		-4-2\sqrt{5}  & 1+\sqrt{5} & 3+\sqrt{5} \\
	\end{bmatrix}_{ij}.\nonumber
\end{eqnarray}
The exact result in Eq.~\eqref{eq:FibSTcorr} shows a distinct behavior compared to dual unitary circuits, whose spatiotemporal correlation functions always vanish inside the lightcone~\cite{Prosen2019DU}. This explicitly demonstrates, at the physical level,  our earlier claim that $\C^*$-weak Hopf algebras are capable of producing solvable unitary circuits that are not dual unitary.  
\subsection{OTOCs}\label{sec:FibOTOC}
\begin{figure}
	\includegraphics[width=.85\linewidth]{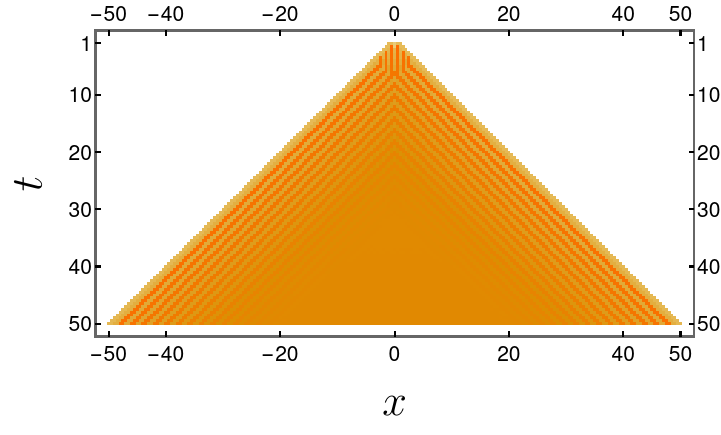}
	\caption{\label{fig:FibOTOC} Exact result for the OTOC $F(\hat{V},\hat{W},x,t)$~[defined in Eq.~\eqref{def:OTOC}] between two randomly chosen local unitary operators $\hat{V},\hat{W}$.}
\end{figure}

We now compute the OTOC $F(\hat{V},\hat{W},x,t)$ for this model defined in Eq.~\eqref{def:OTOC}, where we choose the local unitary operators $\hat{V}$ and  $\hat{W}$ randomly. The exact result is shown in Fig.~\ref{fig:FibOTOC}. From the result we clearly see that $F(\hat{V},\hat{W},x,t)$ saturates to a constant value as $t\to\infty$. This demonstrates a clear physical distinction from solvable Clifford circuits, whose $F(\hat{V},\hat{W},x,t)$ keeps oscillating at late times and does not settle to a constant value even in the thermodynamic limit~\cite{Keyserlingk2018OperatorHydrodynamics}. 

\section{Summary and outlook}\label{sec:concl}%
In summary, we have introduced a new family of solvable unitary circuits based on finite dimensional $\C^*$ (weak) Hopf algebras. We have shown how to systematically construct new solvable models from algebraic data, and how the underlying algebraic structures  enable the exact computation of all the relevant physical quantities in the full evolution dynamics, for any initial MPS. The fundamental building block of our construction is a set of tensors~(constructed from algebraic data) satisfying Eq.~\eqref{eq:MUpentagon}, which leads to an MPO representation of a 2D circuit tensor network in Eq.~\eqref{eq:triangularTNMPO}, which then leads to the MPO representation of time evolved local observables~[Eq.~\eqref{eq:local_observableMPO}] and a 1D tensor network representation of the Renyi entanglement entropy~[via the reduced density matrix in Eq.~\eqref{eq:DMMPOFinSubs}], allowing these physical quantities to be exactly computed. 

We then presented several explicit examples of this construction, including several dual unitary gates constructed from $\C^*$-Hopf algebras in Sec.~\ref{sec:HAexamples}, which are shown to be much more solvable than generic dual unitary gates. Then we presented a non-dual unitary gate constructed from a $\C^*$-weak Hopf algebra in Sec.~\ref{sec:GoldenModel}, and studied in detail the full evolution dynamics of this model and presented exact results for all the physical quantities. We also provided an exact mapping of this model to a PXP-type model generated by an IRF gate defined in Eq.~\eqref{def:FibIRF}, which defines a solvable floquet dynamics for the Rydberg chain and may shed light on the physics of quantum many body scars~\cite{giudici2023unraveling}.  %

We now mention some potential generalizations of this work. First, as we already hinted in Sec.~\ref{sec:GroupHA}, we expect that our construction can be straightforwardly generalized to certain class of infinite dimensional $\C^*$-Hopf algebras. In this generalization, the MPO representation of $\hat{O}(t)$ in Eq.~\eqref{eq:local_observableMPO} is still formally valid, but the bond dimension grows polynomially in the evolution time, which still guarantees efficient solvability of $\braket{\hat{O}(t)}$ for any initial MPS. This could lead to a much bigger family of solvable unitary circuits. In particular, since the Larson-Radford theorem~\cite{LarsonRadford1988} does not apply to infinite dimensional $\C^*$-Hopf algebras, this can potentially generate non dual unitary  gates solvable in the entire Hilbert space~(in contrast to the $\C^*$-weak Hopf algebra examples that are only solvable in a subspace). This generalization can also incorporate some known integrable circuits into this family. For example, it is known that $\hat{O}(t)$ in the quantum Rule 54 model~\cite{Friedman2019Rule54,Klobas2019TDMPA,Klobas2021Rule54,Buca2021Rule54} has an MPO representation~\cite{Klobas2019TDMPA,Alba2019Rule54LOMPO} with bond dimension growing polynomially in $t$, so it is interesting to investigate if the Rule 54 gate can be constructed from a suitable infinite dimensional Hopf algebra. 

Another interesting generalization is to construct solvable unitary circuits from unitary fusion categories~\cite{TenCat_EGNO}, which is an important mathematical structure underlying 2+1-dimensional topological phases~\cite{Kitaev2006}. This is possible due to the close relationship between $\C^*$-(weak) Hopf algebras and unitary fusion categories~\cite{TenCat_EGNO}. Indeed, we already know that the IRF model defined in Eq.~\eqref{def:FibIRF} can be more conveniently constructed from the Fibonacci fusion category, which is the unitary (braided) fusion category  describing Fibonacci anyons. This generalization may also lead to a bigger family of solvable models. %
Further generalization to unitary fusion 2-categories may allow us to construct unitary circuits in 2+1 dimensions where $\hat{O}(t)$ is an exact projected entangled paired operator with finite bond dimension. Although this may not allow a full exact solution in the 2+1D case, some physical properties may already be deduced from this tensor network representation.

Furthermore, the method of constructing solvable tensor networks using bialgebras introduced in Sec.~\ref{sec:general_formalism} may have other applications in (quantum) many body physics where tensor network methods are applied.  For example,
if all elements of the solvable tensor
\begin{tikzpicture}[baseline={([yshift=-0.5ex]current bounding box.center)}, scale=0.63]
	\Wgatered{0}{0}
\end{tikzpicture} are positive, then we obtain a solvable 2D classical statistical model~\cite{baxter2016exactly}. If 
\begin{tikzpicture}[baseline={([yshift=-0.5ex]current bounding box.center)}, scale=0.63]
	\Wgatered{0}{0}
\end{tikzpicture} has a suitable bilayer structure and is completely positive as a matrix, then we obtain a solvable projected entangled pair state~\cite{Cirac2021Matrix}. 
Similarly, if 
\begin{tikzpicture}[baseline={([yshift=-0.5ex]current bounding box.center)}, scale=0.63]
	\Wgatered{0}{0}
\end{tikzpicture} is completely positive and trace preserving, then we obtain a solvable circuit of local quantum channels in 1+1D.
The main technical challenge in these directions is that we do not yet know what kind of additional algebraic structure is needed to guarantee these physical conditions, in a similar way as how the $\C^*$-structure of the underlying Hopf algebra guarantees the unitarity of %
\begin{tikzpicture}[baseline={([yshift=-0.5ex]current bounding box.center)}, scale=0.63]
	\Wgatered{0}{0}
\end{tikzpicture}~(Thm.~\ref{thm:HopfUgate}). 
Finally, we mention that the construction of solutions to Eq.~\eqref{eq:MUpentagon} using bialgebra data, as given in Eq.~\eqref{def:Urhovbialg}, has an unexpected application in the realization of emergent $R$-paraparticles~\cite{wang2023para,wang2024parastatistics} in quantum spin systems~(in spatial dimension $d>1$), as solution to this such an equation plays a key role in the construction of the MPO Jordan-Wigner transformation~\cite{wang2023para} in dimension $d>1$~\footnote{The precise relation is that the Eq.~(S32) in Ref.~\cite{wang2023para} with $R'$ being the swap gate is essentially the first line of Eq.~\eqref{eq:MUpentagon} of this paper with $U$ being the $R$-matrix. }.

\acknowledgments
We thank J. Ignacio Cirac, Bal\'{a}zs Pozsgay, Timothy Hsieh,  Pavel Kos, Georgios Styliaris, and Xiehang Yu for helpful discussions. This work is supported by the Munich Quantum Valley~(MQV), which is supported by the Bavarian state government with funds from the Hightech Agenda Bayern Plus.
\appendix
\section{Bialgebras and the proof of Thm.~\ref{thm:bialgebraTN}}\label{app:BAandTN} %
In this section %
we first review basic concepts of finite dimensional bialgebras in App.~\ref{app:HAbasics} and then in App.~\ref{app:proofBAsolvableTN} and App.~\ref{app:SolvableTNfrompreBA} we prove Thm.~\ref{thm:bialgebraTN} and its generalization to prebialgebras. In App.~\ref{app:reduction} and App.~\ref{app:HAregREpcorep} we give a method to construct all solvable gates from a given bialgebra,  by means of decomposing the regular representation and corepresentation. In App.~\ref{app:dualstructure} we review the duality of bialgebras, which is not necessary for understanding the main text, but is used in later sections of the Appendix. 
\subsection{Bialgebra basics}\label{app:HAbasics}
We begin by reviewing the key concepts of finite dimensional bialgebras and their representation theory used in this paper; to learn more about this subject, we refer the interested readers to textbooks in Hopf algebras~\cite{Majid1995BookFoundationQG,klimyk1997book,kassel2012quantum} and representation theory~\cite{etingof2011introduction}~[see also Ref.~\cite{molnar2022matrix,Ruiz-de-Alarcon2024MPOAHopfII} for the connection to tensor network theory]. 
\begin{definition}\label{def:algebra}{(Algebras and representations)}
A (unital, associative) algebra $\A$  is a vector space equipped with an associative bilinear map $m:\A\otimes\A\to \A$, called multiplication, denoted by juxtaposition $m(x\otimes y)=xy,\forall x,y\in\A$, together with a unit element $1\in \A$ satisfying $1x=x1=x,\forall x\in \A$. A $d$-dimensional representation of  $\A$ is a linear map $\rho:\A\to M_d(\C)$ satisfying
\begin{equation}\label{def:rep}
	\rho(1)=\mathds{1},~~\rho(x)\rho(y)=\rho(xy),~~\forall x,y\in \A.
\end{equation}
Here  $M_d(\C)$ denotes the algebra of $d\times d$ matrices over $\C$, and $\mathds{1}$ denotes the $d$-dimensional identity matrix. Furthermore, $\rho$ is called a faithful representation if it is  injective as a linear map $\rho:\A\to M_d(\C)$. Two $d$-dimensional representations $\rho_1$ and $\rho_2$ are called isomorphic if there is an invertible matrix $V\in  M_d(\C)$ such that $\rho_1(x)=V\rho_2(x)V^{-1},~\forall x\in\A$.
\end{definition}

Let $B(\A)$ be a basis of $\A$ with $1\in B(\A)$ and  $|B(\A)|=d_{\A}$ the dimension of $\A$, and we define the structure constants $\{\Omega_{xy}^z\}$ of $\A$ by writing its multiplication in the basis $B(\A)$ as follows
\begin{eqnarray}\label{def:structureconstalg}
	x\cdot y=\sum_{z\in B(\A)} \Omega_{xy}^z z,\quad\forall x,y \in B(\A).%
\end{eqnarray}
Then the associativity and unitality of $\A$ translates into the following conditions for the structure constants
\begin{eqnarray}\label{eq:algcondition}
\sum_{u\in B(\A)}\Omega^u_{xy}\Omega^w_{uz}&=&\sum_{u\in B(\A)}\Omega^w_{xu}\Omega^u_{yz},\nonumber\\ %
\Omega_{x1}^y&=&\Omega_{1x}^y=\deltak_{x,y},
\end{eqnarray}
for all $x,y,z,w \in B(\A)$. 
For any algebra $\A$, we can use its structure constants 
to define a $d_\A$-dimensional representation as %
\begin{equation}\label{def:regRep}
[\rho(x)]_{zy}=\Omega_{xy}^z,\quad\forall x,y,z \in B(\A).
\end{equation} 
It is straightforward to verify that $\rho$ satisfies Eq.~\eqref{def:rep} as a consequence of Eq.~\eqref{eq:algcondition}. We call this the regular representation of $\A$. 
\begin{definition}{(Coalgebras and corepresentations)}
A (counital, coassociative) coalgebra $\A$ is a vector space equipped with a linear map called comultiplication $\Delta:\A\to\A\otimes\A$,
and a linear map called counit $\epsilon: \A\to\C$ satisfying coassociativity 
\begin{equation}\label{eq:coassociativityaxiom}
(\Delta\otimes\id)\Delta(x)=(\id\otimes\Delta)\Delta(x),~~\forall x\in\A,
\end{equation}
and the counit condition
\begin{equation}\label{eq:counitaxiom}
(\epsilon\otimes\id)\Delta(x)=(\id\otimes \epsilon)\Delta(x)=x,~~\forall x\in\A,
\end{equation}
where $\id$ denotes the identity map on $\A$, i.e., $\id(x):=x$. 
A $d$-dimensional corepresentation of  $\A$ is a $d\times d$ matrix with elements in $\A$, denoted as $v_{ij}\in\A,1\leq i,j\leq d$, satisfying  
\begin{equation}\label{def:corep}
\Delta(v_{ij})=\sum_{k=1}^d v_{ik}\otimes v_{kj},\quad \epsilon(v_{ij})=\deltak_{i,j},~~1\leq i,j\leq d.
\end{equation}
 Two $d$-dimensional corepresentations $v_1$ and $v_2$ are called isomorphic if there is an invertible matrix $V\in  M_d(\C)$ such that $$v_{1,ij}=\sum_{k,l}V_{ik} v_{2,kl} (V^{-1})_{lj}.$$
\end{definition}

Let $B(\A)$ be a basis of $\A$ with $|B(\A)|=d_{\A}$ the dimension of $\A$, and we define the structure constants $\{\Lambda^{xy}_z\}$ of $\A$ by writing its comultiplication in the basis $B(\A)$ as follows
\begin{equation}\label{def:structureconstcoalg}
	\Delta(z)=\sum_{x,y\in B(\A)}\Lambda_z^{xy}x\otimes y,	
\end{equation}
Then the coassociativity and counitality of $\A$ translates into the following conditions for the structure constants
\begin{eqnarray}\label{eq:coalgconditionsc}
	\sum_{u\in B(\A)}\Lambda_u^{xy}\Lambda_w^{uz}&=&\sum_{u\in B(\A)}\Lambda_w^{xu}\Lambda_u^{yz},\nonumber\\ %
\sum_{x\in B(\A)}\Lambda_z^{xy}\epsilon(x)&=&\sum_{x\in B(\A)}\Lambda_z^{yx}\epsilon(x)=\deltak_{y,z},
\end{eqnarray}
for all $x,y,z,w \in B(\A)$. 
For any coalgebra $\A$, we can use the its structure constants to define a $d_\A$-dimensional corepresentation as %
\begin{equation}\label{def:regCorep}
	v_{xz}=\sum_{y\in B(\A)}\Lambda^{xy}_z y,\quad\forall x,z \in B(\A).
\end{equation} 
It is straightforward to verify that $v$ satisfies Eq.~\eqref{def:corep} as a consequence of Eq.~\eqref{eq:coalgconditionsc}. We call this the regular corepresentation of $\A$. 

For a coalgebra $\A$, we follow the standard convention in the literature and use Sweedler notation for comultiplication. For any $x\in \A$, $\Delta(x)\in \A\otimes \A$ can be written as a finite sum~[e.g., by %
using Eq.~\eqref{def:structureconstcoalg}] 
\begin{equation}
	\Delta(x)=\sum_i x_{1 i} \otimes x_{2 i}, \quad x_{1 i}, x_{2 i} \in \A.
\end{equation}
Sweedler notation denotes this finite sum symbolically as
\begin{equation}
	\Delta(x)= x_{(1)} \otimes x_{(2)}.
\end{equation}
We also define the $n$-fold coproduct $\Delta^{(n)}: \A \rightarrow \A^{\otimes( n+1)}$ inductively as
\begin{equation}
	\Delta^{(n)}:=(\operatorname{id} \otimes \Delta) \circ \Delta^{(n-1)}, \quad n>1, \text { and }  \Delta^{(1)}=\Delta,
\end{equation}
and the coassociativity of $\Delta$ allows us to write the $\Delta^{(n)}(x)\in \A^{\otimes (n+1)}$ as
\begin{equation}
	\Delta^{(n)}(x):= x_{(1)} \otimes x_{(2)} \otimes \cdots \otimes x_{(n+1)}.
\end{equation}

\begin{definition}{(Prebialgebras~\cite{nill1998axioms,Bohm2011WBAnMC})}
A prebialgebra $\A$ is an algebra and a coalgebra such that the comultiplication $\Delta$ is multiplicative %
\begin{equation}\label{eq:DeltaMultiplicative}
	\Delta(xy)=\Delta(x)\Delta(y).%
\end{equation}
A representation of a prebialgebra $\A$ is simply a representation of $\A$ as an algebra, and similarly, a corepresentation of $\A$ is a corepresentation of $\A$ as a coalgebra.
\end{definition}

\begin{definition}{(Bialgebras)}
	A bialgebra $\A$ is a prebialgebra such that the unit $1$ and the counit $\epsilon$ satisfy the following axioms %
	\begin{equation}\label{eq:bialgebraunitcounit}
		\Delta(1)=1\otimes1,~~\epsilon(xy)=\epsilon(x)\epsilon(y),~~\forall x,y\in\A. %
	\end{equation}
\end{definition}
Note that for a bialgebra we always have $\epsilon(1)=1$, since, applying the map $\epsilon\otimes\id$ to the first equation in Eq.~\eqref{eq:bialgebraunitcounit} and using Eq.~\eqref{eq:counitaxiom}, we obtain $\epsilon(1)1=1$. 

\subsection{Proof of Thm.~\ref{thm:bialgebraTN}}\label{app:proofBAsolvableTN}
In the following we prove that  the tensors $
\begin{tikzpicture}[baseline={([yshift=-0.5ex]current bounding box.center)}, scale=0.63]
	\Wgatered{0}{0}
\end{tikzpicture}$,
$%
\begin{tikzpicture}[baseline={([yshift=-0.5ex]current bounding box.center)}, scale=0.63]
	\rhotensor{0}{0}%
\end{tikzpicture}$, 
$%
\begin{tikzpicture}[baseline={([yshift=-0.5ex]current bounding box.center)}, scale=0.63]
	\vtensor{0}{0}%
\end{tikzpicture}$, $\begin{tikzpicture}[baseline={([yshift=-0.5ex]current bounding box.center)}, scale=0.63]
\unittensor{0}{0}
\end{tikzpicture}$, and $\begin{tikzpicture}[baseline={([yshift=-0.5ex]current bounding box.center)}, scale=0.63]
\counittensor{0}{0}
\end{tikzpicture}$
constructed from bialgebra data via Eq.~\eqref{def:Urhovbialg} satisfy Eq.~\eqref{eq:MUpentagon}, thereby proving Thm.~\ref{thm:bialgebraTN}. We first prove the first line of  Eq.~\eqref{eq:MUpentagon}. We have
\begin{eqnarray}\label{eq:MUpentagonProof}
	\begin{tikzpicture}[baseline={([yshift=-0.5ex]current bounding box.center)}, scale=0.63]
		\rhotensor{-0.5}{1}%
		\vtensor{0.5}{1}%
		\hindices{{a,i}}{-0.5}{1.5}{above}
		\hindices{{b,j}}{-0.5}{0.5}{below}
		\paraindices{-0.5}{1}{y}{}
		\paraindices{0.5}{1}{}{x}
	\end{tikzpicture}	
	&=&\sum_z (\deltad_y\otimes\rho_{ab} )\Delta(z)\deltad_z(xv_{ij})  \nonumber\\
	&=&   (\deltad_y\otimes\rho_{ab} )\Delta(xv_{ij})  \nonumber\\
	&=&   (\deltad_y\otimes\rho_{ab} )\Delta(x)\Delta(v_{ij})  \nonumber\\
&=&(\deltad_y\otimes\rho_{ab} )\sum_k(x_{(1)} v_{ik}\otimes x_{(2)} v_{kj})\nonumber\\
&=&\sum_k \deltad_y(x_{(1)} v_{ik}) \rho_{ab}(x_{(2)} v_{kj})\nonumber\\
&=&\sum_{k,c }\deltad_y(x_{(1)} v_{ik}) \rho_{ac}(x_{(2)})\rho_{cb}( v_{kj})\nonumber\\
&=&\sum_{k,c,z}\deltad_y(z v_{ik}) (\deltad_z\otimes\rho_{ac})\Delta(x)\rho_{cb}( v_{kj})\nonumber\\
&=&
	\begin{tikzpicture}[baseline={([yshift=-0.5ex]current bounding box.center)}, scale=0.63]
		\rhotensor{0.5}{1}%
		\vtensor{-0.5}{1}%
		\paraindices{-0.5}{1}{y}{}
		\paraindices{0.5}{1}{}{x}
		\Wgatered{0}{0}
		\hindices{{i,a}}{-0.5}{1.5}{above}
		\hindices{{b,j}}{-0.5}{-0.5}{below}
	\end{tikzpicture},
	\end{eqnarray}
where in the first line we apply the definitions of the two tensors in Eq.~\eqref{def:Urhovbialg}, in the third line we use the multiplicativity of $\Delta$ in Eq.~\eqref{eq:DeltaMultiplicative}, in the fourth line we use the definition of a corepresentation in Eq.~\eqref{def:corep}, in the sixth line we use the definition of a representation in Eq.~\eqref{def:rep}, and in the last line we use the definitions of the  tensors in Eq.~\eqref{def:Urhovbialg} again. 
	
The three relations in the second line of  Eq.~\eqref{eq:MUpentagon} are proved as follow. We have
	\begin{eqnarray}
	\begin{tikzpicture}[baseline={([yshift=-0.5ex]current bounding box.center)}, scale=0.63]
		\MYsquareB{-0.5}{0}
		\vtensor{0}{0}%
		\paraindices{0}{0}{}{x}
		\quantumindices{0}{0}{i}{j}
	\end{tikzpicture}&=&\sum_y\epsilon(y)\deltad_y(x v_{ij})\nonumber\\
	&=&\epsilon(x v_{ij})\nonumber\\
	&=&\epsilon(x)\epsilon(v_{ij})\nonumber\\
	&=&
	\begin{tikzpicture}[baseline={([yshift=-0.5ex]current bounding box.center)}, scale=0.63]
		\deltatensor{0}{0}{\colorR}{}{}
		\MYsquareB{0.3}{0}
		\draw[ultra thick] (0.35, 0 ) -- (0.7,0);
			\paraindices{0}{0}{}{x}
		\quantumindices{0}{0}{i}{j}
	\end{tikzpicture}, 
	\end{eqnarray}
	where in the third line we apply the multiplicativity of $\epsilon$ in Eq.~\eqref{eq:bialgebraunitcounit}, and in the last line we use Eq.~\eqref{def:corep}. Similarly, we have
	\begin{eqnarray}
	\begin{tikzpicture}[baseline={([yshift=-0.5ex]current bounding box.center)}, scale=0.63]
		\rhotensor{0}{0}%
		\MYsquare{0.5}{0}
		\paraindices{0}{0}{y}{}
		\quantumindices{0}{0}{a}{b}
	\end{tikzpicture}&=&(\deltad_y\otimes\rho_{ab} )\Delta(1)\nonumber\\
	&=&\deltad_y(1) \rho_{ab}(1) \nonumber\\
	&=&
	\begin{tikzpicture}[baseline={([yshift=-0.5ex]current bounding box.center)}, scale=0.63]
		\deltatensor{0}{0}{\colorL}{}{}
		\draw[ultra thick] (-0.35, 0 ) -- (-0.7,0);
		\MYsquare{-0.3}{0}
		\paraindices{0}{0}{y}{}
		\quantumindices{0}{0}{a}{b}
	\end{tikzpicture}, 		
\end{eqnarray}
where in the second line we use Eq.~\eqref{eq:bialgebraunitcounit}, and in the last line we use Eq.~\eqref{def:rep}. Furthermore, $\epsilon(1)=1$ implies that $%
\begin{tikzpicture}[baseline={([yshift=-0.5ex]current bounding box.center)}, scale=0.63]
	\draw[ultra thick] (-0.25, 0 ) -- (0.25,0);
	\MYsquareB{-0.25}{0}
	\MYsquare{0.25}{0}
\end{tikzpicture}=1$. This completes the proof of  Thm.~\ref{thm:bialgebraTN}.

\subsection{Generalization of   Thm.~\ref{thm:bialgebraTN} to  prebialgebras}\label{app:SolvableTNfrompreBA}
We now generalize Thm.~\ref{thm:bialgebraTN} to prebialgebras, which implies that the derivation  for Eq.~\eqref{eq:triangularTNMPO} given in Eq.~\eqref{eq:triangularTNMPOderive} is still valid for solvable tensors constructed from prebialgebras. If $\A$ is only a prebialgebra, then the tensors defined in Eq.~\eqref{def:Urhovbialg} do not satisfy the second line of  Eq.~\eqref{eq:MUpentagon}. Instead, they satisfy the following:
\begin{eqnarray}\label{eq:MUpentagon-PreBA}
	\begin{tikzpicture}[baseline={([yshift=-0.5ex]current bounding box.center)}, scale=0.63]
		\rhotensor{-0.5}{1}%
		\vtensor{0.5}{1}%
		\hindices{{a,i}}{-0.5}{1.5}{above}
		\hindices{{b,j}}{-0.5}{0.5}{below}
	\end{tikzpicture}	
	&=&
	\begin{tikzpicture}[baseline={([yshift=-0.5ex]current bounding box.center)}, scale=0.63]
		\rhotensor{0.5}{1}%
		\vtensor{-0.5}{1}%
		\Wgatered{0}{0}
		\hindices{{i,a}}{-0.5}{1.5}{above}
		\hindices{{b,j}}{-0.5}{-0.5}{below}
	\end{tikzpicture},\quad %
	\begin{tikzpicture}[baseline={([yshift=-0.5ex]current bounding box.center)}, scale=0.63]
		\rhotensor{-0.5}{1}%
		\vtriangW{0.5}{1}
		\hindices{{a,i}}{-0.5}{1.5}{above}
		\hindices{{b,j}}{-0.5}{0.5}{below}
	\end{tikzpicture}	
	=
	\begin{tikzpicture}[baseline={([yshift=-0.5ex]current bounding box.center)}, scale=0.63]
		\vtriangW{-0.5}{1}
		\Wgatered{0}{0}
		\hindices{{i}}{-0.5}{1.5}{above}
		\hindices{{a}}{+0.5}{0.5}{above}
		\hindices{{b,j}}{-0.5}{-0.5}{below}
	\end{tikzpicture},	\nonumber\\
	\begin{tikzpicture}[baseline={([yshift=-0.5ex]current bounding box.center)}, scale=0.63]
		\rhotriangW{-0.5}{1}
		\vtensor{0.5}{1}%
		\hindices{{a,i}}{-0.5}{1.5}{above}
		\hindices{{b,j}}{-0.5}{0.5}{below}
	\end{tikzpicture}	
	&=&
	\begin{tikzpicture}[baseline={([yshift=-0.5ex]current bounding box.center)}, scale=0.63]
		\rhotriangW{0.5}{1}
		\Wgatered{0}{0}
		\hindices{{i}}{-0.5}{0.5}{above}
		\hindices{{a}}{0.5}{1.5}{above}
		\hindices{{b,j}}{-0.5}{-0.5}{below}
	\end{tikzpicture},\quad 
	\begin{tikzpicture}[baseline={([yshift=-0.5ex]current bounding box.center)}, scale=0.63]
		\rhotriangW{-0.5}{1}
		\vtriangW{0.5}{1}
		\hindices{{a,i}}{-0.5}{1.5}{above}
		\hindices{{b,j}}{-0.5}{0.5}{below}
	\end{tikzpicture}	
	=
	\begin{tikzpicture}[baseline={([yshift=-0.5ex]current bounding box.center)}, scale=0.63]
		\Wgatered{0}{0}
		\hindices{{i,a}}{-0.5}{0.5}{above}
		\hindices{{b,j}}{-0.5}{-0.5}{below}
	\end{tikzpicture}.
\end{eqnarray}
Notice that with Eq.~\eqref{eq:MUpentagon-PreBA}, the derivation in Eq.~\eqref{eq:triangularTNMPOderive} is still valid. The proof for the first relation in Eq.~\eqref{eq:MUpentagon-PreBA} is identical to the bialgebra case in Eq.~\eqref{eq:MUpentagonProof}, and the other three are proved similarly. For example, the second one is proved as follow:
\begin{eqnarray}\label{eq:MUpentagonProofPreBA}
\begin{tikzpicture}[baseline={([yshift=-0.5ex]current bounding box.center)}, scale=0.63]
	\rhotensor{-0.5}{1}%
	\vtriangW{0.5}{1}
	\hindices{{a,i}}{-0.5}{1.5}{above}
	\hindices{{b,j}}{-0.5}{0.5}{below}
	\paraindices{-0.5}{1}{y}{}
\end{tikzpicture}
	&=&  (\deltad_y\otimes\rho_{ab} )\Delta(1_\A\cdot v_{ij})  \nonumber\\
	&=&   (\deltad_y\otimes\rho_{ab} )\Delta(v_{ij})  \nonumber\\
	&=&\sum_k \deltad_y( v_{ik}) \rho_{ab}( v_{kj})\nonumber\\
	&=&
	\begin{tikzpicture}[baseline={([yshift=-0.5ex]current bounding box.center)}, scale=0.63]
		\vtriangW{-0.5}{1}
		\Wgatered{0}{0}
		\hindices{{i}}{-0.5}{1.5}{above}
		\hindices{{a}}{+0.5}{0.5}{above right=-0.1}
		\hindices{{b}}{-0.5}{-0.5}{below left=-0.1}
		\hindices{{j}}{+0.5}{-0.5}{below right=-0.1}
			\paraindices{-0.5}{1}{y}{}
	\end{tikzpicture}.
\end{eqnarray}
The remaining two are proved similarly. 

\subsection{Reduction of a solvable tensor}\label{app:reduction}
We mention that sometimes the solvable tensor \begin{equation}%
	U_{ij}^{ab}=
	\begin{tikzpicture}[baseline={([yshift=-0.5ex]current bounding box.center)}, scale=0.63]
		\WgateredInd{0}{0}{$i$}{$a$}{$b$}{$j$}
	\end{tikzpicture}=\rho_{ab}(v_{ij})
\end{equation}
is reducible in the following sense. 
For any value of the indices $i$ and $j$, let 
$\hat{U}_{ij}$ be an operator  acting on the space $\Hil_\rho$ with matrix elements $[\hat{U}_{ij}]_{ab}=U_{ij}^{ab}$.
If $\Hil_\rho$ has a proper subspace $\Hil'_\rho$ that is left invariant by all the $\hat{U}_{ij}$, i.e. $\hat{U}_{ij}\Hil'_\rho\subseteq\Hil'_\rho, 1\leq i,j\leq d_v$, 
then it is clear that $\hat{U}$ satisfies 
\begin{equation}
	\hat{U}(\Hil'_\rho\otimes\Hil_v)\subseteq\Hil_v\otimes\Hil'_\rho,
\end{equation}
and we can obtain a new solvable tensor by restricting $\hat{U}$ to the subspace $\Hil'_\rho\otimes\Hil_v$. Such a proper invariant subspace exists if the representation $\rho$ is reducible, for example. [However, even if $\rho$ is irreducible, proper invariant subspace may still exist, since it is possible that a irreducible representation $\rho$ of $\A$ is reducible as a representation of the subalgebra generated by all the $v_{ij}$.] The 2D tensor network generated by the reduced tensor $\hat{U}'$ still has an exact MPO representation, as we can simply restrict Eq.~\eqref{eq:triangularTNMPO} to the  subspaces $\Hil'_\rho$ and $\Hil_v$, and all results of Sec.~\ref{sec:solvableQCA} still apply after restricting to this subspace. The above defines reduction in $\Hil_\rho$, but using a similar procedure, one can do reduction in $\Hil_v$ as well. 
This provides a simple way to construct new solvable tensors~(with smaller dimension) by reducing existing ones. 

\subsection{Solvable tensor from regular representations and corepresentations}\label{app:HAregREpcorep}
For any finite dimensional prebialgebra $\A$, we can construct a solvable tensor by taking $\rho$ to be the regular representation of $\A$ in Eq.~\eqref{def:regRep} and $v$ to be the regular corepresentation of $\A$ defined in Eq.~\eqref{def:regCorep}, and we obtain
\begin{equation}%
	\begin{tikzpicture}[baseline={([yshift=-0.5ex]current bounding box.center)}, scale=0.63]
		\WgateredInd{0}{0}{$i$}{$a$}{$b$}{$j$}
	\end{tikzpicture}=\sum_{x\in B(\A)}\Lambda_j^{ix}\Omega_{xb}^a,
\end{equation}
where $a,b,i,j\in B(\A)$. %
We can also rewrite this tensor alternatively as
\begin{eqnarray}\label{eq:UgateRegularRep}
	\hat{U}(\ket{x}\otimes\ket{y})&=&\sum_{u,z,w\in B(\A)}\Lambda_{y}^{zu}\Omega_{ux}^w\ket{z}\otimes\ket{w}\nonumber\\
	&=&\ket{y_{(1)}}\otimes\ket{y_{(2)}x}.
\end{eqnarray}
Since any irreducible representation of a $\C^*$-Hopf algebra is a subrepresentation of the regular representation, any solvable gate constructed from irreducible representations and corepresentations can be obtained by reducing the gate $\hat{U}$ in Eq.~\eqref{eq:UgateRegularRep} to invariant subspaces as described in App.~\ref{app:reduction}. This gives a straightforward way to construct solvable tensors from a given prebialgebra without having to explicitly deal with (co)representations.

\subsection{Dual structure}\label{app:dualstructure}
In this section we review the important concept of duality between algebras and coalgebras, and the self-duality of bialgebras. This is not necessary for understanding the main text, but will be used in App.~\ref{app:HA} and App.~\ref{app:WHA}.
\subsubsection{Duality between algebras and coalgebras}\label{app:prebialgebraduality}
Let $\A$ be a finite dimensional coalgebra, and let $\A^*$ denote the dual vector space of $\A$, i.e., the space of linear functionals  $\A\to\C$. Then it can be checked straightforwardly that  $\A^*$ is an algebra with multiplication defined as 
\begin{equation}\label{eq:dualcoalgebra}
	(f\cdot g)(x):=\sum f\left(x_{(1)}\right) g\left(x_{(2)}\right), \quad x \in \A, \quad f, g \in \A^{*},
\end{equation}
and unit $1_{\A^*}=\epsilon_{\A}$.
In particular, the coassociativity of $\Delta$~[Eq.~\eqref{eq:coassociativityaxiom}] directly translates into associativity of multiplication in $\A^{*}$, and the counit axiom of $\A$~[Eq.~\eqref{eq:counitaxiom}] directly translates into the unit axiom of $\A$. We can write down this algebra structure more explicitly using the structure constants of the coalgebra defined in Eq.~\eqref{def:structureconstcoalg}. Specifically, let $B(\A)$ be a basis of $\A$ and let $B(\A^*)=\{\deltad_x\in\A^*|x\in B(\A)\}$ be the corresponding dual basis of $\A^*$ satisfying $\deltad_x(y)=\deltak_{x,y}$. Then we have
\begin{eqnarray}\label{def:structureconstdualalg}
	\deltad_x\cdot \deltad_y=\sum_{z\in B(\A)} \Lambda^{xy}_z \deltad_z,\quad\forall x,y \in B(\A).%
\end{eqnarray}
In other words, the structure constant of the coalgebra $\A$ becomes the structure constant of the algebra $\A^*$ in the corresponding dual basis. 

Similarly, let $\A$ be a finite dimensional algebra, then  the dual vector space  $\A^*$   is a coalgebra with comultiplication defined as 
\begin{eqnarray}\label{eq:dualalgebra}
	\Delta: \A^*&\to&\A^*\otimes\A^*\cong (\A\otimes\A)^*,\\
	\Delta(f)(x\otimes y)&:=&f(xy), \quad x,y \in \A, \quad f \in \A^{*},\nonumber
\end{eqnarray}
and counit $\epsilon_{\A^*}=1_{\A}\in\A\cong(\A^*)^*$. More explicitly, we can write down the comultiplication of $\A^*$ in the dual basis $B(\A^*)$ as
\begin{equation}\label{def:structureconstdualcoalg}
	\Delta(\deltad_z)=\sum_{x,y\in B(\A)}\Omega^z_{xy}\deltad_x\otimes \deltad_y,	
\end{equation}
where $\Omega^z_{xy}$ is the structure constant of $\A$ defined in Eq.~\eqref{def:structureconstalg}. 

Now let $\A$ be a prebialgebra. Then the dual vector space  $\A^*$ is an algebra and a coalgebra, with comultiplication and multiplication defined in Eq.~\eqref{eq:dualcoalgebra} and Eq.~\eqref{eq:dualalgebra}, respectively. Indeed, it is straightforward to check that the multiplicativity of $\Delta_{\A}$~[Eq.~\eqref{eq:DeltaMultiplicative}] in $\A$ translates into the multiplicativity of  $\Delta_{\A^*}$ in the algebra  $\A^*$, therefore $\A^*$ is a prebialgebra. 

For later convenience, it is useful to introduce a bilinear functional $\braket{,}: \A^*\times \A\to\C$ defined as
\begin{equation}
\braket{f,x}:=f(x),
\end{equation}
called the dual pairing between $\A^*$ and $\A$, satisfying the following conditions %
\begin{eqnarray}\label{eq:duapairingaxiom}
\braket{f\cdot g,x}&=&\braket{f\otimes g,\Delta(x)}, \nonumber\\	
\braket{\Delta(f),x\times y}&=&\braket{f,xy}.
\end{eqnarray}
Eq.~\eqref{eq:duapairingaxiom} is simply a rewritting of Eqs.~\eqref{eq:dualcoalgebra} and \eqref{eq:dualalgebra}.
We also define the canonical element associated to $\A$  as
\begin{eqnarray}\label{def:CE2}
	\mathbf{c}=\sum_{x\in B(\A)} x\otimes \deltad_x\in \A\otimes \A^*,
\end{eqnarray}
which will play an important role in later sections of this Appendix.

\subsubsection{Duality between representations and corepresentations}\label{app:dual_rep_corep}
Let $\rho$ be a representation of a prebialgebra $\A$. In the following we show that $\{\rho_{ab}\in\A^*\}_{1\leq a,b\leq d_\rho}$ is a corepresentation of $\A^*$. We have
\begin{eqnarray}
\Braket{\Delta(\rho_{ab}),x\otimes y}&=&\rho_{ab}(xy)\nonumber\\
&=&\sum_c \rho_{ac}(x)\rho_{cb}(y)\nonumber\\
&=&\Braket{\sum_c \rho_{ac}\otimes\rho_{cb},x\otimes y},
\end{eqnarray}
for all $x,y\in\A$, implying that
\begin{equation}
\Delta(\rho_{ab})=\sum_c \rho_{ac}\otimes\rho_{cb}.
\end{equation}
In addition, we have $\epsilon(\rho_{ab})=\rho_{ab}(1)=\deltak_{a,b}$, therefore, $\rho_{ab}$ is a corepresentation of $\A^*$.
Similarly, if $\{v_{ij}\in \A\}_{1\leq i,j\leq d_v}$ is a corepresentation of $\A$, then 
\begin{eqnarray}\label{eq:repAstarduality}
	v:\A^*&\to& M_{d_v}(\C),\nonumber\\
	{}[v(f)]_{ij}&=&f(v_{ij}),~~1\leq i,j\leq d_v
\end{eqnarray}
is a representation of $\A^*$. Note that it is sometimes convenient to write $f(v_{ij})$ as  $v_{ij}(f)$, where we identify an arbitrary element $x$ of $\A\cong(\A^*)^*$  as a linear functional on $\A^*$ via 
\begin{eqnarray}
	x(f):=f(x),~\forall f\in \A^*.
\end{eqnarray}

In the following we obtain an  alternative expression for the solvable tensor which will be useful later in App.~\ref{app:PBCevol} when we derive the PBC evolution operator and the system revival time. 
It is a basic fact in representation theory~\cite{etingof2011introduction} that if $\rho$ is a representation of an algebra $\A$ and $v$ is a representation of an algebra $\mathcal{B}$, 
then 
\begin{equation}
\rho\otimes v:\A\otimes\mathcal{B}\to M_{d_\rho}(\C)\otimes M_{d_v}(\C)\cong M_{d_\rho d_v}(\C)
\end{equation}
is a $d_\rho d_v$-dimensional representation of the algebra $\A\otimes\mathcal{B}$.  In this way the solvable tensor defined in Eq.~\eqref{def:Urhovbialg} can be viewed as $\rho\otimes v$ applied to the canonical element $\mathbf{c}$:
\begin{eqnarray}\label{def:gate_construction_repcorep}
	\begin{tikzpicture}[baseline={([yshift=-0.5ex]current bounding box.center)}, scale=0.63]
		\WgateredInd{0}{0}{$i$}{$a$}{$b$}{$j$}
	\end{tikzpicture}&=&\rho_{ab}(v_{ij})\nonumber\\
	&=&\sum_{x\in B(\A)} \rho_{ab}(x) v_{ij}(\deltad_x)\nonumber\\
	&=&(\rho_{ab}\otimes v_{ij})\mathbf{c}.
\end{eqnarray}

\section{$\C^*$-Hopf algebras and solvable unitary circuits}\label{app:HA}
In this section we first review the basic definitions of $\C^*$-Hopf algebras~(App.~\ref{app:CHA}) and show that the solvable tensor  
\begin{tikzpicture}[baseline={([yshift=-0.5ex]current bounding box.center)}, scale=0.63]
	\Wgatered{0}{0}
\end{tikzpicture} defined in Eq.~\eqref{def:Urhovbialg} is unitary~(App.~\ref{app:Cstarunitarity}). %
Then in Apps.~\ref{app:groupHA}-\ref{app:HA-XYX}  we provide several families of explicit examples of finite dimensional $\C^*$-Hopf algebras constructed out of finite groups, which are the Hopf algebra structures behind the examples given in Sec.~\ref{sec:HAexamples}. 
\subsection{$\C^*$-Hopf algebra basics}\label{app:CHA}
 We begin by defining Hopf algebras. 
\begin{definition}{(Hopf algebras)}
	A Hopf algebra $\A$ is a bialgebra with an invertible linear map $S:\A\to \A$, called the antipode, satisfying
	\begin{equation}\label{def:antipode_axiom}
		m(S\otimes\id)\Delta(x)=m(\id\otimes S)\Delta(x)=\epsilon(x) 1,~\forall x\in\A.
	\end{equation}
\end{definition}
It can be proved that the antipode of a Hopf algebra $\A$ is always an anti-homomorphism of the algebra and coalgebra structure of $\A$, i.e.,
\begin{equation}\label{eq:S_anti-homomorphism}
	\begin{aligned}
		S(xy)&=S(y) S(x), \quad x, y \in \A, \quad S(1)=1, \\
		\Delta[S(x)]&=\Pi[(S \otimes S)\Delta(x)],\quad x\in \A, \quad \epsilon \circ S=\epsilon,
	\end{aligned}
\end{equation}
where $\Pi:\A\otimes\A\to\A\otimes\A$ is the swap map $\Pi(x\otimes y)=y\otimes x,~~\forall x,y\in\A$.
\begin{remark}{(Invertibility of the canonical element)}\label{rmk:ceinvertible}
The antipode axiom  Eq.~\eqref{def:antipode_axiom} implies that the canonical element $\mathbf{c}$ defined in Eq.~\eqref{def:CE2} is an invertible element of $\A\otimes\A^*$. Specifically, let 
\begin{equation}\label{def:CEinverse}
\mathbf{c}'=(S\otimes \id)\mathbf{c}=\sum_{x\in B(\A)} S(x)\otimes \deltad_x.
\end{equation}
In the following we show that $\mathbf{c}\cdot \mathbf{c}'=1\otimes \epsilon$, which is the unit element of the algebra  $\A\otimes\A^*$. We have
\begin{eqnarray}\label{proof:CEinverse}
	\mathbf{c}\cdot \mathbf{c}'&=&\sum_x(x\otimes \deltad_x)\cdot \sum_y [S(y)\otimes \deltad_y]\nonumber\\
	&=&\sum_{x,y} x S(y)\otimes \deltad_x\deltad_y\nonumber\\
	&=&\sum_{x,y,z} x S(y)\otimes (\deltad_x\deltad_y)(z)\deltad_z\nonumber\\
	&=&\sum_{x,y,z} x S(y)\otimes \deltad_x(z_{(1)})\deltad_y(z_{(2)})\deltad_z\nonumber\\
	&=&\sum_{z} z_{(1)} S(z_{(2)})\otimes \deltad_z%
\end{eqnarray}
where in the third line we used the identity $f=\sum_z f(z)\deltad_z$ for $f\in\A^*$, and in the  fourth line we used the definition of multiplication in $\A^*$,  Eq.~\eqref{eq:dualcoalgebra}. Applying the antipode axiom in Eq.~\eqref{def:antipode_axiom} to the last line of Eq.~\eqref{proof:CEinverse}, we obtain
\begin{equation}
\mathbf{c}\cdot \mathbf{c}'	=\sum_{z} \epsilon(z) 1\otimes\deltad_z= 1\otimes\epsilon.%
\end{equation}
A similar derivation shows that $\mathbf{c}'\cdot \mathbf{c}= 1\otimes\epsilon$. Therefore, $\mathbf{c}'$ is the inverse of $\mathbf{c}$ in  $\A\otimes\A^*$.
\end{remark}
\begin{definition}\label{def:CstarAlg}{($\C^*$-algebras and representations)}
	A finite dimensional algebra $\A$ over $\C$ is a $*$-algebra if there is an
	anti-linear map $*:\A\to\A$ such that it is an involution $(x^*)^*=x,~\forall x\in \A$, and an anti-homomorphism $(xy)^* = y^* x^*,~\forall x,y\in\A$. %
	A representation $\rho$ of $\A$ is called a $*$-representation if 
	\begin{equation}\label{def:star_rep}
		\rho(x^*)=\rho(x)^\dagger,\quad \forall x\in \A.
	\end{equation} 
	A finite dimensional $*$-algebra $\A$ is called a $\C^*$-algebra, if it is a $*$-algebra and has a faithful $*$-representation. 
\end{definition}

\begin{definition}\label{def:CstarpreBA}{($\C^*$-prebialgebras)}
	A finite dimensional prebialgebra $\A$ over $\C$ is a $*$-prebialgebra if it is a $*$-algebra such that the $*$-structure is compatible with comultiplication $\Delta(x^*)=\Delta(x)^*$. It is a $\C^*$-prebialgebra if it is a $*$-prebialgebra and has a faithful $*$-representation. 
\end{definition}

\begin{definition}\label{def:CstarHA}{($\C^*$-Hopf algebras and unitary corepresentations)}	
	A finite dimensional Hopf algebra $\A$ is called a $*$-Hopf algebra if it is also a $*$-prebialgebra as defined in Definition~\ref{def:CstarpreBA}, and it is a 
	$\C^*$-Hopf algebra if it is also a $\C^*$-prebialgebra.
	A corepresentation $v$ of a $*$-Hopf algebra $\A$ is called a unitary corepresentation if 
	\begin{equation}\label{def:unitary_corep}
		S(v_{ij})=v_{ji}^*.
	\end{equation}
\end{definition}

\begin{remark}{(Dual structure)}\label{rmk:dualstructure}
The axioms of prebialgebra, bialgebra, Hopf algebra,  $*$-Hopf algebra, and $\C^*$-Hopf algebra are all self dual, %
in the sense that if $\A$ is a $\C^*$-Hopf algebra~(prebialgebra, bialgebra, Hopf algebra,  $*$-Hopf algebra, respectively), then $\A^*$ is also a $\C^*$-Hopf algebra~(prebialgebra, bialgebra, Hopf algebra,  $*$-Hopf algebra, respectively). 
\end{remark}
We have already seen in App.~\ref{app:dualstructure} that the axioms of prebialgebra is self dual, and it is straightforward to check that under duality, the two additional axioms of bialgebra in Eq.~\eqref {eq:bialgebraunitcounit} are transformed into each other.  For a Hopf algebra $\A$, the antipode of $\A^*$ is defined as
\begin{equation}\label{def:dualantipode}
\braket{S_{\A^*}(f),x}:=\braket{f,S(x)},~\forall f\in \A^*,~x\in \A,
\end{equation}
and it is straightforward to check that $S_{\A^*}$ along with $m_{\A^*},\Delta_{\A^*},1_{\A^*},\epsilon_{\A^*}$ satisfy the antipode axiom, Eq.~\eqref{def:antipode_axiom}. 

For a $*$-Hopf algebra $\A$,  $\A^*$ naturally has a $*$-structure  defined as 
\begin{equation}\label{eq:dualstarstructure}
	\braket{f^*,x}:=\braket{f, S(x)^*}^*, ~\forall f\in \A^*,~x\in \A,
\end{equation}
which can be straightforwardly shown to satisfy the $*$-axioms in Definition~\ref{def:CstarHA}.
The proof that the axiom of $\C^*$-Hopf algebra is self dual can be found in textbooks on $\C^*$-Hopf algebras, such as Ref.~\cite{enock2013kac}. Note that finite dimensional $\C^*$-Hopf algebras are also called Kac algebras in the mathematics literature. 

We now give some different perspectives to understand the definition of a unitary corepresentation in Eq.~\eqref{def:unitary_corep}. A unitary corepresentation of a $\C^*$-Hopf algebra $\A$ is simply a $*$-representation of $\A^*$. Specifically, let $v$ be a $*$-representation of $\A^*$. We have already shown in App.~\ref{app:dual_rep_corep} that $\{v_{ij}\in \A\}_{1\leq i,j\leq d_v}$ is a corepresentation of $\A$. Being a $*$-representation means that 
\begin{equation}
v(f^*)=[v(f)]^\dagger,~~\forall f\in\A^*.
\end{equation}
Expanding this equation in matrix elements, we have
\begin{alignat}{3}\label{eq:proofdualunitarycorep}
	 &&v_{ij}(f^*)&=[v(f)^\dagger]_{ij},\nonumber\\
\Leftrightarrow\quad&&	\braket{f^*,v_{ij}}  &=v_{ji}(f)^*,\nonumber\\
\Leftrightarrow\quad&&	\braket{f,S(v_{ij})^*} &=\braket{f,v_{ji}},~~\forall f\in\A^*,%
\end{alignat}
leading to $S(v_{ij})=v_{ji}^*$, therefore $\{v_{ij}\in \A\}_{1\leq i,j\leq d_v}$ is a unitary corepresentation of $\A$. %

The representation theory of finite dimensional $\C^*$-algebras~\cite{etingof2011introduction} says that every finite dimensional representation of $\A$ is isomorphic to a $*$-representation. Therefore, by duality, every finite dimensional corepresentation of a $\C^*$-Hopf algebra $\A$ is isomorphic to a unitary corepresentation. The reason for the name ``unitary corepresentation'' is seen as follows. Applying Eq.~\eqref{def:antipode_axiom} to a unitary corepresentation $v$  we have
\begin{eqnarray}
	m(S\otimes\id)\Delta(v_{ij})	&\equiv &\sum_{k} S(v_{ik}) v_{kj}\nonumber\\
	&=&\sum_{k} v_{ki}^* v_{kj}\nonumber\\
	&=&\epsilon(v_{ij})1 \nonumber\\
	&=&\deltak_{i,j}1,
\end{eqnarray}
where in the first and last lines we used Eq.~\eqref{def:corep}, and in the second line we used the definition of a unitary corepresentation in Eq.~\eqref{def:corep}. Therefore, we obtain
\begin{equation}\label{eq:reason_unitary_corep}
	\sum_{k} v_{ki}^* v^{\vphantom{*}}_{kj}=\deltak_{i,j}1.
\end{equation}
Eq.~\eqref{eq:reason_unitary_corep} can be viewed as a generalization of unitary matrices to matrices whose matrix elements are elements of a $*$-algebra. Similarly, the identity $m(\id\otimes S)\Delta(v_{ij})=\epsilon(v_{ij}) 1$ gives us
\begin{equation}\label{eq:reason_unitary_corep2}
	\sum_{k} v^{\vphantom{*}}_{ik} v_{jk}^*=\deltak_{i,j}1.
\end{equation}
\begin{remark}\label{rmk:cunitary}
For a $\C^*$-Hopf algebra, the canonical element $\mathbf{c}$ is a unitary element of $\A\otimes\A^*$, i.e., $\mathbf{c}'=\mathbf{c}^{-1}=\mathbf{c}^*$. To prove this, we %
show that 
\begin{equation}\label{eq:cc'proof}
(\id\otimes y)\mathbf{c}'=(\id\otimes y)\mathbf{c}^*,~\forall y\in\A\cong(\A^*)^*,
\end{equation}
where $\mathbf{c}'$ is defined in Eq.~\eqref{def:CEinverse}.
We have
\begin{eqnarray}
(\id\otimes y)\mathbf{c}^*%
&=&\sum_{x\in B(\A)} x^*  \braket{\deltad_x^*,y}\nonumber\\
&=&\sum_{x\in B(\A)} x^* \braket{\deltad_x,S(y)^*}^*\nonumber\\
&=&[S(y)^*]^*\nonumber\\
&=&S(y),
\end{eqnarray}
where in the second line we use Eq.~\eqref{eq:dualstarstructure} and in the last line we use the fact that $*$ is an involution. On the other side, we have
\begin{eqnarray}
	(\id\otimes y)\mathbf{c}'=\sum_{x\in B(\A)} S(x ) \braket{\deltad_x,y}=S(y).
\end{eqnarray}
This proves Eq.~\eqref{eq:cc'proof}. Therefore $\mathbf{c}'=\mathbf{c}^*$. Combining with $\mathbf{c}'=\mathbf{c}^{-1}$ proved in Remark~\ref{rmk:ceinvertible}, we conclude that  $\mathbf{c}$ is a unitary element of $\A\otimes\A^*$.
\end{remark}

\subsection{$\C^*$-structure and unitarity}\label{app:Cstarunitarity}
Now let $\rho$ be a $*$-representation and $v$ a unitary corepresentation of a $\C^*$-Hopf algebra $\A$. Then %
Eqs.~(\ref{eq:reason_unitary_corep},\ref{eq:reason_unitary_corep2}) guarantee that 
$%
\begin{tikzpicture}[baseline={([yshift=-0.5ex]current bounding box.center)}, scale=0.63]
\WgateredInd{0}{0}{$i$}{$a$}{$b$}{$j$}
\end{tikzpicture}$ defined in Eq.~\eqref{def:Urhovbialg} is a unitary matrix if we group $b,j$ as input indices and $i,a$ as output indices, since, applying $\rho$ on both sides of Eq.~\eqref{eq:reason_unitary_corep}, we obtain
\begin{equation}\label{eq:proof_unitary_gate}
	\sum_{k,c} \rho(v_{ki})^*_{ca} \rho(v_{kj})_{cb}=\deltak_{i,j}\deltak_{a,b}.
\end{equation}
In fact, an important theorem in Hopf algebra theory implies that the gate $
\begin{tikzpicture}[baseline={([yshift=-0.5ex]current bounding box.center)}, scale=0.63]
\WgateredInd{0}{0}{$i$}{$a$}{$b$}{$j$}
\end{tikzpicture}$ constructed in Eq.~\eqref{def:Urhovbialg} is a dual-unitary gate~\cite{Prosen2019DU}. %
The theorem says that
\begin{theorem}\label{thm:LarsonRadford}{(Larson-Radford~\cite{LarsonRadford1988})}
The antipode of a finite dimensional $\C^*$-Hopf algebra is involutive, i.e.  $S^2=\id$. 
\end{theorem}
Applying $S$ on both sides of Eq.~\eqref{eq:reason_unitary_corep2}, we have 
\begin{eqnarray}\label{eq:DU_proof}
	\deltak_{i,j}1&=&\sum_{k}S(v_{jk}^*) S(v^{\vphantom{*}}_{ik})\nonumber\\
	&=&\sum_{k}S^2[v_{kj}]\vphantom{*} v^{*}_{ki}\nonumber\\
	&=&\sum_{k}v_{kj}\vphantom{*} v^{*}_{ki},
\end{eqnarray}
where in the first line we used that $S$ is an anti-homomorphism of algebra, Eq.~\eqref{eq:S_anti-homomorphism}, in the second line we use Eq.~\eqref{def:unitary_corep}, and in the last line we use $S^2=\id$. Now apply $\rho$ on both sides of Eq.~\eqref{eq:DU_proof} as we did in Eq.~\eqref{eq:proof_unitary_gate} we obtain
\begin{equation}\label{eq:DU_proof2}
	\sum_{k,c} \rho(v_{kj})_{ac} \rho(v_{ki})^*_{bc} =\deltak_{i,j}\deltak_{a,b},
\end{equation}
which means that 
$%
\begin{tikzpicture}[baseline={([yshift=-0.5ex]current bounding box.center)}, scale=0.63]
\WgateredInd{0}{0}{$i$}{$a$}{$b$}{$j$}
\end{tikzpicture}$ defined in Eq.~\eqref{def:Urhovbialg} is also a unitary matrix when viewed horizontally, i.e., if we group $a,j$ as input indices and $i,b$ as output indices, therefore, it is a dual unitary gate. The tensor graphical representations of Eqs.~(\ref{eq:proof_unitary_gate},\ref{eq:DU_proof2}) are 
\begin{equation}\label{eq:DUgraphical}
	\begin{tikzpicture}[baseline={([yshift=-0.5ex]current bounding box.center)}, scale=0.63]
		\WgateblueInv{0}{1}
		\Wgatered{0}{0}
	\end{tikzpicture}=	
	\begin{tikzpicture}[baseline={([yshift=-0.5ex]current bounding box.center)}, scale=0.63]
		\draw[very thick, draw=red] (0,-1) -- (0,+1);
		\draw[very thick, draw=blue] (1,-1) -- (1,+1);
	\end{tikzpicture}~,\quad
		\begin{tikzpicture}[baseline={([yshift=-0.5ex]current bounding box.center)}, scale=0.63]
		\WgateblueHInv{0}{0}
		\Wgatered{1}{0}
	\end{tikzpicture}=	
	\begin{tikzpicture}[baseline={([yshift=-0.5ex]current bounding box.center)}, scale=0.63]
		\draw[very thick, draw=red] (-1,0) -- (+1,0);
		\draw[very thick, draw=blue] (-1,-1) -- (+1,-1);
	\end{tikzpicture}~.
\end{equation}

\subsection{Example: finite group algebras $\C[G]$}\label{app:groupHA}
The simplest examples of finite dimensional $\C^*$-Hopf algebras are group algebras of finite groups. Let $G$ be a finite group, and let $\A=\C[G]$ denote the finite dimensional vector space spanned by  the elements of $G$, with basis
\begin{equation}
B(\A)=\{g | g\in G\}.
\end{equation}
Any element $x\in\C[G]$ can be written as formal linear combinations of group elements 
\begin{equation}
x=\sum_{g \in G} x_g g,
\end{equation}
where $x_g\in\C$ are coefficients. The group multiplication of $G$ induces an algebra structure on $\C[G]$ via
\begin{eqnarray}\label{def:CGalgebra}
	x\cdot y&=&\left(\sum_{g \in G} x_g g\right)\cdot \left(\sum_{h \in G} y_h h\right)\nonumber\\
	&=&\sum_{g,h \in G} x_g y_h (gh),\quad\forall x,y\in\C[G],
\end{eqnarray}
where $gh$ is the product of $g$ and $h$ in $G$. One can straightforwardly check that the multiplication in Eq.~\eqref{def:CGalgebra} defines an associative algebra structure on  $\C[G]$, where the algebra unit is $1_{\C[G]}=1_{G}$, the group unit~(we will omit the subscript of the unit element when it is clear from context which group or algebra it belongs to). 

In addition, $\C[G]$ has the structure of a $*$-Hopf algebra where the comultiplication, counit, antipode, and $*$-operation are defined as
\begin{eqnarray}
\Delta(x)&=&\sum_{g \in G} x_g g\otimes g,\nonumber\\
\epsilon(x)&=&\sum_{g \in G} x_g,\nonumber\\
S(x)&=&\sum_{g \in G} x_g g^{-1},\nonumber\\
x^* &=& \sum_{g \in G} x^*_g g^{-1}.
\end{eqnarray}
The regular representation $\rho_{\mathrm{reg}}$ of the group algebra $\C[G]$ is defined on a $d_{\A}=|G|$ dimensional  vector space where the basis states are labeled by the group elements $\{\ket{g}| g\in G\}$, and $\rho_{\mathrm{reg}}$ is defined by 
\begin{equation}
\rho_{\mathrm{reg}}(x)\ket{g}=\sum_{h \in G} x_h \ket{hg}.
\end{equation}
Note that this coincides with the definition of regular representation in Eq.~\eqref{def:regRep} for the case of $\C[G]$. 

All  irreducible corepresentations of $\C[G]$ are one dimensional, labeled by a specific element $g\in G$: $v^{(g)}=g$, and an arbitrary $d$-dimensional corepresentation of $\C[G]$ can be written as a direct sum of 1-dimensional corepresentations, as in Eq.~\eqref{def:Gcorep}. The group algebra $\C[G]$ is the Hopf algebra structure behind the solvable gates constructed in Sec.~\ref{sec:GroupHA}. 
\subsection{Example: the Hopf algebra $\C^{Z_n}\bltimes \C[G^{\times n}]$ for the solvable gate in  Sec.~\ref{sec:twistedperm}}\label{app:HA-TP}
In this section we provide the $\C^*$-Hopf algebra structure underlying the solvable gate in  Sec.~\ref{sec:twistedperm}. Let $G$ be a finite group. The  Hopf algebra $\A=\C^{Z_n}\bltimes \C[G^{\times n}]$ is spanned by elements of the form $\deltad_{\sigma^k}({h_1,  \ldots, h_n}) $ for $h_1,\ldots,h_n\in G$ and $0\leq k\leq n-1$.
The multiplication is defined by
\begin{equation}
	\deltad_{\sigma^k}(f_1,  \ldots, f_n)\cdot\deltad_{\sigma^l}(h_1,  \ldots, h_n) = \deltak_{k,l}	\deltad_{\sigma^k}(f_1h_1,  \ldots, f_nh_n).
\end{equation}
The comultiplication is given by
\begin{eqnarray}
\Delta[\deltad_{\sigma^k}(h_1,  \ldots, h_n)]=\!\!\sum_{l+m=k}&&\deltad_{\sigma^l}(h_{1+m},  \ldots, h_{n+m})\nonumber\\
&&{}\otimes \deltad_{\sigma^m}(h_1,  \ldots, h_n),
\end{eqnarray}
where the subscripts $1+m,\ldots,n+m$ are understood modolo $n$. 
The unit, counit, antipode, and the $*$-structure are defined by
\begin{eqnarray}
1_\A&=&\sum_k \deltad_{\sigma^k}(1,\ldots,1),\nonumber\\
\epsilon[\deltad_{\sigma^k}(h_1,  \ldots, h_n)]&=&\deltak_{k,0},\nonumber\\
S[\deltad_{\sigma^k}(h_1,  \ldots, h_n)]&=&\deltad_{\sigma^{-k}}(h_{1+k}^{-1},  \ldots, h_{n+k}^{-1}),\nonumber\\
{}[\deltad_{\sigma^k}(h_1,  \ldots, h_n)]^*&=&[\deltad_{\sigma^k}(h^{-1}_1,  \ldots, h^{-1}_n)].
\end{eqnarray}

Now  let $\rho$ be a $d$-dimensional representation of $G$ and let $g_1,g_2,\ldots, g_d$ be arbitrary elements of $G$ as in Sec.~\ref{sec:twistedperm}.
It can be verified straightforwardly that the following defines a representation of $\C^{Z_n}\bltimes \C[G^{\times n}]$
\begin{equation}
	\rho[\deltad_{\sigma^k} (h_1,\ldots, h_n)]=\deltak_{k,1}\rho(h_1),
\end{equation}
and the following defines a corepresentation
\begin{equation}
	v_{ij}=\deltad_{\sigma^{i-j}} (g_j,\ldots, g_{j+n-1}).
\end{equation}
Inserting $\rho,v$ into Eq.~\eqref{def:Urhovbialg} we obtain the gate in Eq.~\eqref{eq:GroupUgatePermute} as well as all the tensors in Eq.~\eqref{def:UrhovTP}. Note that in Eq.~\eqref{def:UrhovTP},
the list $(k, h_1,\ldots, h_n)$ represents the basis element $\deltad_{\sigma^k} (h_1,\ldots, h_n)$.	

\subsection{Example: the Hopf algebra  $\C[G\times Z_2]\bowtiewb \C^G $ for the solvable gate in Sec.~\ref{sec:XYXgate}}\label{app:HA-XYX}
Let $G$ be a finite group. We define a $2|G|^2$ dimensional Hopf algebra $\A=\C[G\times Z_2]\bowtiewb \C^G $ as follows. 
$\A$ is spanned by basis elements of the form $az^s\deltad_g$ where $g,a\in G$ and $s=0,1$. The multiplication of $\A$ is defined by
\begin{eqnarray}
(az^s\deltad_g)\cdot (b z^r\deltad_h)&=&(abz^{s+r}\deltad_h)\deltak_{g^{1-2r}b,bh},
\end{eqnarray}
where the exponent $s+r$ in $z^{s+r}$ is understood modulo 2. 
The comultiplication of $\A$ is given by
\begin{eqnarray}
\Delta(az^s\deltad_g)&=&\sum_{hf=g}az^s\deltad_h\otimes ah^sz^s\deltad_f.
\end{eqnarray}
The unit, counit, antipode, and the $*$-structure are defined by
\begin{eqnarray}
	1_\A&=&\sum_g \deltad_g,\nonumber\\
	\epsilon(az^s\deltad_g)&=&\deltak_{g,1},\nonumber\\
	S(az^s\deltad_g)&=& (ag^s)^{-1}z^s\deltad_{ag^{2s-1}a^{-1}},\nonumber\\
	(az^s\deltad_g)^*&=&a^{-1}z^{s}\deltad_{ag^{1-2s}a^{-1}}.
\end{eqnarray}

The comultiplication of $\A$ induces the multiplication of the dual Hopf algebra $\A^*$ on the dual basis $\deltad_{az^s} g$ via the duality in App.~\ref{app:prebialgebraduality}:
\begin{eqnarray}
(\deltad_{az^s} g)\cdot(\deltad_{bz^r} h)&=&\deltak_{ag^r,b}\deltak_{s,r}(\deltad_{bg^{-r}z^r} gh).
\end{eqnarray}
The canonical element of $\A\otimes \A^*$ is
\begin{equation}\label{eq:ceXYX}
\mathbf{c}%
=\sum_{a,g\in G} (a\deltad_g\otimes \deltad_{a} g+az\deltad_g\otimes \deltad_{az} g).
\end{equation}
To construct a solvable gate, we use the following representation of $\A$ 
\begin{eqnarray}\label{eq:repXYX}
az^s\deltad_g\ket{h}=\deltak_{g,h}\ket{ah^{1-2s} a^{-1}},
\end{eqnarray}
and the following representation of $\A^*$
\begin{eqnarray}\label{eq:repAdualXYXgate}
\deltad_{az^s}g\ket{h}=\deltak_{s,1}\deltak_{ag,h}\ket{hg^{-1}},
\end{eqnarray}
where in the two equations above the representation space is $|G|$-dimensional, with basis states labeled by elements of the group $h\in G$. 
Eq.~\eqref{eq:repAdualXYXgate} also defines a corepresentation of $\A$ via the duality in App.~\ref{app:dual_rep_corep}. 
The solvable gate in Eq.~\eqref{eq:finitegroupExtSeth} is obtained using Eq.~\eqref{def:gate_construction_repcorep}, where $\rho,v$ and $\mathbf{c}$ are given by Eqs.~\eqref{eq:repXYX}, \eqref{eq:repAdualXYXgate}, and \eqref{eq:ceXYX}, respectively. 
The tensors in Eq.~\eqref{def:UrhovXYX}	 are obtained by inserting these representations into Eq.~\eqref{def:Urhovbialg}, and in Eq.~\eqref{def:UrhovXYX} we use $(a,g,s)$ to represent the basis element $az^s\deltad_g\in\A$.

\section{$\C^*$-Weak Hopf algebras and solvable unitary circuits in constrained subspaces}\label{app:WHA}
In this section  we first review the basic definitions of weak Hopf algebras~(App.~\ref{app:WHAdefinitions}) and prove the generalization of Thm~\ref{thm:HopfUgate} to $\C^*$-weak Hopf algebras~(App.~\ref{app:WHAprojector}). Then in App.~\ref{app:dimension_solvable_space} we prove a claim about the dimension of the solvable subspace we made at the end of Sec.~\ref{sec:circuitdef}. In App.~\ref{app:FibonacciWHA} we give the detailed definition of the weak Hopf algebra used in the example in Sec.~\ref{sec:GoldenModel}. 
\subsection{Definitions and basic properties}\label{app:WHAdefinitions}
\begin{definition}{(Weak bialgebras~\cite{Bohm1999WHAI})}
A weak bialgebra $\A$ is a prebialgebra such that 
the unit $1$ and the counit $\epsilon$ satisfy the following axioms
\begin{eqnarray}\label{eq:WBAunitcounitaxiom}
	\Delta^2(1)&=&[1 \otimes \Delta(1)] \cdot[\Delta(1) \otimes 1]=[\Delta(1) \otimes 1] \cdot[1 \otimes \Delta(1)],\nonumber\\
	\epsilon(x y z)&=&%
	\epsilon(x y_{(1)}) \epsilon(y_{(2)} z)=
	\epsilon(x y_{(2)}) \epsilon(y_{(1)} z),%
\end{eqnarray}
for all $x,y,z\in\A$. 
\end{definition}
Using Sweedler's notation, we can rewrite the first line of Eq.~\eqref{eq:WBAunitcounitaxiom} as follows
\begin{eqnarray}\label{eq:WBAunitaxiom2}
	\!\!\!\!\!\!\!\!\!%
	1_{(1)} \otimes 1_{(2)} \otimes 1_{(3)}&=&%
	1_{(1)} \otimes 1_{(2)} 1_{(1^{\prime})} \otimes 1_{(2^{\prime})}\nonumber\\
	&=&%
	1_{(1)} \otimes 1_{(1^{\prime})} 1_{(2)} \otimes 1_{(2^{\prime})},
\end{eqnarray}
where the prime symbol distinguishes different coproducts of $1\in\A$. 
Similarly, the second line of Eq.~\eqref{eq:WBAunitcounitaxiom} can be rewritten as  the weak unit axiom of the  dual weak bialgebra $\A^*$
\begin{eqnarray}\label{eq:WBAcounitaxiom2}
	\!\!\!%
	\epsilon_{(1)} \otimes \epsilon_{(2)} \otimes \epsilon_{(3)}&=&%
	\epsilon_{(1)} \otimes \epsilon_{(2)} \epsilon_{(1^{\prime})} \otimes \epsilon_{(2^{\prime})}\nonumber\\
	&=&%
	\epsilon_{(1)} \otimes \epsilon_{(1^{\prime})} \epsilon_{(2)} \otimes \epsilon_{(2^{\prime})}. 
\end{eqnarray}

For a weak bialgebra $\A$, we define two linear maps $\epsilon_s, \epsilon_t:\A\to\A$ as
\begin{eqnarray}\label{def:sourcetargetmaps}
\epsilon_s(x)&:=&%
1_{(1)} \epsilon(x 1_{(2)}),\\
\epsilon_t(x)&:=&%
\epsilon(1_{(1)} x) 1_{(2)},~~\forall x\in\A.\nonumber
\end{eqnarray}
In the literature~\cite{Bohm1999WHAI,Nikshych2003InvarianntKnotWHA,BOHM2001MultiplicativeIsometry}, $\epsilon_s$ and $\epsilon_t$ are called the source and target counital maps~(note that in Ref.~\cite{Bohm1999WHAI}, $\epsilon_s$ and $\epsilon_t$ are denoted as $\sqcap^R$ and $\sqcap^L$, respectively). 

We now gather some useful properties of the unit and counit of a weak bialgebra $\A$, which we will use later.
The following identity follows directly from the definition of $\epsilon_s$ and $\epsilon_t$  in Eq.~\eqref{def:sourcetargetmaps} and the second line in Eq.~\eqref{eq:WBAunitcounitaxiom}
\begin{equation}\label{eq:counitproj}
\epsilon[x \epsilon_t(y)] =\epsilon(x y)=\epsilon[\epsilon_s(x) y],~~\forall x,y\in  \A.
\end{equation}
The following identity directly follows from the axioms of a prebialgebra
\begin{equation}\label{eq:counitcoproduct}
	\epsilon(x_{(1)}y_{(1)})\epsilon(x_{(2)}y_{(2)})=\epsilon(xy),~~\forall x,y\in \A,
\end{equation}
since we have
\begin{eqnarray}
\epsilon(x_{(1)}y_{(1)})\epsilon(x_{(2)}y_{(2)})&=&\braket{\epsilon\otimes\epsilon,x_{(1)}y_{(1)}\otimes x_{(2)}y_{(2)}}\nonumber\\
&=&\braket{\epsilon\otimes\epsilon,\Delta(xy)}\nonumber\\
&=&\braket{\epsilon\cdot\epsilon,xy}\nonumber\\
&=&\epsilon(xy),
\end{eqnarray}
where in the second line we use the multiplicativity of $\Delta$~[Eq.~\eqref{eq:DeltaMultiplicative}], in the third line we use the dual pairing axiom in Eq.~\eqref{eq:duapairingaxiom}, and in the last line we use the fact that $\epsilon$ is the unit of $\A^*$. 
Eq.~\eqref{eq:counitcoproduct} can be rewritten as the following relations in $\A^*$
\begin{eqnarray}\label{eq:epsilonprop}
	\epsilon(x_{(1)}?)\epsilon(x_{(2)}?)&=&\epsilon(x?),\nonumber\\
	\epsilon(?x_{(1)})\epsilon(?x_{(2)})&=&\epsilon(?x), \forall x\in \A,
\end{eqnarray}
where we use $\epsilon(x?)$ to denote the linear functional $f(y):= \epsilon(xy)$ and similarly for $\epsilon(?x)$. 
For example, to prove the first line in Eq.~\eqref{eq:epsilonprop}, we show that it is true when applied to an arbitrary element of $\A$
\begin{eqnarray}\label{eq:epsilonpropproof}
\braket{\epsilon(x_{(1)}?)\epsilon(x_{(2)}?),y}&=&\braket{\epsilon(x_{(1)}?)\otimes \epsilon(x_{(2)}?),\Delta(y)},\nonumber\\
&=&\epsilon(x_{(1)}y_{(1)})\epsilon(x_{(2)}y_{(2)})\nonumber\\
&=&\epsilon(xy)\nonumber\\
&=&\braket{\epsilon(x?),y},~~\forall y\in\A.
\end{eqnarray}

The following identity is proved in Lemma~2.3 and Proposition~2.4 in Ref.~\cite{Bohm1999WHAI}
\begin{eqnarray}\label{eq:unitcombination}
	1_{(1)} 1_{(1')} \otimes 1_{(2)} \otimes 1_{(2')}&=&1_{(1)} \otimes \epsilon_t(1_{(2)}) \otimes 1_{(3)}, \nonumber\\
	1_{(1)} \otimes 1_{(1')} \otimes 1_{(2)} 1_{(2')}&=&1_{(1)} \otimes \epsilon_s(1_{(2)}) \otimes 1_{(3)}.
\end{eqnarray}

We now define weak Hopf algebras.
\begin{definition}{(Weak Hopf algebras~\cite{Bohm1999WHAI})}
A weak Hopf algebra $\A$ is a weak bialgebra together  with a linear map $S:\A\to \A$, called the antipode, satisfying
\begin{eqnarray}\label{def:WHAantipode_axiom}
	S(x_{(1)}) x_{(2)}&=&\epsilon_s(x), \nonumber\\
	x_{(1)} S(x_{(2)})&=&\epsilon_t(x), \nonumber\\
	S(x_{(1)}) x_{(2)} S(x_{(3)})&=&S(x).
\end{eqnarray}
\end{definition}

\begin{definition}{($\C^*$-structure)}\label{def:CWHA}
A finite dimensional weak Hopf algebra $\A$ is a $*$-weak Hopf algebra if it is also a $*$-prebialgebra as defined in Definition~\ref{def:CstarpreBA}, and it is a 
$\C^*$-weak Hopf algebra if it is also a $\C^*$-prebialgebra.
\end{definition} 

\begin{remark}{(Duality of weak Hopf algebras)}\label{rmk:dualstructureWHA}
	The axioms of weak bialgebra, weak Hopf algebra, $*$-weak Hopf algebra, and $\C^*$-weak Hopf algebra are all self dual, %
	in the sense that if $\A$ is a $\C^*$-weak Hopf algebra~(weak bialgebra, weak Hopf algebra, $*$-weak Hopf algebra, respectively), then $\A^*$ is also a $\C^*$-weak Hopf algebra~(weak bialgebra, weak Hopf algebra, $*$-weak Hopf algebra, respectively). 
\end{remark}
For a $*$-weak Hopf algebra $\A$, the antipode and $*$-operation of $\A^*$ are defined in Eq.~\eqref{def:dualantipode} and Eq.~\eqref{eq:dualstarstructure}, similar to the case of $*$-Hopf algebras. 
Refs.~\cite{Bohm1999WHAI,Bohm2000WHAII} proves that the axiom of $\C^*$-weak Hopf algebra is self dual.

Furthermore,  a $*$-representation $v$ of $\A^*$ is a corepresentation $\{v_{ij}\in \A\}_{1\leq i,j\leq d_v}$ of $\A$ satisfying Eq.~\eqref{def:unitary_corep}, the derivation for this in Eq.~\eqref{eq:proofdualunitarycorep} is still valid here. However, in this case, we do not call $v_{ij}$ a unitary corepresentation of $\A$, since  %
in the case of $*$-weak Hopf algebra, Eq.~\eqref{def:unitary_corep} does not lead to Eqs.~(\ref{eq:reason_unitary_corep},\ref{eq:reason_unitary_corep2}), and  $v_{ij}$ and the canonical element $\mathbf{c}$ satisfy a weaker unitarity condition that we will derive in the next section. 

\subsection{The projection operator to the solvable subspace}\label{app:WHAprojector}
The goal of this section is to prove Eqs.~(\ref{eq:WHAUP}-\ref{eq:UPCommuteInv}) of Sec.~\ref{sec:circuitdef}, which essentially generalize Thm~\ref{thm:HopfUgate} to $\C^*$-weak Hopf algebras. To this end we first prove a number of abstract algebraic relations in the $\C^*$ weak Hopf algebra $\A$ and then use the $*$-algebra representations $\rho$ and $v$ to map these algebraic relations into operator equations in the physical Hilbert space. 

For a weak Hopf algebra, the derivation in Eq.~\eqref{proof:CEinverse} is still valid, but in this case, we apply the weak antipode axiom Eq.~\eqref{def:WHAantipode_axiom} to the last line of Eq.~\eqref{proof:CEinverse}, and obtain
\begin{eqnarray}
\mathbf{c}\cdot \mathbf{c}'
&=&\sum_{z} z_{(1)} S(z_{(2)})\otimes \deltad_z\nonumber\\
&=&\sum_{z} \epsilon(1_{(1)} z) 1_{(2)}\otimes\deltad_z\nonumber\\
&=&%
1_{(2)}\otimes\epsilon(1_{(1)} ?).
\end{eqnarray}
Similarly,
\begin{eqnarray}
	\mathbf{c}'\cdot \mathbf{c}=%
	1_{(1)}\otimes\epsilon(? 1_{(2)}).
\end{eqnarray}
Now define 
\begin{eqnarray}\label{def:pqidempotent}
\mathbf{p}&=&\Pi(\mathbf{c}\cdot \mathbf{c}')=%
\epsilon(1_{(1)} ?)\otimes 1_{(2)}\in \A^*\otimes\A,\nonumber\\
\mathbf{q}&=&\mathbf{c}'\cdot \mathbf{c}=%
1_{(1)}\otimes\epsilon(? 1_{(2)})\in \A\otimes\A^*,
\end{eqnarray}
where $\Pi:\A\otimes\A^*\to\A^*\otimes\A$ is the swap operator, i.e. $\Pi(x\otimes f)=(f\otimes x),x\in\A,f\in \A^*$. In the following we prove that both $\mathbf{p}$ and $\mathbf{q}$ are Hermitian idempotent, i.e., 
\begin{eqnarray}\label{eq:Hermidemp}
\mathbf{p}^2&=&\mathbf{p}=\mathbf{p}^*,\nonumber\\
\mathbf{q}^2&=&\mathbf{q}=\mathbf{q}^*.
\end{eqnarray}
We first prove $\mathbf{p}^2=\mathbf{p}$. We have
\begin{eqnarray}\label{proof:pidempotent}
\mathbf{p}^2&=&%
\epsilon(1_{(1)} ?) \epsilon(1_{(1')} ?)\otimes1_{(2)}1_{(2')}\nonumber\\
&=&%
\epsilon(1_{(1)} ?) \epsilon[\epsilon_s(1_{(2)}) ?]\otimes1_{(3)}\nonumber\\
&=&%
\epsilon(1_{(1)} ?) \epsilon(1_{(2)} ?)\otimes1_{(3)}\nonumber\\
&=&%
\epsilon(1_{(1)} ?) \otimes1_{(2)}\nonumber\\
&=&\mathbf{p},
\end{eqnarray}
where in the second line we use Eq.~\eqref{eq:unitcombination}, in the third line we use Eq.~\eqref{eq:counitproj},  and in the fourth line we use Eq.~\eqref{eq:epsilonprop}.  Similarly, $\mathbf{q}^2=\mathbf{q}$ is proved as follows
\begin{eqnarray}\label{proof:qidempotent}
	\mathbf{q}^2&=&%
	1_{(1)}1_{(1')}\otimes \epsilon(?1_{(2)} ) \epsilon(?1_{(2')})\nonumber\\
	&=&%
	1_{(1)}\otimes  \epsilon[?\epsilon_t(1_{(2)})]\epsilon(?1_{(3)} )\nonumber\\
	&=&%
	1_{(1)}\otimes  \epsilon[?1_{(2)}]\epsilon(?1_{(3)})\nonumber\\
	&=&%
	1_{(1)}\otimes\epsilon(?1_{(2)})\nonumber\\
	&=&\mathbf{q}.
\end{eqnarray}
Next we prove that $\mathbf{q}=\mathbf{q}^*$. Indeed, for a $\C^*$-weak Hopf algebra, the proof that $\mathbf{c}'=\mathbf{c}^*$ in Remark~\ref{rmk:cunitary} is still valid, therefore, we have $\mathbf{q}^*=(\mathbf{c}^*\mathbf{c})^*%
=\mathbf{c}^*\mathbf{c}=\mathbf{q}$. Similarly, we have $\mathbf{p}=\mathbf{p}^*$.

Now we show the following commutativity properties between $\mathbf{p}$ and $\mathbf{q}$:
\begin{eqnarray}\label{eq:pqcommute}
(\mathbf{p}\otimes \epsilon)\cdot(\epsilon\otimes\mathbf{q})&=&(\epsilon\otimes\mathbf{q})\cdot(\mathbf{p}\otimes \epsilon),\nonumber\\
	(\mathbf{q}\otimes 1)\cdot(1\otimes\mathbf{p})&=&(1\otimes\mathbf{p})\cdot(\mathbf{q}\otimes 1).
\end{eqnarray}
Indeed, Eq.~\eqref{eq:pqcommute} directly follows from Eqs.~(\ref{eq:WBAunitaxiom2},\ref{eq:WBAcounitaxiom2}). For example, the first line of Eq.~\eqref{eq:pqcommute} is proved as follows
\begin{eqnarray}
(\mathbf{p}\otimes \epsilon)\cdot(\epsilon\otimes\mathbf{q})&=&\epsilon(1_{(1)} ?) \otimes1_{(2)}1_{(1')}\otimes\epsilon(?1_{(2')})\nonumber\\
&=&\epsilon(1_{(1)} ?) \otimes1_{(1')}1_{(2)}\otimes\epsilon(?1_{(2')})\nonumber\\
&=&(\epsilon\otimes\mathbf{q})\cdot(\mathbf{p}\otimes \epsilon),
\end{eqnarray}
where in the second line we used  Eq.~\eqref{eq:WBAunitaxiom2}. Similarly, the second line of Eq.~\eqref{eq:pqcommute} follows from  Eq.~\eqref{eq:WBAcounitaxiom2}. 

Next we prove the following identity, which will be important later:
\begin{eqnarray}\label{eq:UPCommute}
&&\sum_{x,y\in B(\A)} 1_{(1')} x \otimes \deltad_x \epsilon(1_{(1)}?) \otimes y 1_{(2)} \otimes \epsilon(? 1_{(2')}) \deltad_y\nonumber\\
&=&\sum_{x,y\in B(\A)}  x \otimes \deltad_x \epsilon(1_{(1)}?) \otimes y 1_{(2)} \otimes  \deltad_y.
\end{eqnarray}
To prove this identity, it is enough to show that both sides are equal after we pair them with $z\otimes w$ in the second and the fourth tensor factors, for any $z,w\in\A$. That is, Eq.~\eqref{eq:UPCommute} is equivalent to 
\begin{eqnarray}\label{eq:UPCommutezwform}
	&&  1_{(1')} z_{(1)} \epsilon(1_{(1)}z_{(2)}) \otimes w_{(2)} 1_{(2)}  \epsilon(w_{(1)}  1_{(2')}) \nonumber\\
	&=&  z_{(1)} \epsilon(1_{(1)}z_{(2)}) \otimes w  1_{(2)}
\end{eqnarray}
for any $z,w\in\A$. In the following we show that both sides of Eq.~\eqref{eq:UPCommutezwform} simplify to  $1_{(1)} z \otimes w  1_{(2)}$. We start with the RHS of Eq.~\eqref{eq:UPCommutezwform}. We have
\begin{eqnarray}\label{eq:UPCommuteRHSderiv}
	&&  z_{(1)} \epsilon(1_{(1)}z_{(2)}) \otimes w  1_{(2)} \nonumber\\
	&=& 1_{(1')} z_{(1)} \epsilon(1_{(1)}1_{(2')}z_{(2)}) \otimes w  1_{(2)} \nonumber\\
	&=& 1_{(1)} z_{(1)} \epsilon(1_{(2)}z_{(2)}) \otimes w  1_{(3)} \nonumber\\
	&=& (1_{(1)} z)_{(1)} \epsilon[(1_{(1)}z)_{(2)}] \otimes w  1_{(2)} \nonumber\\
	&=& 1_{(1)} z \otimes w  1_{(2)}, 
\end{eqnarray}
where in the second line we use $1\cdot z=z$ and the multiplicativity of $\Delta$, in the third line we use Eq.~\eqref{eq:WBAunitaxiom2}, in the fourth line we use the coassociativity and multiplicativity of $\Delta$, and in the last line we use the counit condition in Eq.~\eqref{eq:counitaxiom}. The LHS of Eq.~\eqref{eq:UPCommutezwform} is simplified as 
\begin{eqnarray}
	&&  1_{(1')} z_{(1)} \epsilon(1_{(1)}z_{(2)}) \otimes w_{(2)} 1_{(2)}  \epsilon(w_{(1)}  1_{(2')}) \nonumber\\
	&=& 1_{(1')} 1_{(1)}z  \otimes  \epsilon(w_{(1)}  1_{(2')}) w_{(2)} 1_{(2)}  \nonumber\\
	&=&  1_{(1)}z  \otimes  \epsilon[w_{(1)}  \epsilon_t(1_{(2)})] w_{(2)} 1_{(3)}  \nonumber\\
	&=&  1_{(1)}z  \otimes  \epsilon(w_{(1)}  1_{(2)}) w_{(2)} 1_{(3)}  \nonumber\\
	&=&  1_{(1)}z  \otimes  \epsilon[(w  1_{(2)})_{(1)}] (w 1_{(2)} )_{(2)} \nonumber\\
	&=&  1_{(1)}z  \otimes  w 1_{(2)}  \nonumber\\
\end{eqnarray}
where going from the first to the second line we apply Eq.~\eqref{eq:UPCommuteRHSderiv}, in the third line we use Eq.~\eqref{eq:unitcombination}, in the fourth line we use Eq.~\eqref{eq:counitproj}, in the fifth line we use the coassociativity and multiplicativity of $\Delta$, and in the last line we use the counit condition in Eq.~\eqref{eq:counitaxiom}. This completes the proof of Eq.~\eqref{eq:UPCommute}.

Now let $\rho$ be a $*$-representation  of a $\C^*$-weak Hopf algebra $\A$,  %
and let $v$ be a $*$-representation of the dual weak Hopf algebra $\A^*$. %
By the duality relation in App.~\ref{app:dual_rep_corep}, $v$ is a corepresentation of  $\A$. Let 
\begin{tikzpicture}[baseline={([yshift=-0.5ex]current bounding box.center)}, scale=0.63]
	\Wgatered{0}{0}
\end{tikzpicture}
be the solvable tensor constructed from $\rho$ and $v$ via Eq.~\eqref{def:Urhovbialg}. 
We define the projection operators $\hat{P},\hat{Q}$ in  Eq.~\eqref{eq:WHAUP} as follows
\begin{eqnarray}
\hat{P}&=&(v\otimes\rho)\mathbf{p}\in M_{d_v}(\C)\otimes M_{d_\rho}(\C),\nonumber\\
\hat{Q}&=&(\rho\otimes v)\mathbf{q} \in M_{d_\rho}(\C)\otimes M_{d_v}(\C),
\end{eqnarray}
where in the first line $v\otimes\rho$ is considered as a $*$-representation of $\A^*\otimes\A$, %
and similarly $\rho\otimes v$ is a $*$-representation of $\A\otimes\A^*$. %
More explicitly, their matrix elements are defined as follow
\begin{eqnarray}
	[\hat{P}]^{ia}_{jb}&=&
\begin{tikzpicture}[baseline={([yshift=-0.5ex]current bounding box.center)}, scale=0.63]
	\projectorP{0}{0}
	\quantumindices{-0.5}{0}{i}{j}
	\quantumindices{+0.5}{0}{a}{b}
\end{tikzpicture}=(v_{ij}\otimes\rho_{ab})\mathbf{p}=%
\epsilon(1_{(1)} v_{ij})\rho_{ab} (1_{(2)}),\nonumber\\
{}[\hat{Q}]^{ai}_{bj}&=&
\begin{tikzpicture}[baseline={([yshift=-0.5ex]current bounding box.center)}, scale=0.63]
	\projectorQ{0}{0}
	\quantumindices{+0.5}{0}{i}{j}
	\quantumindices{-0.5}{0}{a}{b}
\end{tikzpicture}=(\rho_{ab}\otimes v_{ij})\mathbf{q}=%
\rho_{ab} (1_{(1)})\epsilon( v_{ij}1_{(2)}). \nonumber\\
\end{eqnarray}
Applying the $*$-representations $v\otimes\rho$ and $\rho\otimes v$ to the first and second lines of Eq.~\eqref{eq:Hermidemp}, respectively, we obtain
\begin{eqnarray}
\hat{P}^2=\hat{P}=\hat{P}^\dagger,\nonumber\\
\hat{Q}^2=\hat{Q}=\hat{Q}^\dagger,
\end{eqnarray}
i.e., $\hat{P}$ and $\hat{Q}$ are both Hermitian projection operators. 
Similarly, applying  $v\otimes\rho$ and $\rho\otimes v$ to the first and second lines of Eq.~\eqref{def:pqidempotent}, respectively, we obtain
\begin{equation}\label{eq:WHAPolarDecomp}
\begin{tikzpicture}[baseline={([yshift=-0.5ex]current bounding box.center)}, scale=0.63]
	\WgateblueInv{0}{0}
	\Wgatered{0}{1}
\end{tikzpicture}=	
\begin{tikzpicture}[baseline={([yshift=-0.5ex]current bounding box.center)}, scale=0.63]
	\projectorP{0}{0}
\end{tikzpicture},\quad 
	\begin{tikzpicture}[baseline={([yshift=-0.5ex]current bounding box.center)}, scale=0.63]
		\WgateblueInv{0}{1}
		\Wgatered{0}{0}
	\end{tikzpicture}=	
	\begin{tikzpicture}[baseline={([yshift=-0.5ex]current bounding box.center)}, scale=0.63]
	\projectorQ{0}{0}
	\end{tikzpicture},
\end{equation}
and Eq.~\eqref{eq:WHAUP} can then be proved using the polar decomposition of %
\begin{tikzpicture}[baseline={([yshift=-0.5ex]current bounding box.center)}, scale=0.63]
	\Wgatered{0}{0}
\end{tikzpicture}.

Furthermore, applying the $*$-representations $v\otimes \rho\otimes v$ and $\rho\otimes v\otimes \rho$ on the first and the second lines of Eq.~\eqref{eq:pqcommute}, respectively, we obtain the  commutativity relation between $\hat{P}$ and $\hat{Q}$ in Eq.~\eqref{eq:UPcommute}. 

We now prove Eq.~\eqref{eq:UPCommuteInv}. We only need to prove the first line, since the second line is obtained by translating the first line by half a unit cell. So we need to prove that
\begin{eqnarray}\label{eq:UPCommuteInvapp}
\hat{\mathbf{Q}}_e\hat{\mathbf{P}}_o\hat{\mathbf{U}}_o\hat{\mathbf{P}}=\hat{\mathbf{U}}_o\hat{\mathbf{P}},
\end{eqnarray}
where we have inserted the definition of $\hat{\mathbf{P}}_{\frac{1}{2}}$ in Eq.~\eqref{eq:total_projector}. Since  $\hat{P}\hat{U}=\hat{U}\hat{Q}$, we have
\begin{eqnarray}\label{eq:UPCommuteInvproof1}
\hat{\mathbf{P}}_o\hat{\mathbf{U}}_o\hat{\mathbf{P}}=\hat{\mathbf{U}}_o\hat{\mathbf{Q}}_o\hat{\mathbf{P}}=\hat{\mathbf{U}}_o\hat{\mathbf{P}},
\end{eqnarray}
where we used Eqs.~(\ref{eq:UPcommute},\ref{eq:total_projector}) and $\hat{\mathbf{Q}}_o^2=\hat{\mathbf{Q}}_o$. 
So we are left with 
\begin{eqnarray}\label{eq:UPCommuteInvapp2}
	\hat{\mathbf{Q}}_e \hat{\mathbf{U}}_o\hat{\mathbf{P}}=\hat{\mathbf{U}}_o\hat{\mathbf{P}}.
\end{eqnarray}
Applying  $\rho\otimes v\otimes \rho\otimes v$~(as a $*$-representation of $\A\otimes\A^*\otimes\A\otimes\A^*$) on both sides of the Eq.~\eqref{eq:UPCommute}, we obtain
\begin{eqnarray}
	\begin{tikzpicture}[baseline={([yshift=-0.5ex]current bounding box.center)}, scale=0.63]
		\projectorP{0}{0}
		\Wgatered{1}{1}
		\Wgatered{-1}{1}
		\projectorQ{0}{2}
	\end{tikzpicture} =%
	\begin{tikzpicture}[baseline={([yshift=-0.5ex]current bounding box.center)}, scale=0.63]
		\projectorP{0}{0}
		\Wgatered{1}{1}
		\Wgatered{-1}{1}
	\end{tikzpicture}.
\end{eqnarray} 
It follows that 
\begin{eqnarray}\label{eq:UPCommuteInvapp3}
\hat{Q}_{j,j+1/2} \hat{\mathbf{U}}_o\hat{\mathbf{P}}=\hat{\mathbf{U}}_o\hat{\mathbf{P}}.
\end{eqnarray}
Applying Eq.~\eqref{eq:UPCommuteInvapp3} for $j=1,2,\ldots,L$, we prove Eq.~\eqref{eq:UPCommuteInvapp2}. This completes the proof of Eq.~\eqref{eq:UPCommuteInv}. 
\subsection{The dimension of the solvable subspace}\label{app:dimension_solvable_space} 
In this section we argue that the dimension of the solvable subspace $D_\N=\mathrm{dim}[\mathrm{P}]=\Tr[\hat{\mathbf{P}}]$ for a length-$\N$ chain grows asymptotically  as 
\begin{equation}\label{eq:DNasymptotic}
D_\N\sim (D_\rho D_v)^\N, \text{ as } L\to\infty,
\end{equation}
where $D_\rho$ and $ D_v$ are the quantum dimension of the representations $\rho$ and $v$ which we define below. 
For the specific model we construct in Sec.~\ref{sec:GoldenModel}, Eq.~\eqref{eq:DNasymptotic} can be proved straightforwardly using recursion relations, and in this case we have $D_\rho=D_v=(\sqrt{5}+1)/2$.
To show Eq.~\eqref{eq:DNasymptotic} in the general case %
requires some knowledge about the representation theory of weak Hopf algebras and the relation to fusion categories, which can be found in Ref.~\cite{Bohm2000WHAII}. In the following we assume this knowledge without giving the basic definitions. In addition, we also need to use some results in App.~\ref{app:othersolvableshapes}. 

The first step is to establish the following equality %
\begin{equation}\label{eq:DLeqUdiamond}
	D_\N=\mathrm{rank}(\hat{U}^{(L)2}_{\Diamond}),
\end{equation}
where the operator $\hat{U}^{(L)}_{\Diamond}$ is defined by the square tensor network in Eq.~\eqref{def:Udiamond}. 
Using Eq.~\eqref{eq:U(t)PBCtkApp} and the first line of Eq.~\eqref{eq:MPOPBCUt} for  the case $k=3$, we see that
\begin{eqnarray}\label{eq:DLleqUdiamond}
D_\N&=&\mathrm{rank}(\hat{\mathbf{P}})\nonumber\\
&=&\mathrm{rank}[\hat{\mathbf{P}}\hat{\mathbf{U}}(t_3)]\nonumber\\
&=&\mathrm{rank}\{\hat{U}^{(L)}_{\mytriangledown}~   
	\hat{U}^{(L)2}_{\Diamond}~
	\hat{U}^{(L)}_{\triangle}\}\nonumber\\
&\leq &	\mathrm{rank}(\hat{U}^{(L)2}_{\Diamond}).
\end{eqnarray}
Meanwhile, looking at the dashed rectangle in Eq.~\eqref{eq:U(t)PBCtkApp}, we see that $\hat{U}^{(n)2}_{\Diamond}$ can be decomposed as
\begin{equation}
\hat{U}^{(L)2}_{\Diamond}=\hat{U}^{(L)}_{\triangle}~\hat{U}^{(L)}_{\Parallelogramm}~\hat{U}^{(L)}_{\mytriangledown},
\end{equation}
where $\hat{U}^{(L)}_{\Parallelogramm}$ corresponds to a tensor network of a parallelogram shape, which turns out to be equal to $\hat{\mathbf{P}}\hat{\mathbf{U}}(t)$ at $t=(L-1)/2$.
Therefore,
\begin{equation}\label{eq:DLgeqUdiamond}
\mathrm{rank}(\hat{U}^{(L)2}_{\Diamond})\leq \mathrm{rank}(\hat{U}^{(L)}_{\Parallelogramm})=\mathrm{rank}(\hat{\mathbf{P}})= D_\N.
\end{equation}
Combining Eqs.~\eqref{eq:DLleqUdiamond} and \eqref{eq:DLgeqUdiamond}, we prove Eq.~\eqref{eq:DLeqUdiamond}. 
To compute the RHS of Eq.~\eqref{eq:DLeqUdiamond}, we apply
the first line of Eq.~\eqref{eq:UdiamondCEk} for the case $k=2$ and obtain
\begin{equation}\label{eq:DLeqUdiamondrankCE2}
	D_\N=\mathrm{rank}(\hat{U}^{(L)2}_{\Diamond})=\mathrm{rank}[(\rho^{(L)}\otimes v^{(L)})(\mathbf{c}^2)].
\end{equation}
As an object of the fusion category Rep$(\A)$~(the category of finite dimensional representations of $\A$~\cite{TenCat_EGNO}), the tensor product representation $\rho^{(L)}$ can always be decomposed as a direct sum 
\begin{equation}
\rho^{(L)}=\rho\otimes\rho\otimes\ldots\otimes \rho\cong\bigoplus_{\sigma\in \mathrm{Rep}(\A)}N^\sigma_{\rho\rho\ldots\rho}\sigma, 
\end{equation}
where $N^\sigma_{\rho\rho\ldots\rho}$ are non-negative integers called fusion multiplicities. Likewise, the fusion category structure of Rep$(\A^*)$ gives a similar decomposition of $v^{(L)}$: 
\begin{equation}
	v^{(L)}=v\otimes v\otimes\ldots\otimes v\cong\bigoplus_{w\in \mathrm{Rep}(\A^*)}\tilde{N}^w_{vv\ldots v}w.
\end{equation}
In general, at large $\N$, the fusion multiplicities grow asymptotically as 
\begin{equation}\label{def:Qdimension}
	N^\sigma_{\rho\rho\ldots\rho}\sim D_\rho^\N, \quad \tilde{N}^{w}_{vv\ldots v}\sim D_v^\N,
\end{equation}
which is indeed the definition of the quantum dimensions $D_\rho$ and $D_v$. The combination of Eqs.~(\ref{eq:DLeqUdiamondrankCE2}~\ref{def:Qdimension}) %
imply the asymptotic behavior of $D_\N$ in Eq.~\eqref{eq:DNasymptotic}. 
\subsection{Example: a $\C^*$-weak Hopf algebra with Fibonacci anyon fusion rules}\label{app:FibonacciWHA}
In the following we present the detailed definition of the $\C^*$-weak Hopf algebra we use to construct the non-dual unitary solvable  circuit in Sec.~\ref{sec:GoldenModel}. This weak Hopf algebra was first introduced in Ref.~\cite{Bohm1996WHALeeYang}, and was used in  Ref.~\cite{Ruiz-de-Alarcon2024MPOAHopfII} to study topological tensor network states. As an algebra, 
$\A_{\mathrm{Fib}}$ is the direct sum
$M_2(\C)\oplus M_3(\C)$, where $M_n(\C)$ is the  algebra of $n\times n$ matrices.
Let $\zeta\in\mathbb{R}$ be the unique positive solution to $z^4 + z^2 -1 = 0$~[i.e. $\zeta^2=(\sqrt{5}-1)/2$ is the golden ratio]
and let $\{e_{1}^{ij}|1\leq i,j\leq 2\}$ be a basis of $M_2(\C)$, where $[e_{1}^{ij}]_{kl}=\deltak_{i,k}\deltak_{j,l}$, and similarly let
$\{e_2^{kl}|1\leq k,l\leq 3\}$ be a basis of $M_3(\C)$. Then,
the comultiplication of $\A_{\mathrm{Fib}}$ is defined by the expressions
\begin{align*}
	\Delta(e_{1}^{11}) := { } & { }%
	e_{1}^{11} \otimes e_{1}^{11} +
	e_{2}^{11} \otimes e_{2}^{22}, \\
	\Delta(e_{1}^{12}) := { } & { }%
	e_{1}^{12} \otimes e_{1}^{12} + 
	\zeta^2 e_{2}^{12} \otimes e_{2}^{21} +
	\zeta   e_{2}^{13} \otimes e_{2}^{23}, \\
	\Delta(e_{1}^{22}) := { } & { }%
	e_{1}^{22} \otimes e_{1}^{22} +
	\zeta^4 e_{2}^{22} \otimes e_{2}^{11} + \\ & 
	\zeta^3 e_{2}^{23} \otimes e_{2}^{13} +
	\zeta^3 e_{2}^{32} \otimes e_{2}^{31} +
	\zeta^2 e_{2}^{33} \otimes e_{2}^{33}, \\
	\Delta(e_{2}^{11}) := { } & { }%
	e_{1}^{11} \otimes e_{2}^{11} +
	e_{2}^{11} \otimes e_{1}^{22} +
	e_{2}^{11} \otimes e_{2}^{33}, \\
	\Delta(e_{2}^{12}) := { } & { }%
	e_{1}^{12} \otimes e_{2}^{12} +
	e_{2}^{12} \otimes e_{1}^{21} +
	e_{2}^{13} \otimes e_{2}^{32}, \\
	\Delta(e_{2}^{13}) := { } & { }%
	e_{1}^{12} \otimes e_{2}^{13} +
	e_{2}^{11} \otimes e_{1}^{22} +
	\zeta   e_{2}^{12} \otimes e_{2}^{31} -
	\zeta^2 e_{2}^{13} \otimes e_{2}^{33}, \\
	\Delta(e_{2}^{22}) := { } & { }%
	e_{0}^{22} \otimes e_{2}^{22} +
	e_{2}^{22} \otimes e_{0}^{11} +
	e_{2}^{33} \otimes e_{2}^{22}, \\
	\Delta(e_{2}^{23}) := { } & { }%
	e_{1}^{22} \otimes e_{2}^{23} +
	e_{2}^{23} \otimes e_{1}^{21} +
	\zeta   e_{2}^{32} \otimes e_{2}^{21} -
	\zeta^2 e_{2}^{33} \otimes e_{2}^{23}, \\
	\Delta(e_{2}^{33}) := { } & { }%
	e_{1}^{22} \otimes e_{2}^{33} + 
	e_{2}^{33} \otimes e_{1}^{22} + 
	\zeta^2 e_{2}^{22}\otimes e_{2}^{11} - \\ &
	\zeta^3 e_{2}^{23}\otimes e_{2}^{13} -
	\zeta^3 e_{2}^{32}\otimes e_{2}^{31} +
	\zeta^4 e_{2}^{33}\otimes e_{2}^{33}.
\end{align*}
The counit $\epsilon$ and the antipode $S$ are defined as
\begin{eqnarray}
\epsilon( e_1^{i j})&=&1, \nonumber\\
\epsilon(e_2^{k l})&=&0,\nonumber\\
 S(e_1^{i j})&=&e_1^{j i}, \nonumber\\
  S\left(e_2^{kl}\right)&=&\zeta^{\mu(k)-\mu(l)} e_2^{\sigma(l) \sigma(k)},
\end{eqnarray}
for all $i, j \in\{1,2\}$ and $k,l \in\{1,2,3\}$. Here $\sigma,\mu$ are permutations of the set $\{1,2,3\}$, where $\sigma$ swaps $1\leftrightarrow2$, and $\mu$ swaps $2\leftrightarrow3$.  
One can check that all weak Hopf algebra axioms in App.~\ref{app:WHAdefinitions} are satisfied. 
Furthermore, $\A_{\mathrm{Fib}}$ is a $*$-algebra where the $*$-involution is given by Hermitian conjugation of matrices. One can check that the $*$-involution defined this way is compatible with the multiplication and comultiplication of $\A_{\mathrm{Fib}}$, turning $\A_{\mathrm{Fib}}$ into a $*$-weak Hopf algebra in the sense of Definition~\ref{def:CWHA}. %
Let $\rho_3$ be the 3-dimensional representation of $\A_{\mathrm{Fib}}$ that maps $e^{ij}_{2}$ to $0$ and $e^{kl}_{3}$ to the corresponding $3\times 3$ matrix in $M_3(\C)$, similarly let $\rho_2$ be the 2-dimensional representation of $\A_{\mathrm{Fib}}$ that maps $e^{kl}_{3}$  to $0$ and $e^{ij}_{2}$  to the corresponding $2\times 2$ matrix in $M_2(\C)$, and let $\rho_5=\rho_2\oplus\rho_3$. Then one can check that $\rho_5$ is a faithful $*$-representation of $\A_{\mathrm{Fib}}$, therefore $\A_{\mathrm{Fib}}$ is a $\C^*$-weak Hopf algebra in the sense of Definition~\ref{def:CWHA}.

Now let $v_{ij}$ be a 3-dimensional corepresentation  of $\A_{\mathrm{Fib}}$ defined by
\begin{equation}
	v=\begin{pmatrix}
		e^{21}_{1} & e^{21}_{2} & e^{31}_2 \\
		\zeta^2 e_2^{12} & e_1^{12} & \zeta e_2^{13} \\
		\zeta e_2^{32} & e_2^{23} & e_1^{22}-\zeta^2 e_2^{33}
	\end{pmatrix}.
\end{equation}
One can check that $v$ is indeed a corepresentation of $\A_{\mathrm{Fib}}$, i.e. $v$ satisfies Eq.~\eqref{def:corep}.  Inserting the representation $\rho_3$ and the corepresentation $v$ into Eq.~\eqref{def:Urhovbialg}, we obtain the solvable gate $\hat{U}$ and all other tensors we need for the exact solution. 

All the algebraic data mentioned above are available in the accompanying Mathematica code~\cite{HAQCACode}, which also automatically constructs all the tensors in Eq.~\eqref{def:Urhovbialg} and solves the quench dynamics of the model.

\section{Transfer matrices for the computation of Renyi entanglement entropy}\label{app:TMRenyi}
In this section we provide explicit expressions for the Renyi entanglement entropy we studied in Sec.~\ref{sec:RenyiEE}, using the transfer matrix formalism. These expressions are used in the accompanying Mathematica code~\cite{HAQCACode} for the exact numerical computation of the Renyi entanglement entropy.

We first consider the late time case $l\leq 2t$, as drawn in Eq.~\eqref{eq:DMMPOFinSubs}, where $l=L_A$ is the subsystem size.
For a small subsystem, at any index $\alpha$, we use the first method in Sec.~\ref{sec:RenyiFinite}. For example, for $l=2$, we have
\begin{eqnarray}\label{eq:ReducedDMTMSmallsystem}
	[\hat{\rho}'_A(t)]^{a_1,a_2,i_1,i_2}_{b_1,b_2,j_1,j_2}&=&\llparenthesis K_L|[\tilde{T}_\rho]^{a_1b_1} T_v [\tilde{T}_\rho]^{a_2b_2} T_v (T_\rho T_v)^{2t-l}\nonumber\\
	&&{}\cdot T_\rho[\tilde{T}_v]^{i_1j_1} T_\rho [\tilde{T}_v]^{i_2j_2} |K_R\rrparenthesis,
\end{eqnarray}\label{eq:RenyiTMSmallsystem}
where the transfer matrices are defined as
\begin{equation}
	[\tilde{T}_{\rho}]^{ab}=
	\begin{tikzpicture}[baseline={([yshift=-0.5ex]current bounding box.center)}, scale=0.7]
		\rhotensor{0}{0}
		\rhotensorb{0}{2}
		\MYtriangle{0}{-0.5}
		\MYtriangleI{0}{2.5}
\node at (0,1.5)[below=-0.05] {\footnotesize $b$};
\node at (0,0.5)[above=-0.05] {\footnotesize $a$};
	\end{tikzpicture},\quad
		[\tilde{T}_{v}]^{ij}=
	\begin{tikzpicture}[baseline={([yshift=-0.5ex]current bounding box.center)}, scale=0.7]
		\vtensor{0}{0}
		\vtensorb{0}{2}
		\MYtriangle{0}{-0.5}
		\MYtriangleI{0}{2.5}
\node at (0,1.5)[below=-0.05] {\footnotesize $j$};
\node at (0,0.5)[above=-0.05] {\footnotesize $i$};
	\end{tikzpicture},
\end{equation}
and $T_\rho=\sum_a[\tilde{T}_{\rho}]^{aa}$, $T_v=\sum_i[\tilde{T}_{v}]^{ii}$. The boundary vectors $\llparenthesis K_L|$ and $|K_R\rrparenthesis$ are the same as given  in Eq.~\eqref{eq:OtTMTTKK} where $\Lambda_L=\Lambda_R=1$ for a product state. 
Eq.~\eqref{eq:ReducedDMTMSmallsystem} is straightforwardly generalized to arbitrary $l$.

For large subsystem, with a small index $\alpha$, we use the second method in Sec.~\ref{sec:RenyiFinite} and obtain
\begin{eqnarray}
	\mathrm{Tr}[\hat{\rho}_A(t)^\alpha]&=&\llparenthesis K_L^{[\alpha]}|(T^{\prime [\alpha]}_\rho T^{[\alpha]}_v)^{l}  (T^{[\alpha]}_\rho T^{[\alpha]}_v)^{2t-l}\nonumber\\
	&&{}\cdot(T^{[\alpha]}_\rho T^{\prime[\alpha]}_v)^{l}|K_R^{[\alpha]}\rrparenthesis,
\end{eqnarray}
Here the various transfer matrices are operators acting on the virtual space $\A^{\otimes 2\alpha}$,  defined as follows
\begin{eqnarray}
	T_\rho^{[\alpha]}&=&(T_{\rho})^{\otimes\alpha},\nonumber\\
	T_v^{[\alpha]}&=&(T_{v})^{\otimes\alpha}, \nonumber\\
	T^{\prime [\alpha]}_\rho&=&\mathcal{T} (T'_{\rho})^{\otimes\alpha}\mathcal{T}^{-1},\nonumber\\
	T^{\prime [\alpha]}_v&=&\mathcal{T} (T'_{v})^{\otimes\alpha}\mathcal{T}^{-1},
\end{eqnarray}
where $\mathcal{T}$ translates the virtual space by one site~(the virtual space $\A^{\otimes 2\alpha}$ has $2\alpha$ sites in total), and 
\begin{alignat}{2}
	T'_{\rho}&=\sum_a
	\begin{tikzpicture}[baseline={([yshift=-0.5ex]current bounding box.center)}, scale=0.7]
		\rhotensor{0}{1-.25}
		\MYtriangle{0}{0.5-.25}
		\rhotensorb{0}{-1+.25}
		\MYtriangleI{0}{-0.5+.25}
		\node at (0,1.5-.25) [above] {\footnotesize $a$};
		\node at (0,-1.5+.25) [below] {\footnotesize $a$};
	\end{tikzpicture}=
	\begin{tikzpicture}[baseline={([yshift=-0.5ex]current bounding box.center)}, scale=0.7]
		\rhotensor{0}{-0.5}
		\rhotensorb{0}{0.5}
		\MYtriangle{0}{-1}
		\MYtriangleI{0}{1}
		\swaptensor{1.}{0.}{0.5}
		\swaptensor{-1.}{0.}{0.5}
	\end{tikzpicture},\nonumber\\
	T'_{v}&=\sum_i
	\begin{tikzpicture}[baseline={([yshift=-0.5ex]current bounding box.center)}, scale=0.7]
		\vtensor{0}{1-.25}
		\MYtriangle{0}{0.5-.25}
		\vtensorb{0}{-1+.25}
		\MYtriangleI{0}{-0.5+.25}
		\node at (0,1.5-.25) [above] {\footnotesize $i$};
		\node at (0,-1.5+.25) [below] {\footnotesize $i$};
	\end{tikzpicture}=
	\begin{tikzpicture}[baseline={([yshift=-0.5ex]current bounding box.center)}, scale=0.7]
		\vtensor{0}{-0.5}
		\vtensorb{0}{0.5}
		\MYtriangle{0}{-1}
		\MYtriangleI{0}{1}
		\swaptensor{1.}{0.}{0.5}
		\swaptensor{-1.}{0.}{0.5}
	\end{tikzpicture}.%
\end{alignat}
where $
\begin{tikzpicture}[baseline={([yshift=-0.5ex]current bounding box.center)}, scale=0.7]
	\swaptensor{0}{0.}{0.5}
	\gateInd{0}{0}{$x$}{$y$}{$z$}{$w$}
\end{tikzpicture}=\deltak_{y,z}\deltak_{x,w}$ is the swap tensor. Furthermore, the two boundary vectors 
 $\llparenthesis K_L^{[\alpha]}|$ and $|K_R^{[\alpha]}\rrparenthesis $ are defined as
\begin{eqnarray}
\llparenthesis K_L^{[\alpha]}|&=&\llparenthesis K_L|^{\otimes\alpha},\nonumber\\
|K_R^{[\alpha]}\rrparenthesis&=&|K_R\rrparenthesis^{\otimes\alpha}.
\end{eqnarray}

We now consider the early time case $l\geq 2t$. In this case Eq.~\eqref{eq:DMMPOFinSubs} has the following tensor network representation instead
\begin{eqnarray}\label{eq:DMMPOFinSubsSmallt}
	\hat{\rho}^\prime_A(t)&=&\hat{U}_A\hat{\rho}_A(t)\hat{U}^\dagger_A\\
	&=&~~
	\begin{tikzpicture}[baseline={([yshift=-0.5ex]current bounding box.center)}, scale=0.6]
		\lowertriangularTN{-1}{0}{\Wgatered}{\numt}
		\uppertriangularTN{-1}{0}{\WgateblueInv}{\numt}	
		\legindices{{b_1,b_2,\ldots,b_l}}{0.5}{0.5}{1 }{1}{below right=-0.2}
		\legindices{{a_1,a_2,\ldots,a_l}}{0.5}{-0.5}{1 }{-1}{above right=-0.2}
		\legindices{{i_l,\ldots,i_2,i_1}}{-0.5}{-0.5}{-1 }{-1}{above left=-0.2}
		\legindices{{j_l,\ldots,j_2,j_1}}{-0.5}{0.5}{-1 }{1}{below left=-0.2}
		\foreach \i in {2,...,\numm}{%
			\draw[very thick, draw=\colorR] (-2.5-\i, -2.51-\i) -- (-2.5-\i,+2.51+\i);
			\draw[very thick, draw=\colorL] (2.5+\i, -2.51-\i) -- (2.5+\i,+2.51+\i);
		}
		\foreach \i in {0,...,\num}{
			\MYtriangle{-0.5-\i}{-\num-1.6}
			\MYtriangle{+0.5+\i}{-\num-1.6}
			\MYtriangleI{-0.5-\i}{\num+1.6}
			\MYtriangleI{+0.5+\i}{\num+1.6}
		}
	\end{tikzpicture}\nonumber\\
	&=&\!
	\begin{tikzpicture}[baseline={([yshift=-0.5ex]current bounding box.center)}, scale=0.6]
		\manyrhov{-0.5}{-4}{\numt}
		\manyrhovb{-0.5}{-2}{\numt}
		\legindices{{b_1,b_2,\ldots,b_l}}{-0.5}{-2.5}{2 }{0}{below =-0.15}
		\legindices{{a_1,a_2,\ldots,a_l}}{-0.5}{-3.5}{2 }{0}{above =-0.15}
		\legindices{{i_1,i_2,\ldots,i_l}}{\numx-6.5}{-3.5}{2}{0}{above =-0.15}
		\legindices{{j_1,j_2,\ldots,j_l}}{\numx-6.5}{-2.5}{2}{0}{below =-0.15}
			\MYsquare{\numx}{-2}
			\MYsquareB{-1}{-2}
		\MYsquare{\numx}{-4}
		\MYsquareB{-1}{-4}
		
		\foreach \i in {0,...,\numx}{
			\MYtriangle{-0.5+\i}{-4.6}
			\MYtriangleI{-0.5+\i}{-1.4}
		}
		\foreach \i in {2,...,\numm}{
			\draw[very thick, draw=\colorR] (2*\numm+1-0.5-2*\i, -3.51) -- (2*\numm+1-0.5-2*\i,+0.51-3);
			\draw[very thick, draw=\colorL] (3.5+2*\i, -3.51) -- (3.5+2*\i,+0.51-3);
		}
	\end{tikzpicture}.\nonumber
\end{eqnarray}
For large subsystem, with a small index $\alpha$, we use the second method in Sec.~\ref{sec:RenyiFinite} and obtain
\begin{eqnarray}
	\mathrm{Tr}[\hat{\rho}_A(t)^\alpha]&=&\llparenthesis K_L^{[\alpha]}|(T^{\prime [\alpha]}_\rho T_v^{[\alpha]})^{2t}  (T^{\prime [\alpha]}_\rho T^{\prime [\alpha]}_v)^{l-2t}\nonumber\\
	&&(T_\rho^{[\alpha]} T^{\prime [\alpha]}_v)^{2t}|K_R^{[\alpha]}\rrparenthesis.
\end{eqnarray} 
We do not provide the expression for $\hat{\rho}'_A(t)$ for a small subsystem at early time, since that case can be easily computed using exact diagonalization. 

For a semi-infinite chain studied in Sec.~\ref{sec:RenyiSemiinfinite}, the tensor network representation in Eq.~\eqref{eq:reducedDMMPO} leads to
\begin{eqnarray}
	\mathrm{Tr}[\hat{\rho}_A(t)^\alpha]=\llparenthesis K_L^{[\alpha]}|(T^{\prime [\alpha]}_\rho T_v^{[\alpha]})^{2t} |K_R^{[\alpha]}\rrparenthesis.
\end{eqnarray} 
\section{The PBC evolution operator}\label{app:PBCevol}
The goal of this section is to prove the MPO representation of the PBC evolution operator in Eq.~\eqref{eq:MPOPBCUt0} of Sec.~\ref{sec:PBCsolution}, and  derive the finite system revival time in Eq.~\eqref{eq:trevHA-0} in the process. As a preparation step, in App.~\ref{app:othersolvableshapes} we first generalize Eq.~\eqref{eq:triangularTNMPO} to obtain MPO representations of 2D tensor networks of some other shapes, and then in App.~\ref{app:MPOPBCUt} we prove  Eq.~\eqref{eq:MPOPBCUt0}, and in App.~\ref{app:recurrencetime} we relate the revival time to the exponent of the underlying Hopf algebra and prove Eq.~\eqref{eq:trevHA-0}.
\subsection{More on MPO representation of 2D tensor networks}\label{app:othersolvableshapes}
\subsubsection{The diamond shape}\label{app:squareshape}
We first derive the analog of Eq.~\eqref{eq:triangularTNMPO} for the following diamond-shaped 2D tensor network
\begin{eqnarray}\label{def:Udiamond}
[\hat{U}^{(n)}_{\Diamond}]^{\mathbf{a},\mathbf{i}}_{\mathbf{b},\mathbf{j}}
&=& 
\begin{tikzpicture}[baseline={([yshift=-0.5ex]current bounding box.center)}, scale=0.63]
	\squareWTN{-1}{-5}{4}{4}
	\legindices{{i_1,i_2,\ldots,i_n}}{-4.5}{-3.5-4}{1}{1}{above left=-0.14}
	\legindices{{a_1,a_2,\ldots,a_n}}{-0.5}{-4.5}{1}{-1}{above right=-0.14}
	\legindices{{j_1,j_2,\ldots,j_n}}{-0.5}{-11.5}{1}{1}{below right=-0.14}
	\legindices{{b_1,b_2,\ldots,b_n}}{-4.5}{-8.5}{1}{-1}{below left=-0.14}
\end{tikzpicture}\nonumber\\
&=&%
\begin{tikzpicture}[baseline={([yshift=-0.5ex]current bounding box.center)}, scale=0.63]
	\manyrho{-3.5}{-4}{4}
	\manyv{0.5}{-4}{4}
	\MYsquare{4}{-4}
	\MYsquareB{-4}{-4}
	\hindices{{a_1,a_2,\ldots,a_n,i_1,i_2,\ldots,i_n}}{-3.5}{-3.5}{above}
	\hindices{{b_1,b_2,\ldots,b_n,j_1,j_2,\ldots,j_n}}{-3.5}{-4.5}{below}
\end{tikzpicture}.~
\end{eqnarray}
Eq.~\eqref{def:Udiamond} is proved using elementary tensor network manipulations similar to the derivation in Eq.~\eqref{eq:triangularTNMPOderive}:
\begin{eqnarray}\label{eq:rectangleTNMPOderive}
	&&\begin{tikzpicture}[baseline={([yshift=-0.5ex]current bounding box.center)}, scale=0.63]
		\rhotriang{-3.5}{-4}
		\vtriang{3.5}{-4}
		\manyrho{-2.5}{-4}{3}
		\manyv{0.5}{-4}{3}
		\hindices{{a_1,a_2,\ldots,a_n,i_1,i_2,\ldots,i_n}}{-3.5}{-3.5}{above}
		\hindices{{b_1,b_2,\ldots,b_n,j_1,j_2,\ldots,j_n}}{-3.5}{-4.5}{below}
	\end{tikzpicture}\nonumber\\
	&=&
	\begin{tikzpicture}[baseline={([yshift=-0.5ex]current bounding box.center)}, scale=0.63]
		\rhotriangB{-2.5}{-4 }
		\vtriangB{2.5}{-4 }
		\manyvrho{-1.5}{-4}{2}
		\uppertriangularTN{-1}{-9}{\Wgatered}{4}
		\hindices{{ ,a_1,i_2,a_2,\ldots,\ldots,i_n, }}{-3.5}{-3.5}{above}
		\hindices{{i_1~, , , , , , ,~a_{n}}}{-3.5}{-4.6}{above}
		\legindices{{b_1,b_2,\ldots,b_n}}{-3.5}{-5.5}{1}{-1}{below left=-0.14}
		\legindices{{j_1,j_2,\ldots,j_n}}{0.5}{-8.5}{1}{1}{below right=-0.14}
	\end{tikzpicture}\nonumber\\
	&=&
	\begin{tikzpicture}[baseline={([yshift=-0.5ex]current bounding box.center)}, scale=0.63]
		\squareWTN{-1}{-5}{4}{4}
		\legindices{{i_1,i_2,\ldots,i_n}}{-4.5}{-3.5-4}{1}{1}{above left=-0.14}
		\legindices{{a_1,a_2,\ldots,a_n}}{-0.5}{-4.5}{1}{-1}{above right=-0.14}
		\legindices{{j_1,j_2,\ldots,j_n}}{-0.5}{-11.5}{1}{1}{below right=-0.14}
		\legindices{{b_1,b_2,\ldots,b_n}}{-4.5}{-8.5}{1}{-1}{below left=-0.14}
	\end{tikzpicture}.
\end{eqnarray}

We now derive an algebraic representation for the second line of Eq.~\eqref{def:Udiamond}, %
which will be useful later. %
We have
\begin{eqnarray}\label{eq:Udiamondderivation}
	&&%
	\begin{tikzpicture}[baseline={([yshift=-0.5ex]current bounding box.center)}, scale=0.63]
		\manyrho{-3.5}{-4}{4}
		\manyv{0.5}{-4}{4}
		\MYsquare{4}{-4}
		\MYsquareB{-4}{-4}
		\hindices{{a_1,a_2,\ldots,a_n,i_1,i_2,\ldots,i_n}}{-3.5}{-3.5}{above}
		\hindices{{b_1,b_2,\ldots,b_n,j_1,j_2,\ldots,j_n}}{-3.5}{-4.5}{below}
	\end{tikzpicture}\\
	&=&\sum_{x\in B(\A)}
	\begin{tikzpicture}[baseline={([yshift=-0.5ex]current bounding box.center)}, scale=0.63]
		\manyrho{-3.5}{-4}{4}
		\manyv{1.5}{-4}{4}
		\MYsquare{5}{-4}
		\node at (0,-4) [right=-0.1] {\footnotesize $x$};
		\node at (1,-4) [left=-0.1] {\footnotesize $x$};
		\MYsquareB{-4}{-4}
		\hindices{{a_1,a_2,\ldots,a_n,,i_1,i_2,\ldots,i_n}}{-3.5}{-3.5}{above}
		\hindices{{b_1,b_2,\ldots,b_n,,j_1,j_2,\ldots,j_n}}{-3.5}{-4.5}{below}
	\end{tikzpicture}.\nonumber
\end{eqnarray}
The first part is simplified as
\begin{eqnarray}\label{eq:rhoMPOderivation}
	&&
	\begin{tikzpicture}[baseline={([yshift=-0.5ex]current bounding box.center)}, scale=0.63]
		\manyrho{-3.5}{-4}{4}
		\MYsquareB{-4}{-4}
		\hindices{{a_1,a_2,\ldots,a_n}}{-3.5}{-3.5}{above}
		\hindices{{b_1,b_2,\ldots,b_n}}{-3.5}{-4.5}{below}
			\node at (0,-4) [right=-0.1] {\footnotesize $x$};
	\end{tikzpicture}\nonumber\\
	&=&\left(\epsilon\otimes\rho_{a_1 b_1}\otimes\ldots \otimes\rho_{a_n b_n}\right)\left[\Delta^{(n)}(x)\right]\nonumber\\
	&=&\left(\rho_{a_1 b_1}\otimes\ldots \otimes\rho_{a_n b_n}\right)\left[\Delta^{(n-1)}(x)\right]\nonumber\\
&\equiv &\rho^{(n)}_{\mathbf{a}\mathbf{b}}(x)
\end{eqnarray}
where  $\rho^{(n)}$ is the representation of $\A$ constructed from the $n$-fold tensor product of $\rho$. Similarly, for the second part in the RHS of Eq.~\eqref{eq:Udiamondderivation}, we have
\begin{eqnarray}\label{eq:vMPOderivation}
	&&
	\begin{tikzpicture}[baseline={([yshift=-0.5ex]current bounding box.center)}, scale=0.63]
		\manyv{0.5}{-4}{4}
		\MYsquare{4}{-4}
		\hindices{{i_1,i_2,\ldots,i_n}}{0.5}{-3.5}{above}
		\hindices{{j_1,j_2,\ldots,j_n}}{0.5}{-4.5}{below}
		\node at (0,-4) [left=-0.1] {\footnotesize $x$};
	\end{tikzpicture}\nonumber\\
	&=&\deltad_x(v_{i_n j_n}\ldots v_{i_2 j_2}v_{i_1 j_1})\nonumber\\
	&=&\Delta^{(n-1)}(\deltad_x)(v_{i_n j_n}\otimes\ldots\otimes v_{i_1 j_1})\nonumber\\
&\equiv&v^{(n)}_{\mathbf{i}\mathbf{j}}(\deltad_x),
\end{eqnarray}
where %
in the third line we use Eq.~\eqref{eq:dualalgebra}, and in the last line we use Eq.~\eqref{eq:repAstarduality}. Here the corepresentation $v_{ij}$ of $\A$ is regarded as a representation of $\A^*$ via the duality introduced in App.~\ref{app:dual_rep_corep}, and $v^{(n)}$ is the representation of $\A^*$ constructed from the $n$-fold tensor product  of $v$. 
Therefore,
\begin{eqnarray}\label{eq:UdiamondCE}
[\hat{U}^{(n)}_{\Diamond}]^{\mathbf{a},\mathbf{i}}_{\mathbf{b},\mathbf{j}}&=&\sum_{x\in B(\A)} \rho^{(n)}_{\mathbf{a}\mathbf{b}}(x)\otimes v^{(n)}_{\mathbf{i}\mathbf{j}}(\deltad_x)\nonumber\\
&=&(\rho^{(n)}_{\mathbf{a}\mathbf{b}}\otimes v^{(n)}_{\mathbf{i}\mathbf{j}})\mathbf{c}.
\end{eqnarray}

It turns out that the $k$-th matrix power of $\hat{U}^{(n)}_{\Diamond}$, $\hat{U}^{(n)k}_{\Diamond}$ for any positive integer $k$ also has a simple MPO representation, which will be useful later. To derive this, notice that since $\rho^{(n)}\otimes v^{(n)}$ is an algebra homomorphism, we have   
\begin{eqnarray}\label{eq:UdiamondCEk}
	[\hat{U}^{(n)k}_{\Diamond}]^{\mathbf{a},\mathbf{i}}_{\mathbf{b},\mathbf{j}} &=&(\rho^{(n)}_{\mathbf{a}\mathbf{b}}\otimes v^{(n)}_{\mathbf{i}\mathbf{j}})(\mathbf{c}^k)\\
	&=&\sum_{x,y\in B(\A)} c^k_{xy} \rho^{(n)}_{\mathbf{a}\mathbf{b}}(x)\otimes v^{(n)}_{\mathbf{i}\mathbf{j}}(\deltad_y)\nonumber\\
&=&\sum_{x,y}c^k_{xy}
\begin{tikzpicture}[baseline={([yshift=-0.5ex]current bounding box.center)}, scale=0.63]
	\manyrho{-3.5}{-4}{4}
	\manyv{1.5}{-4}{4}
	\MYsquare{5}{-4}
	\node at (0,-4) [right=-0.1] {\footnotesize $x$};
	\node at (1,-4) [left=-0.1] {\footnotesize $y$};
	\MYsquareB{-4}{-4}
	\hindices{{a_1,a_2,\ldots,a_n,,i_1,i_2,\ldots,i_n}}{-3.5}{-3.5}{above}
	\hindices{{b_1,b_2,\ldots,b_n,,j_1,j_2,\ldots,j_n}}{-3.5}{-4.5}{below}
\end{tikzpicture}\nonumber\\
&=&	\sum_{x,y}c^k_{xy} 
\begin{tikzpicture}[baseline={([yshift=-0.5ex]current bounding box.center)}, scale=0.63]
	\foreach \i in {0,...,3}
	{
		\rhootimesv{2*\i-3}{0}
	}
	\MYsquare{4}{-0.2}
	\MYsquareB{-4}{0.2}
	\hindices{{i_1,a_1,i_2,a_2,\ldots,\ldots,i_n,a_n}}{-3.5}{0.5}{above}
	\hindices{{b_1,j_1,b_2,j_2,\ldots,\ldots,b_n,j_n}}{-3.5}{-0.5}{below}
	\node at (4,0.2) [right=-0.1] {\footnotesize $x$};
	\node at (-4,-0.2) [left=-0.1] {\footnotesize $y$};
\end{tikzpicture},\nonumber
\end{eqnarray}
where the coefficients $c^k_{xy}$ are defined by 
\begin{equation}\label{def:ckxy}
\mathbf{c}^k=\sum_{x,y\in B(\A)}c^k_{xy} x\otimes \deltad_y,
\end{equation}
which can be efficiently computed in the algebra $\A\otimes\A^*$ for any $k$, and 
\begin{equation}
	\begin{tikzpicture}[baseline={([yshift=-0.5ex]current bounding box.center)}, scale=0.63]
		\rhootimesv{0}{0}
		\quantumindices{-0.5}{0}{i}{b}
		\quantumindices{0.5}{0}{a}{j}
		\node at (-1,0.2) [left=-0.1] {\footnotesize $x$};
		\node at (1,0.2) [right=-0.1] {\footnotesize $y$};
		\node at (-1,-0.2) [left=-0.1] {\footnotesize $z$};
		\node at (1,-0.2) [right=-0.1] {\footnotesize $w$};
	\end{tikzpicture}=
	\begin{tikzpicture}[baseline={([yshift=-0.5ex]current bounding box.center)}, scale=0.63]
		\rhotensor{0}{0}%
		\quantumindices{0}{0}{a}{b}
		\paraindices{0}{0}{x}{y}
	\end{tikzpicture}\times 
	\begin{tikzpicture}[baseline={([yshift=-0.5ex]current bounding box.center)}, scale=0.63]
		\vtensor{0}{0}%
		\quantumindices{0}{0}{i}{j}
		\paraindices{0}{0}{z}{w}
	\end{tikzpicture}.
\end{equation}
(We arrange the indices in this particular way for later convenience.)

\subsubsection{The inverted triangle shape}\label{app:invertedtriangleshape}
Next we derive an MPO representation for the following 2D tensor network of an inverted triangle shape
\begin{equation}\label{eq:inverttriangularTNdef}
[\hat{U}^{(n)}_{\mytriangledown}]^{\mathbf{a},\mathbf{i}}_{\mathbf{b},\mathbf{j}}:=
\begin{tikzpicture}[baseline={([yshift=-0.5ex]current bounding box.center)}, scale=0.63]
		\uppertriangularTN{-1}{-9}{\Wgatered}{4}
		\hindices{{i_1,a_1,i_2,a_2,\ldots,\ldots,i_n,a_{n}}}{-3.5}{-4.5}{above}
		\legindices{{b_1,b_2,\ldots,b_n}}{-3.5}{-5.5}{1}{-1}{below left=-0.14}
		\legindices{{j_1,j_2,\ldots,j_n}}{0.5}{-8.5}{1}{1}{below right=-0.14}
	\end{tikzpicture}.
\end{equation}
The trick is to construct two tensors 	
\begin{tikzpicture}[baseline={([yshift=-0.5ex]current bounding box.center)}, scale=0.63]
	\rhoptensor{0}{0}%
\end{tikzpicture} and 
\begin{tikzpicture}[baseline={([yshift=-0.5ex]current bounding box.center)}, scale=0.63]
\vptensor{0}{0}%
\end{tikzpicture} satisfying the following equation similar to Eq.~\eqref{eq:MUpentagon}
\begin{equation}\label{eq:MUpentagonL}
	\begin{tikzpicture}[baseline={([yshift=-0.5ex]current bounding box.center)}, scale=0.63]
		\rhoptensor{0.5}{1}%
		\vptensor{-0.5}{1}%
		\hindices{{i,a}}{-0.5}{1.5}{above}
		\hindices{{j,b}}{-0.5}{0.5}{below}
	\end{tikzpicture}	
	=
	\begin{tikzpicture}[baseline={([yshift=-0.5ex]current bounding box.center)}, scale=0.63]
		\rhoptensor{-0.5}{0}%
		\vptensor{0.5}{0}%
		\Wgatered{0}{1}
		\hindices{{i,a}}{-0.5}{1.5}{above}
		\hindices{{b,j}}{-0.5}{-0.5}{below}
	\end{tikzpicture},\quad %
	\begin{tikzpicture}[baseline={([yshift=-0.5ex]current bounding box.center)}, scale=0.63]
		\rhoptensor{0}{0}%
			\MYsquare{-0.5}{0}
	\end{tikzpicture}=
	\begin{tikzpicture}[baseline={([yshift=-0.5ex]current bounding box.center)}, scale=0.63]
		\deltatensor{0}{0}{\colorL}{}{}
		\draw[ultra thick] (0.35, 0 ) -- (0.7,0);
		\MYsquare{0.3}{0}
	\end{tikzpicture}, \quad
	\begin{tikzpicture}[baseline={([yshift=-0.5ex]current bounding box.center)}, scale=0.63]
		\MYsquareB{0.5}{0}
		\vptensor{0}{0}%
	\end{tikzpicture}=
	\begin{tikzpicture}[baseline={([yshift=-0.5ex]current bounding box.center)}, scale=0.63]
		\deltatensor{0}{0}{\colorR}{}{}
		\MYsquareB{-0.3}{0}
		\draw[ultra thick] (-0.35, 0 ) -- (-0.7,0);
	\end{tikzpicture}.			
\end{equation}
Using a derivation similar to that in Eq.~\eqref{eq:triangularTNMPOderive}, we arrive at the following MPO representation
\begin{equation}\label{eq:inverttriangularTNMPO}
	[\hat{U}^{(n)}_{\mytriangledown}]^{\mathbf{a},\mathbf{i}}_{\mathbf{b},\mathbf{j}}=
	\begin{tikzpicture}[baseline={([yshift=-0.5ex]current bounding box.center)}, scale=0.63]
		\manyvrhop{-3.5}{-4}{4}
		\MYsquareB{4}{-4}
		\MYsquare{-4}{-4}
		\hindices{{i_1,a_1,i_2,a_2,\ldots,\ldots,i_n,a_{n}}}{-3.5}{-3.5}{above}
		\hindices{{j_1,b_1,j_2,b_2,\ldots,\ldots,j_n,b_n}}{-3.5}{-4.5}{below}%
	\end{tikzpicture}.
\end{equation}

The tensors 	
\begin{tikzpicture}[baseline={([yshift=-0.5ex]current bounding box.center)}, scale=0.63]
	\rhoptensor{0}{0}%
\end{tikzpicture} and 
\begin{tikzpicture}[baseline={([yshift=-0.5ex]current bounding box.center)}, scale=0.63]
	\vptensor{0}{0}%
\end{tikzpicture} in Eq.~\eqref{eq:MUpentagonL} are constructed in a similar way as  $
\begin{tikzpicture}[baseline={([yshift=-0.5ex]current bounding box.center)}, scale=0.63]
\rhotensor{0}{0}%
\end{tikzpicture}$ and 
$
\begin{tikzpicture}[baseline={([yshift=-0.5ex]current bounding box.center)}, scale=0.63]
\vtensor{0}{0}%
\end{tikzpicture}$ in Eq.~\eqref{def:Urhovbialg}: 
\begin{eqnarray}\label{def:pinkgreentensors}
	\begin{tikzpicture}[baseline={([yshift=-0.5ex]current bounding box.center)}, scale=0.63]
		\rhoptensor{0}{0}%
		\paraindices{0}{0}{y}{x}
		\quantumindices{0}{0}{a}{b}
	\end{tikzpicture}&=&(\rho_{ab}\otimes\deltad_y )\Delta(x) ,\nonumber\\ %
	\begin{tikzpicture}[baseline={([yshift=-0.5ex]current bounding box.center)}, scale=0.63]
		\vptensor{0}{0}%
		\paraindices{0}{0}{y}{x}
		\quantumindices{0}{0}{i}{j}
	\end{tikzpicture}&=&\deltad_y(v_{ij}x),%
\end{eqnarray}
and the	vectors $\begin{tikzpicture}[baseline={([yshift=-0.5ex]current bounding box.center)}, scale=0.63]
	\unittensor{0}{0}
\end{tikzpicture}$, $\begin{tikzpicture}[baseline={([yshift=-0.5ex]current bounding box.center)}, scale=0.63]
	\counittensor{0}{0}
\end{tikzpicture}$ are the same as that given in Eq.~\eqref{def:Urhovbialg}. 
One can prove that the tensors constructed in Eq.~\eqref{def:pinkgreentensors} indeed satisfy Eq.~\eqref{eq:MUpentagonL}, using a similar method as we use in the proof of Thm.~\ref{thm:bialgebraTN} in App.~\ref{app:proofBAsolvableTN}.

\subsection{MPO representation of the PBC evolution operator at $t=t_k$}\label{app:MPOPBCUt}
We are now ready to compute the PBC evolution operator $\hat{\mathbf{U}}(t)$ at $t=t_k=(k\N+1)/2$, for any positive integer $k$. We have
\begin{eqnarray}\label{eq:U(t)PBCtkApp}
\hat{\mathbf{U}}(t_k)&=&\sum_{\mathbf{i}}\hspace{-.6in}
\begin{tikzpicture}[baseline={([yshift=5ex]current bounding box.center)}, scale=0.5]
	\PBCevol{2}{9}{{i_{1,1},\cdots,i_{1,L},i_{2,1},\cdots,i_{2,L},i_{k,1},\cdots,i_{k,L}}}
	\draw[thick, dashed] (-2,1) -- (-2-6,1-6)--(-2-6+5,1-6-5)--(-2+5,1-5)-- cycle;
\end{tikzpicture}%
\end{eqnarray}
The dashed rectangle breaks the tensor network into 3 parts: the upper triangle $\hat{U}^{(L)}_{\mytriangledown}$, the central block, which turns out to be equal to $[\hat{U}^{(L)}_{\Diamond}]^{k-1}$, and the lower triangle $\hat{U}^{(L)}_{\triangle}$. 
Applying Eqs.~(\ref{eq:triangularTNMPO},\ref{def:Udiamond},\ref{eq:inverttriangularTNMPO}), we obtain
\begin{eqnarray}\label{eq:MPOPBCUt}
	\hat{\mathbf{U}}(t_k)
&=&\hat{\mathcal{T}}^k_{L/2} 
\hat{U}^{(L)}_{\mytriangledown}~   
[\hat{U}^{(L)}_{\Diamond}]^{k-1}~
\hat{U}^{(L)}_{\triangle}\\
&=&\!\hat{\mathcal{T}}_{L/2}^k	\!\sum_{x,y}c^{k-1}_{xy}\!  %
\begin{tikzpicture}[baseline={([yshift=-0.5ex]current bounding box.center)}, scale=0.6]
\manyvrhop{-3.5}{1.}{4}
\MYsquareB{4}{1.}
\MYsquare{-4}{1.}

\foreach \i in {0,...,3}
{
\rhootimesv{2*\i-3}{0}
}
\MYsquare{4}{-0.2}
\MYsquareB{-4}{0.2}

\manyrhov{-3.5}{-1.}{4}
\MYsquare{4}{-1.}
\MYsquareB{-4}{-1.}

\node at (4,0.2) [right=-0.1] {\footnotesize $x$};
\node at (-4,-0.2) [left=-0.1] {\footnotesize $y$};
\end{tikzpicture}\nonumber\\
&=&\!\hat{\mathcal{T}}_{L/2}^k \!	\sum_{x,y}c^{k-1}_{xy}\! 
\begin{tikzpicture}[baseline={([yshift=-0.5ex]current bounding box.center)}, scale=0.6]
\foreach \i in {0,...,3}
{
	\rhorhootimesvv{2*\i-3}{0}
}
\MYsquare{4}{-0.2}
\MYsquareB{-4}{0.2}

\MYsquare{4}{-0.6}
\MYsquareB{-4}{-0.6}

\node at (4 ,0.2) [right=-0.1] {\footnotesize $x$};
\node at (-4,-0.2) [left=-0.1] {\footnotesize $y$};

\MYsquareB{4}{0.6}
\MYsquare{-4}{0.6}
\end{tikzpicture},\nonumber
\end{eqnarray}
where  $\hat{\mathcal{T}}_{L/2}$ is the  operator that translates the 1D chain by $L/2$ unit cells~(i.e. $L$ sites) satisfying $\hat{\mathcal{T}}_{L/2}^2=1$, the sum in $x,y$ runs over a basis $B(\A)$ of $\A$, and the MPO tensor in the last line is defined as
\begin{equation}
\begin{tikzpicture}[baseline={([yshift=-0.5ex]current bounding box.center)}, scale=0.63]
		\rhorhootimesvv{0}{0}
\end{tikzpicture}=
\begin{tikzpicture}[baseline={([yshift=-0.5ex]current bounding box.center)}, scale=0.63]
\manyvrhop{0}{1.}{1}
\rhootimesv{0.5}{0}
\manyrhov{0}{-1.}{1}
\end{tikzpicture}.
\end{equation}
This completes the proof of Eq.~\eqref{eq:MPOPBCUt0}. 
\subsection{Recurrence time and the exponent of a (weak) Hopf algebra}\label{app:recurrencetime}
The MPO representation for the PBC evolution operator in Eq.~\eqref{eq:MPOPBCUt} allows us to compute the revival time $t_{\mathrm{rev}}$ of the finite system, which is defined as the period of the evolution operator $\hat{\mathbf{U}}(t)$, i.e., the smallest $t$ such that 
\begin{equation}
\hat{\mathbf{U}}(t_0)=\hat{\mathbf{U}}(t+t_0),
\end{equation}
for some $t_0> 0$. 
In our models, the revival time is determined by the exponent of the underlying Hopf algebra~\cite{etingof1999exponent}:
\begin{theorem}%
\label{thm:HAexponent}
For any finite dimensional $\mathbb{C}^* $-Hopf algebra  $\mathcal{A}$, the canonical element $\mathbf{c}$ defined in Eq.~\eqref{def:CE2} has finite order, i.e., there exists a positive integer $n$ such that 
\begin{equation}\label{eq:CEorder}
\mathbf{c}^{\eta}=1\otimes \epsilon.
\end{equation}
The smallest such $n$ is called the exponent of the Hopf algebra $\A$. Equivalently, $\eta$ is the smallest integer such that
\begin{equation}
	\sum_{(z)} z_{(1)}z_{(2)}\ldots z_{(\eta)}=\epsilon(z)1,\quad\forall z\in \mathcal{A}.
\end{equation}
\end{theorem}
See Ref.~\cite{etingof1999exponent} for a proof. Inserting Eq.~\eqref{eq:CEorder} into Eq.~\eqref{eq:MPOPBCUt}, we obtain~\footnote{In general, the arguments in this section only provides an upper bound for the revival time $t_{\mathrm{rev}}$: the actual revival time of the system must be a divisor of $\eta L$.
In particular, if $\eta$ is even, then  $t_{\mathrm{rev}}$ is actually a divisor of $\eta \N/2$. 
For the specific model we studied in Sec.~\ref{sec:GoldenModel}, we have $\eta=5$, and $t_{\mathrm{rev}}=\eta \N$ indeed gives the actual revival time, as we verified numerically for small system sizes up to $\N=4$.
}
\begin{equation}\label{eq:trevHA}
t_{\mathrm{rev}}=\eta \N,
\end{equation}
that is, the revival time is proportional to the system size, which is often a signature of (super)integrability.  

The generalization of Thm.~\ref{thm:HAexponent} to finite dimensional $\C^*$ weak Hopf algebras does not seem to have appeared in the literature. In this case, we conjecture that there exists a positive integers $\eta,\nu$ such that 
\begin{equation}\label{eq:CEorderWHA}
	\mathbf{c}^{\eta+\nu}=\mathbf{c}^{\nu}.%
\end{equation} 
This conjecture is verified to be correct for the   $\C^*$-WHA in App.~\ref{app:FibonacciWHA}, where we have $\eta=5$ and $\nu=2$~(see the accompanying code~\cite{HAQCACode}). If this conjecture is true in general, then Eq.~\eqref{eq:trevHA} also applies to the solvable unitary circuits constructed from $\C^*$-WHAs. 

\bibliography{/home/lagrenge/Documents/Mendeley_bib/library,HAQCACode}
\end{document}